\documentclass[
reprint,
superscriptaddress,
amsmath,
amssymb,
aps,
prl,
twocolumn]{revtex4-2}
\usepackage{graphicx}
\usepackage{dcolumn}
\usepackage{bm}
\usepackage{physics,mathrsfs}
\usepackage{appendix}
\usepackage[normalem]{ulem}
\usepackage{subfigure}
\usepackage{svg}
\usepackage{csquotes}
\usepackage{bbm}
\usepackage{dsfont}
\usepackage{hyperref}
\usepackage[capitalise]{cleveref}
\usepackage{CJKutf8,orcidlink}
\usepackage{longtable}
\usepackage{amsthm}
\usepackage{algorithm}
\usepackage{algorithmic}

\newtheorem{theorem}{Theorem}
\usepackage{xcolor}
\begin{document}
\begin{CJK*}{UTF8}{gbsn}
\preprint{APS/123-QED}

\title{Optimizing quantum violation yields facet Bell inequalities}

\author{Jin-Fu Chen (陈劲夫)\orcidlink{0000-0002-7207-969X}}
\email{jinfuchen@lorentz.leidenuniv.nl}
\affiliation{Instituut-Lorentz, Universiteit Leiden, P.O. Box 9506, 2300 RA Leiden, The Netherlands}
\affiliation{$\langle aQa^L\rangle$ Applied Quantum Algorithms, Universiteit Leiden, The Netherlands}
\author{Mengyao Hu (胡梦瑶)\orcidlink{0000-0003-2621-3365}}
\email{mengyao@lorentz.leidenuniv.nl}
\affiliation{Instituut-Lorentz, Universiteit Leiden, P.O. Box 9506, 2300 RA Leiden, The Netherlands}
\affiliation{$\langle aQa^L\rangle$ Applied Quantum Algorithms, Universiteit Leiden, The Netherlands}
\affiliation{Centre for Quantum Technologies, National University of Singapore, 3 Science Drive 2, Singapore 117543, Singapore}
\author{Jordi Tura\orcidlink{0000-0002-6123-1422}}
\email{tura@lorentz.leidenuniv.nl}
\affiliation{Instituut-Lorentz, Universiteit Leiden, P.O. Box 9506, 2300 RA Leiden, The Netherlands}
\affiliation{$\langle aQa^L\rangle$ Applied Quantum Algorithms, Universiteit Leiden, The Netherlands}
\date{\today}

\begin{abstract}
Nonlocality is revealed through violations of Bell inequalities, whose complete characterization defines the local polytope. 
For multipartite systems, however, the local polytope becomes rapidly intractable as the number of parties increases. 
A natural strategy is therefore to optimize Bell inequalities with respect to the ratio of their quantum value to their classical bound. 
We prove that a nondegenerate local maximum of this ratio gives a facet Bell inequality, and we propose a gradient-based method that searches for such inequalities iteratively.
Applied to permutation-invariant scenarios, our method produces tight Bell inequalities with robust quantum violations. 
Our method enables certification of nonlocality without complete knowledge of the local polytope. 
More generally, it applies to optimization problems in quantum information involving a classical polytope and a quantum convex set.
\end{abstract}
\maketitle
\end{CJK*}

\emph{Introduction.}---Bell inequalities serve as fundamental tools for detecting quantum nonlocality, the inability of local hidden variable theories to reproduce all quantum predictions \cite{einsteinCanQuantumMechanicalDescription1935,bellEinsteinPodolskyRosen1964}. 
Among them, the Clauser-Horne-Shimony-Holt (CHSH) inequality is the most studied and experimentally tested Bell inequality \cite{clauserProposedExperimentTest1969}. 
It provides a clear signature of nonlocal correlations in bipartite systems~\cite{landauEmpiricalTwopointCorrelation1988,masanesNecessarySufficientCondition2003}, where recent breakthroughs have enabled loophole-free Bell tests~\cite{hensenLoopholefreeBellInequality2015,giustinaSignificantLoopholeFreeTestBells2015,shalmStrongLoopholeFreeTest2015,rosenfeldEventReadyBellTest2017,storzLoopholefreeBellInequality2023}.
Beyond its foundational significance, nonlocality has emerged as a critical resource for device-independent quantum information protocols, including quantum key distribution~\cite{acinDeviceIndependentSecurityQuantum2007, pironioDeviceindependentQuantumKey2009, vaziraniFullyDeviceIndependentQuantum2014, xuSecureQuantumKey2020}, certified randomness generation~\cite{pironioRandomNumbersCertified2010,fyrillasCertifiedRandomnessTight2024,zhangRandomnessNonlocalityMultipleInput2025}, and quantum system verification through self-testing~\cite{supicSelftestingQuantumSystems2020b,liNecessarySufficientCondition2025,baccariScalableBellInequalities2020}. 

Generalizing Bell nonlocality to the multipartite regime poses significant conceptual and practical challenges.
As the number of parties increases, while the settings per party remain fixed, the number of correlation terms grows exponentially, and the complexity of facet enumeration is doubly exponential in $N$~\cite{chazelleOptimalConvexHull1993},
rendering a complete characterization of multipartite nonlocality intractable~\cite{pitowskyQuantumProbabilityQuantum1989, pitowskyCorrelationPolytopesTheir1991, bancalDetectingGenuineMultipartite2011}.
The local polytope, defined as the intersection of all valid Bell inequalities~\cite{froissartConstructiveGeneralizationBells1981}, can be characterized either by its vertices or its facets.
Although generating the vertices is straightforward, the number grows exponentially with the number of parties. 
Deriving the corresponding facets quickly becomes computationally intractable~\cite{chazelleOptimalConvexHull1993,pitowskyOptimalTestsQuantum2001}. 
A practical approach is to restrict to few-body correlators and exploit symmetries, thereby greatly reducing the number of free coefficients in Bell inequalities and constraining the structure of correlations.
For instance, permutation symmetry among parties allows the construction of symmetric Bell inequalities that detect nonlocality in many-body systems~\cite{turaDetectingNonlocalityManybody2014c,  turaNonlocalityManybodyQuantum2015,guoDetectingBellCorrelations2023a,marconiSymmetricQuantumStates2025}.
Symmetry thus provides a promising route to construct scalable and experimentally relevant Bell inequalities for many-body systems~\cite{schmiedBellCorrelationsBoseEinstein2016, engelsenBellCorrelationsSpinSqueezed2017, wangMultidimensionalQuantumEntanglement2018,wangProbingManyBodyBell2025}.

To characterize multipartite Bell nonlocality, it is essential to optimize Bell inequalities to enhance the gap between the classical bound and the quantum value~\cite{jungeOperatorSpaceTheory2010}.
Such optimized inequalities are particularly useful in experiments, as larger separation generally implies greater robustness to noise and improved feasibility in experimental implementations~\cite{wangProbingManyBodyBell2025, gomezOptimalStrategyCertify2021, liImprovedNonlocalityCertification2024a}. 
In addition to enhancing the quantum–classical gap, identifying tight Bell inequalities that define facets of the local polytope is also important, as they set the fundamental limits of classical correlations, and all facets together form the minimal description of correlations consistent with a local hidden-variable model.
However, current methods for identifying these robust and  tight Bell inequalities face significant limitations. 
Recent approaches leveraging machine learning techniques \cite{dengMachineLearningDetection2018} and iterative optimization methods \cite{liImprovedNonlocalityCertification2024a} have shown potential, but these methods remain limited in scalability and cannot systematically explore facets for complex polytopes.

In this Letter, we prove that optimizing the ratio between the quantum value and the classical bound can yield tight Bell inequalities, and introduce a gradient-based method for optimizing Bell inequalities by maximizing this ratio at fixed Hilbert-space dimension.
Applied to permutation-invariant Bell inequalities with dichotomic outcomes and up
to two-body correlators, the method identifies tight inequalities without full
facet enumeration, recovers CHSH in the symmetrized bipartite case, and provides
new tight inequalities for scenarios with more measurement settings. 
The asymptotic behavior and optimization in the thermodynamic limit $N\to\infty$ are presented in the follow-up work~\cite{chenOptimizingCertifyingMultipartite2026}.
More broadly, our method is useful for optimization problems involving a classical polytope and a quantum convex set beyond Bell nonlocality, for instance to contextuality~\cite{budroniKochenSpeckerContextuality2022}, causality~\cite{branciardSimplestCausalInequalities2015},
and magic~\cite{heinrichRobustnessMagicSymmetries2019}.

\emph{Setup.}---We consider the $(N,m,d)$ Bell scenario with $N$ parties, $m$ measurement settings per party, and $d$ outcomes per measurement. A Bell inequality is a linear functional of the probability distribution~\cite{brunnerBellNonlocality2014},
\begin{align}
    I=\vec{T}\cdot\vec{P}=\sum_{\boldsymbol{a},\boldsymbol{x}}T_{\boldsymbol{x}}^{\boldsymbol{a}}P(\boldsymbol{a}\vert\boldsymbol{x})\geq\beta_{C},\label{eq:IPvec}
\end{align}
where the coefficients $T_{\boldsymbol{x}}^{\boldsymbol{a}}\in\mathbb{R}$ specify the inequality, and $P(\boldsymbol{a}\vert\boldsymbol{x})$ denotes the conditional probability of obtaining outcomes $\boldsymbol{a}=(a_1,\dots,a_N)$ with $a_i\in[d]=\{0,\dots,d-1\}$ given measurement settings $\boldsymbol{x}=(x_1,\dots,x_N)$ with $x_i\in[m]=\{0,\dots,m-1\}$.

The local polytope $\mathcal L$ is the convex hull of all deterministic local
strategies. The classical bound is
\begin{align}
\beta_C
=
\min_{\vec P\in\mathcal L} I(\vec P)
=
\min_{\vec v\in\mathrm{vert}(\mathcal L)} I(\vec v),
\end{align}
where $\vec P=\{P(\boldsymbol a|\boldsymbol x)\}_{\boldsymbol a,\boldsymbol x}$
denotes a local behavior and $\mathrm{vert}(\mathcal L)$ denotes the set of
vertices of $\mathcal L$. Since $I$ is linear, its minimum over the polytope is
always attained at one or more vertices. The inequality is tight if the
vertices saturating $\beta_C$ span a facet of $\mathcal L$. Both the number of
vertices and the dimension grow exponentially with $N$~\cite{brunnerBellNonlocality2014},
thus complete facet enumeration is infeasible for large $N$.

For the quantum case, the Bell operator is
\begin{align}
\begin{aligned}
    \hat{I}(\boldsymbol{\theta})&=\sum_{\boldsymbol{a},\boldsymbol{x}}T_{\boldsymbol{x}}^{\boldsymbol{a}}\,\hat{M}_{\boldsymbol{a}|\boldsymbol{x}}(\boldsymbol{\theta}),\\\hat{M}_{\boldsymbol{a}|\boldsymbol{x}}(\boldsymbol{\theta})&=\bigotimes_{j=1}^{N}\hat{M}_{a_{j}|x_{j}}^{(j)}(\theta_{x_{j}}^{(j)}),
\end{aligned}
\end{align}
where $\{\hat{M}_{a_{j}|x_{j}}^{(j)}(\theta_{x_{j}}^{(j)})\}_{a_{j}}$ is a local POVM parameterized by $\theta_{x_j}^{(j)}$. 
Using Born's rule, the probability distribution is given by $P(\boldsymbol{a}|\boldsymbol{x})=\mathrm{tr}(\hat{M}_{\boldsymbol{a}|\boldsymbol{x}}(\boldsymbol{\theta})\rho )$, where $\rho$ is the density matrix of the system.
The set of probability distributions that admit such a representation is known as the quantum set $\mathcal{Q}$~\cite{pitowskyRangeQuantumProbability1986}, and the minimum is the quantum bound  
$\beta_Q^{\mathrm{full}}=\min_{\vec{P}\in \mathcal{Q}}I$.
We restrict the search to probability distributions generated by a variationally parameterized family of states \(\rho\) and measurements \(\hat M_{\boldsymbol a|\boldsymbol x}(\boldsymbol\theta)\), and
the corresponding optimization gives the variational quantum value 
\begin{align}
    \beta_Q
=
\min_{\rho,\boldsymbol{\theta}}
\mathrm{tr}( \hat I(\boldsymbol{\theta}) \rho).\label{eq:variationalbetaQ}
\end{align}
For given $\hat{M}_{\boldsymbol{a}|\boldsymbol{x}}(\boldsymbol{\theta})$, it is sufficient to minimize over pure-state projectors $\rho=\ket{\psi}\bra{\psi}$, since the minimum of the corresponding Hermitian operator is attained by its ground state.

\textit{Ratio optimization.}---We define the quantum-to-classical ratio as
\begin{align}
    \Delta:=\beta_Q/\beta_C,\label{eq:ratio=betaqbetac}
\end{align}
which characterizes the robustness of a Bell inequality to detect nonlocal correlations.  
In Eq.~\eqref{eq:IPvec}, we choose the coefficients $\vec T$ such that the uniform behavior $P_\mathrm{uni}(\boldsymbol{a}|\boldsymbol{x})=1/d^N$ satisfies $I=\sum_{\boldsymbol{a},\boldsymbol{x}}T_{\boldsymbol{x}}^{ \boldsymbol{a}}/d^N=0$. This provides a natural reference
corresponding to white noise. 
The hyperplane $I=0$ divides both $\mathcal{L}$ and $\mathcal{Q}$ into two parts $I>0$ and $I<0$. Under this convention, the quantum value and the classical bound are negative, $\beta_Q\leq \beta_C<0$; $\Delta >1$ implies quantum violation.
For $\vec P_v=v\vec P_Q+(1-v)\vec P_{\mathrm{uni}}$, where $\vec P_Q$ attains $\beta_Q$, the critical visibility for violating the inequality is $v_c=\beta_C/\beta_Q=1/\Delta$.
The variational quantum value is an upper bound on the exact minimum 
$\beta_Q\geq \beta_Q^{\mathrm{full}}$ and gives a lower bound on the exact ratio.

\begin{figure}
    \centering
    \includegraphics[width=\linewidth]{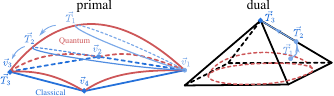}
    \caption{
   Diagram of the optimization method in the primal space of $\vec{P}$ and the dual space of $\vec{T}$. In the primal space, the facet is represented by the blue quadrilateral at the bottom, while the quantum convex set is shown as the red curved surface extending beyond it. The initial inequality $\vec T_1$ is saturated by $\vec{v}_1$, the vertex $\vec{v}_2$ is identified by $\vec{T}_2$. The final inequality $T_3$ becomes a facet and identifies the vertices $\vec{v}_3$ and $\vec v_4$. In the dual space, each inequality is represented by a point, corresponding to a hyperplane in the primal space.}
    \label{fig:diagram}
\end{figure}

Since complete facet enumeration rapidly becomes infeasible for multipartite scenarios as $N$ increases, a natural strategy is to optimize the ratio $\Delta$ directly while exploring the relevant saturating vertices without requiring full knowledge of $\mathcal{L}$.
The following theorem explains why local maxima of the ratio are expected to define facets:
\begin{theorem}\label{theorem1}
A nondegenerate local maximum of the quantum-to-classical ratio implies tightness of the Bell inequality.
\end{theorem}
\noindent Here nondegeneracy means that $\Delta$ is not locally
constant along any nonzero coefficient direction preserving the current
saturating vertices. The formal statement and proof are given
in~\cite{SupplementalMaterial}. At a degenerate local maximum, the ratio is locally constant along directions
that preserve the currently saturating vertices. We follow such a direction
until a new vertex reaches the classical bound. Iterating this update
increases the affine rank of the saturating vertices without decreasing the
ratio until a tight Bell inequality is reached. Figure~\ref{fig:diagram} illustrates this
procedure: the gradient $\partial\Delta/\partial\vec T$ determines the update
direction subject to the current saturation constraints, and successive updates
from $\vec T_1$ through $\vec T_2$ to $\vec T_3$ add saturated vertices until
the final hyperplane is a facet. Since trivial positivity facets have
$\Delta=1$, any facet reached from an initial point with $\Delta>1$ is automatically a nontrivial Bell inequality.
The same argument applies in a restricted coefficient space. 
Coefficients of these facets provide natural initial points for optimization in a larger coefficient space, allowing tight Bell inequalities in the projected space to be refined into more robust ones.
Details of the update rule are provided
in the End Matter.

\begin{figure}
    \centering
    \includegraphics[width=\linewidth]{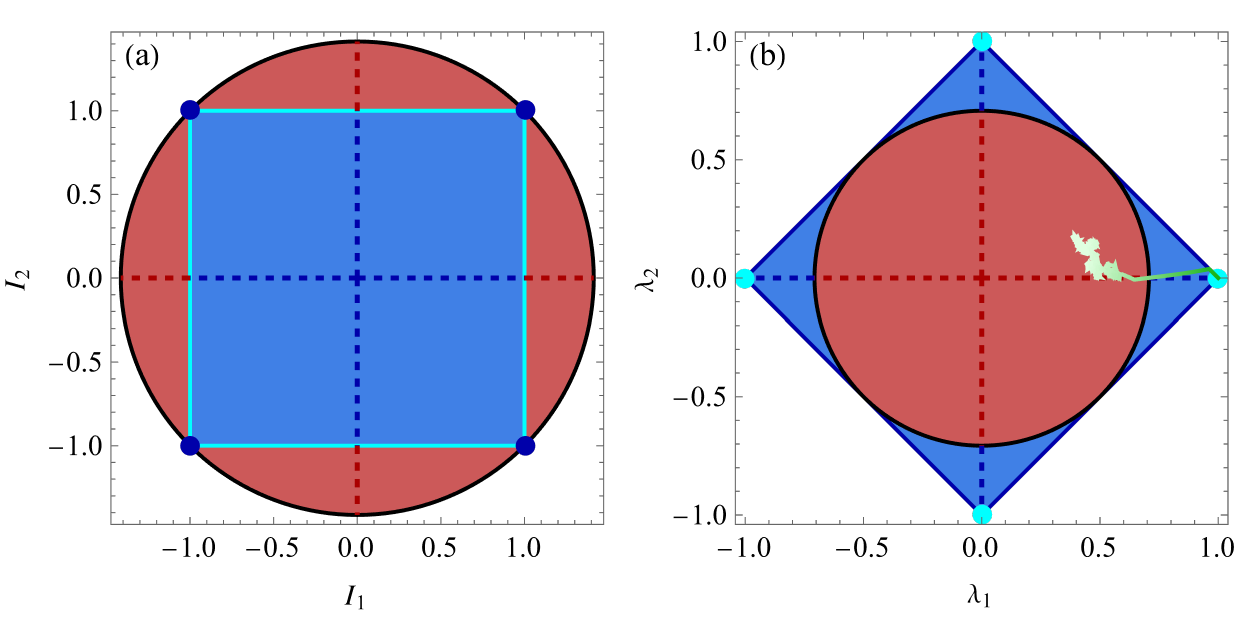}
     \caption{Illustration of the ratio optimization for a two-body bipartite Bell inequality. The maximum ratio identifies a tight Bell inequality. 
    (a) Local (blue) polytope and quantum (red) convex set in the primal space, projected onto a two-dimensional plane spanned by the linear functionals $I_1$ and $I_2$. Blue dots represent the vertices of the local polytope after projection.
    (b) Corresponding objects in the dual-space cross-section, where cyan dots represent tight Bell inequalities. In both panels, the dashed curve indicates the tight Bell inequality achieving the largest ratio. 
    The light-green curve shows the projected trajectory of the coefficients $\alpha_{kl}$ during the optimization, plotted in the $(\lambda_1,\lambda_2)$ cross-section, demonstrating convergence to a CHSH inequality.}
    \label{fig:illustration of the optimization}
\end{figure}

\emph{Benchmarking CHSH inequality.}---
We illustrate our optimization method with a bipartite dichotomic-outcome scenario, where each party performs two measurements.
\space The linear functional can be written equivalently in correlator form. 
We denote the dichotomic observables of the first and second parties by 
$A_x$ and $B_y$, respectively, with $x,y\in\{0,1\}$ labeling the local measurement settings. 
With the mapping $a,b\mapsto (-1)^a,(-1)^b$, the correlator is
\[
\langle A_xB_y\rangle
=
\sum_{a,b=0}^1(-1)^{a+b}P(a,b|x,y).
\]
If the local measurements are described by POVM elements 
$\{M_{a|x}^{A}\}_a$ and $\{M_{b|y}^{B}\}_b$, then $P(a,b|x,y)=
\mathrm{tr}\!\big[
\rho_{AB}
\big(M_{a|x}^{A}\otimes M_{b|y}^{B}\big)
\big]
$, and the associated observables are $A_x=M_{0|x}^{A}-M_{1|x}^{A}$ and  $B_y=M_{0|y}^{B}-M_{1|y}^{B}$.

Using this correlator representation, we write the linear functional as
\[
I=\sum_{x,y=0}^1\alpha_{x,y}\langle A_xB_y\rangle,
\]
where the outcomes of $A_x$ and $B_y$ are $\pm1$. 
For simplicity, we consider the symmetric inequality with $\alpha_{1,0}=\alpha_{0,1}$. 
To evaluate the quantum value, we assign the symmetric measurement setting 
$\hat{A}_0=\hat{\sigma}_z^A$, 
$\hat B_0=\hat \sigma_z^B$,  
$\hat A_1=\cos\theta \hat \sigma_z^A+\sin\theta \hat\sigma _x^A$, 
$\hat B_{1}=\cos\theta \hat \sigma_{z}^{B}+\sin\theta\hat \sigma_{x}^{B}$ 
and obtain the Bell operator $\hat{I}(\theta)=\sum_{x,y=0}^1\alpha_{x,y} \hat {A}_x \hat {B}_y$. We optimize the coefficients $\alpha_{x,y}$ to maximize the quantum-to-classical ratio $\Delta=\beta_Q/\beta_C$.

Figure~\ref{fig:illustration of the optimization}(a) shows a two-dimensional affine projection of the local polytope and the quantum convex set in the primal space of correlators, using the two linear functionals
$I_{1}=(\langle A_{0}B_{0}\rangle+\langle A_{0}B_{1}\rangle+\langle A_{1}B_{0}\rangle-\langle A_{1}B_{1}\rangle)/2$ and $I_{2}=(-\langle A_{0}B_{0}\rangle+\langle A_{0}B_{1}\rangle+\langle A_{1}B_{0}\rangle+\langle A_{1}B_{1}\rangle)/2$ \cite{branciardDetectionLoopholeBell2011,brunnerBellNonlocality2014,gohGeometrySetQuantum2018}.
Figure~\ref{fig:illustration of the optimization}(b) shows the corresponding dual-space cross-section (the space of the functionals up to scale).
In general, dual extreme points are represented by coefficients $\lambda_{x,y}=\alpha_{x,y}/\beta$, so that the inequality reads $\sum_{x,y=0}^{1}\lambda_{x,y}\langle A_{x}B_{y}\rangle\ge-1$.
Facet Bell inequalities correspond to vertices of the dual polytope at $\beta=\beta_C$; the dual quantum boundary is obtained at $\beta=\beta_Q$.
We evaluate $\beta_C$ and $\beta_Q$ for the family $I=\lambda_1 I_1+\lambda_2 I_2$ spanned by $I_1$ and $I_2$.
In the dual space, the admissible region is the set of $(\lambda_1,\lambda_2)$ such that every point in the primal space satisfies $\lambda_1 I_1+\lambda_2 I_2\ge -1$.

We then apply our gradient-based optimization to the coefficients $\alpha_{x,y}$.
The light-green curve shows the optimization trajectory (projected onto the dual cross-section), which converges to the CHSH facet, attaining the maximum ratio $\Delta^{\max}=\sqrt{2}$ ( $\beta_Q=-2\sqrt2$ vs. $\beta_C=-2$).
The initial linear functional is $I=0.3\langle A_{0}B_{1}\rangle+0.3\langle A_{1}B_{0}\rangle-0.5\langle A_{1}B_{1}\rangle$, where the gradient vanishes; to escape this flat region we add small random perturbations to the coefficients $\alpha_{x,y}$. We provide a full characterization of the local polytope and the quantum set in~\cite{SupplementalMaterial}.

\emph{Permutation-invariant Bell inequalities}.---
For multipartite scenarios, restricting to low-order correlators further simplifies the problem and is also more accessible to experimental implementation.
As an illustration of our method, we consider permutation-invariant (PI) multipartite Bell inequalities involving only one- and two-body correlators. The corresponding linear functional is
\begin{align}
I&=\sum_{k=0}^{m-1} \alpha _k \mathcal{S}_{k}+\frac{1}{2}\sum_{k,l=0}^{m-1}\alpha _{k,l}\mathcal{S}_{k,l}
    \label{eq:PIbellineqtwobody}
\end{align}
where the coefficients satisfy $\alpha_{k,l}=\alpha_{l,k}$, and the correlators are $\mathcal S_k:=\sum_{i=1}^{N}A_k^{(i)}$ and $\mathcal S_{k,l}:=\sum_{i\ne j}A_k^{(i)}A_l^{(j)}$.
Here $i,j$ label the sites, $k,l$ label the measurement settings.

To evaluate the classical bound $\beta_{C}$, we consider the deterministic local strategies with $\mathcal{S}_{k,l}=\{\mathcal{S}_{k},\mathcal{S}_{l}\}/2-\mathcal{Z}_{k,l}$
and
$\mathcal{Z}_{k,l}:=\sum_{i=1}^{N}(A_{k}^{(i)}A_{l}^{(i)}+A_{l}^{(i)}A_{k}^{(i)})/2$.
Due to permutation invariance, the vertices of the local polytope are characterized by the occupation numbers of deterministic local strategies.
For the quantum value $\beta_Q$, we adopt the symmetric measurement setting with the observables $\hat{\mathcal{S}}_{k}=2\cos\theta_{k}\hat{S}_{z}+2\sin\theta_{k}\hat{S}_{x}$ and  $\hat{\mathcal{Z}}_{k,l}=N\cos(\theta_{k}-\theta_{l})$. We restrict to the symmetric subspace as a spin-$N/2$ irreducible representation of SU(2), with the collective spin operators $\hat{S}_x$ and $\hat{S}_z$. 
The details of the derivation of the classical bound and quantum value are given in~\cite{SupplementalMaterial}.

We apply our gradient-based optimization to the $(N,2,2)$ scenario. By virtue of Jordan's lemma, 
each party's two dichotomic observables can be restricted to real qubit measurements for $m=2$~\cite{tonerMonogamyBellCorrelations2006}.
In our permutation-invariant variational search, we impose identical measurements for all parties, so the variational quantum value is an upper bound on the exact minimum $\beta_Q^{\mathrm{full}}$.
\space For each fixed Bell inequality, the  variational ratio is a lower bound on the exact ratio of that same inequality. Theorem~\ref{theorem1} also applies to this variational quantum value,
because its proof uses only concavity under coefficient interpolation. In
the restricted coefficient space, tightness refers to facets of the
corresponding projected local polytope.
For finite $N$, we seek the coefficient vector $\vec\alpha=(\alpha_0,\alpha_1,\alpha_{0,0},\alpha_{0,1},\alpha_{1,1})$ that maximizes 
the variational ratio $\Delta_{N,2}$ within our permutation-invariant ansatz, and denote the
maximum ratio obtained for given $N$ by $\Delta_{N,2}^{\mathrm{max}}$.
We show the results of the maximum ratio in Table~\ref{tab:finiteNwithm=2}. It is interesting to point out that $\Delta_{3,2}^{\mathrm{max}}=\Delta_{4,2}^{\mathrm{max}}=(6\sqrt{65}-25)/21=1.11303$.

\begin{figure}
    \centering
    \includegraphics[width=\linewidth]{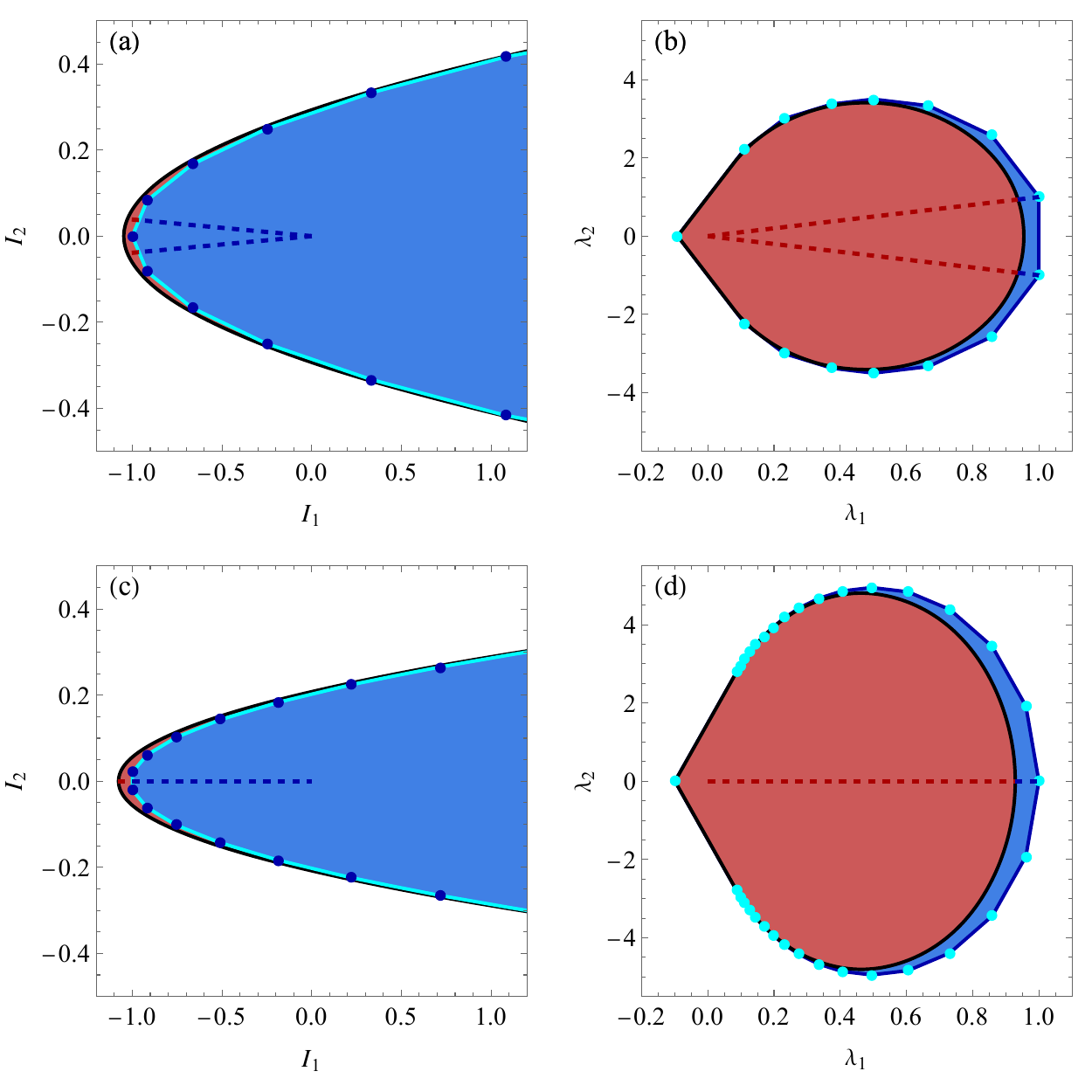}
\caption{Two-dimensional affine projections (left) and corresponding cross-sections (right) for the $(N,m,2)$ scenarios.
(a) and (b) $N=12$, $m=2$. The projection plane is spanned by the two functionals with coefficients $\vec\alpha_{1}=(1,-1,1,1,1)/24$ and $\vec\alpha_{2}=(1,1,0,0,0)/24$.
(c) and (d) $N=11$, $m=3$. The affine projection is spanned by $\vec \alpha_{1}=(2,0,-2,1,1,1,1,1,1)/49$ and $\vec \alpha_{2}=(1,1,1,0,0,0,0,0,0)/49$.
Dashed lines indicate the largest ratio in the affine projections and cross sections.
Blue dots (left panels) show the vertices in the primal affine projections and correspond to blue lines (right panels) in the dual cross-sections.
Cyan lines (left panels) denote tight Bell facets, which map to cyan vertices (right panels).}
    \label{fig:crosssectionNm2}
\end{figure}

For the $m=2$ scenario, we find a family of tight Bell inequalities with coefficient vector
\begin{align}
\vec{\alpha}=\vec{\alpha}_{1}+n\vec{\alpha}_{2},
\label{eq:afamilyofbellineq}
\end{align}
where $\vec{\alpha}_{1}=(1,-1,1,1,1)$ and $\vec{\alpha}_{2}=(1,1,0,0,0)$ with $n\in\{-2N+7,\,-2N+9,\,\ldots,\,2N-9,\,2N-7\},\;N\geq4$.  Within this family, the maximum ratio is reached by $n=\pm1$. 
Figure~\ref{fig:crosssectionNm2}(a) and (b) show the affine projection and the cross section for these inequalities for the $(12,2,2)$ scenario, respectively.
The blue and red dashed lines show the classical bound and the quantum value, respectively.  
This family includes the Bell inequality with coefficients $(2,0,1,1,1)$ which achieves the largest ratio as $N$ increases.

\begin{center}
\setlength{\LTcapwidth}{\linewidth}  
\begin{longtable}{c|c|c|c}
    \caption{Maximum ratio $\Delta^\mathrm{max}_{N,2}$, classical bound $\beta_C$, and corresponding coefficients for the $(N,2,2)$ scenario.\label{tab:finiteNwithm=2}}\\
    \hline
    $N$ & $\Delta^\mathrm{max}_{N,2}$  & $\beta_C$ & $\vec{\alpha}$ \\
    \hline
    \endfirsthead

    \hline
    \multicolumn{4}{l}{\textit{(continued from previous page)}} \\
    \hline
    $N$ & $\Delta^\mathrm{max}_{N,2}$  & $\beta_C$ & $\vec{\alpha}$ \\
    \hline
    \endhead

    \hline
    \multicolumn{4}{r}{\textit{(continued on next page)}} \\
    \endfoot

    \hline
    \endlastfoot

    $2$  & $\sqrt{2}$  & $-2$   & $(0, 0, 1, 1, -1)$ \\
    $3$  & $1.11303$   & $-18$  & $(6, 2, 6, 3, -2)$ \\
    $4$  & $1.11303$   & $-18$  & $(0, 0, 6, 2, -1)$ \\
    $5$  & $1.05904$   & $-80$  & $(20, 4, 20, 5, -2)$ \\
    $6$  & $1.05884$   & $-60$  & $(0, 0, 15, 3, -1)$ \\
    $7$  & $1.03384$   & $-14$  & $(2, 0, 1, 1, 1)$ \\
    $8$  & $1.04058$   & $-16$  & $(2, 0, 1, 1, 1)$ \\
    $10$ & $1.05528$   & $-20$  & $(2, 0, 1, 1, 1)$ \\
    $20$ & $1.09814$   & $-40$  & $(2, 0, 1, 1, 1)$ \\
    $40$ & $1.13479$   & $-80$  & $(2, 0, 1, 1, 1)$ \\
    $\infty$ &  $5/4$       & $-2N$ &$(1,-1,1,1,1)$ \\ 
\end{longtable}
\end{center}

We now apply our method to the $(N,3,2)$ scenario. 
Restricting the local Hilbert-space dimension to 2 gives a variational upper bound on $\beta_Q$ and a lower bound on $\Delta$ since $\beta_C<0$. 
Adding a third measurement setting can increase the achievable ratio, and the corresponding optimal inequalities are therefore more robust
to white noise.
We verify this improvement for $N=3$ and $N=4$ by enumerating all
tight Bell inequalities for $N\leq4$ using
\texttt{cddlib}~\cite{fukudaCddlibReferenceManual2003}, as detailed
in~\cite{SupplementalMaterial}.
For larger $N$, a complete facet enumeration is infeasible, but our gradient-based method remains applicable to optimize the coefficients $\vec{\alpha}=(\alpha_0,\alpha_1,\alpha_2,\alpha_{00},\alpha_{01},\alpha_{02},\alpha_{11},\alpha_{12},\alpha_{22})
$ to identify inequalities with larger ratios. 
Table \ref{tab:finiteNwithm=3} lists selected tight inequalities distinct from those in the $m=2$ scenario.

For the $m=3$ scenario, we find a family of tight Bell inequalities by Eq.~\eqref{eq:afamilyofbellineq} where $\vec{\alpha}_{1}=(2,0,-2,1,1,1,1,1,1)$ and $\vec{\alpha}_2=(1,1,1,0,0,0,0,0,0)$ with $n\in\{-3N+11,\,-3N+13,\,\ldots,\,3N-13,\,3N-11\}$, $N\geq 4$, where $\vec\alpha_{1}$ corresponds to the Bell inequality considered in Ref. \cite{wagnerBellCorrelationsManyBody2017}.
Figure \ref{fig:crosssectionNm2}(c) and (d) show the affine projection and the cross section for $N=11$. 
Interestingly, for sufficiently large $N$ the best inequality found in our search does not belong to this family.
For example, for $N=141$, we find $\vec{\alpha}=(694,0,-694,385,309,385,248,309,385)$ with $\Delta_{141,3}=1.21485$, exceeding $1.21482$ obtained by $\vec{\alpha}_1$.

\begin{center}
\setlength{\LTcapwidth}{\linewidth}
\begin{longtable}{c|c|c|c}
    \caption{
    Selected Bell inequalities for $m=3$ with their ratios and classical bounds. The search is initialized either from the $N\to\infty$
inequality~\cite{chenOptimizingCertifyingMultipartite2026} or from a random
inequality with $\Delta>1$. These inequalities are tight for finite $N$ and are distinct from those in the $m=2$ scenario.}
    \label{tab:finiteNwithm=3} \\
    \hline
    $N$ & $\Delta_{N,3}$ & $\beta_C$ & $\vec{\alpha}$ \\
    \hline
    \endfirsthead

    \hline
    \multicolumn{4}{l}{\textit{(continued from previous page)}} \\
    \hline
    $N$ & $\Delta_{N,3}$ & $\beta_C$ & $\vec{\alpha}$ \\
    \hline
    \endhead

    \hline
    \multicolumn{4}{r}{\textit{(continued on next page)}} \\
    \endfoot

    \hline
    \endlastfoot
    $2$  & $5/4$  & $-8$    & $(2, 1, 1, 2, 2, 2, -2, 1, -2)$ \\
    $3$  & $1.10033$  & $-42$   & $(6, 6, 4, 6, 6, 3, 0, 3, -4)$ \\
    $4$  & $1.11760$  & $-30$   & $(0, 0, 0, 3, 3, 2, 3, 0, -1)$ \\

          $5$  & $1.00538$  & $-56$   & $(10, 4, 0, 3, 3, 2, 3, 2, 1)$ \\

    $6$  & $1.06750$  & $-150$   & $(0, 0, 0, 15, 10, 5, 5, 1, -2)$ \\

         $7$  & $1.02686$  & $-113$  & $(8, 6, 0, 5, -4, 4, 3, -3, 3)$ \\

    $8$  & $1.04123$  & $-120$  & $(7, 0, -7, 4, 3, 4, 2, 3, 4)$ \\
   
        $10$ & $1.06249$  & $-360$  & $(25, 7, -8, 10, 8, 9, 6, 7, 8)$ \\
    $20$ & $1.11557$  & $-840$  & $(19, 0, -19, 10, 9, 10, 8, 9, 10)$ \\

    $40$ & $1.15932$  & $-3480$ & $(39,0,-39,20,19,20,18,19,20)$ \\

    $\infty$ &  $9/7$       & $-98 N$ &$(45,0,-45,25,20,25,16,20,25)$ \\
\end{longtable}
\end{center}

In Ref.~\cite{chenOptimizingCertifyingMultipartite2026}, we optimized the ratio $\Delta$ for a family of
PI Bell inequalities in the thermodynamic limit $N\to\infty$. 
This yields
nontrivial optimized ratios $\Delta_{\infty,m}$ for finite $m$, such as $\Delta_{\infty,2}=5/4$ and $\Delta_{\infty,3}=9/7$ in Tables~\ref{tab:finiteNwithm=2} and \ref{tab:finiteNwithm=3}. In the
double limit $N\to\infty$ and $m\to\infty$ the ratio converges to
$\Delta=\coth(1)$. See Ref.~\cite{chenOptimizingCertifyingMultipartite2026} for the full derivation and finite-size analysis. For PI Bell inequalities with up to two-body correlators, the ratio is lower than the CHSH ratio but can be improved by including higher-body correlators~\cite{guoDetectingBellCorrelations2023a}, where our optimization method is still applicable but the dimension becomes larger.

\emph{Conclusion.}---
We prove that nondegenerate local maxima of the ratio $\Delta$ determine facet Bell inequalities. At a degenerate local maximum, flat directions can be followed to reach a tight Bell inequality without decreasing the ratio.  This result connects robustness and tightness, and provides a gradient-based method for searching robust Bell inequalities.
Our method is not restricted to the specific examples studied here, but provides a general framework for constructing tight and robust Bell inequalities, for example,  permutation-invariant inequalities with higher-body correlators~\cite{guoDetectingBellCorrelations2023a}, translationally invariant Bell inequalities~\cite{turaEnergyDetectorNonlocality2017,wangEntanglementNonlocalityInfinite2017,huTropicalContractionTensor2026a,huCharacterizingTranslationinvariantBell2026a,turaTranslationallyInvariantMultipartite2014,emontsEffectsTopologicalBoundary2024,wangTwodimensionalTranslationinvariantProbability2018}, and more general Bell scenarios~\cite{sliwaSymmetriesBellCorrelation2003,lopez-rosaMaximumNonlocality3222016,wangProbingManyBodyBell2025,bernardsFindingOptimalBell2021,bernardsGeneralizingOptimalBell2020}.

This Letter also points to several promising directions for future work. The framework naturally extends to certifying specific states~\cite{guhneBellInequalitiesGraph2005,supicSelftestingQuantumSystems2020b,liNecessarySufficientCondition2025,baccariScalableBellInequalities2020,tavakoliBellNonlocalityNetworks2022}, including inverse optimization of the classical bound for a fixed Bell operator. 
In the present implementation, the quantum value is obtained as the ground-state energy of the associated Hamiltonian within the symmetric measurement setting, giving a variational upper bound at fixed Hilbert-space dimension. Complementary certified lower bounds from SDP relaxations can be further explored~\cite{navascuesBoundingSetQuantum2007,navascuesConvergentHierarchySemidefinite2008,fadelBoundingSetClassical2017a,tavakoliSemidefiniteProgrammingRelaxations2024}. 
More broadly, our optimization method may be useful beyond nonlocality, for instance in contextuality~\cite{budroniKochenSpeckerContextuality2022}, causality~\cite{branciardSimplestCausalInequalities2015},
and magic~\cite{heinrichRobustnessMagicSymmetries2019}. When combined with Frank–Wolfe algorithms, the gradient-based method may also be used to improve the bound on the three-dimensional Grothendieck constant~\cite{designolleImprovedLocalModels2023}.

\emph{Acknowledgments.}---
We thank Owidiusz Makuta and Eloïc Vallée 
for reading the manuscript and offering helpful comments. We thank Pawe{\l} Cie\'sli\'nski for the helpful discussion of Bell inequalities in $(N,3,2)$ scenario. We thank Hao Wang for the helpful discussion of the proof of Theorem 1.
J.F.C., M.H., and J.T. acknowledge the support received from the European Union's Horizon Europe research and innovation programme through the ERC StG FINE-TEA-SQUAD (Grant No.~101040729). 
This publication is part of the `Quantum Inspire - the Dutch Quantum Computer in the Cloud' project (with project number [NWA.1292.19.194]) of the NWA research program `Research on Routes by Consortia (ORC)', which is funded by the Netherlands Organization for Scientific Research (NWO). 
This project is supported by the National Research Foundation, Singapore through the National Quantum Office, hosted in A*STAR, under its Centre for Quantum Technologies Funding Initiative (S24Q2d0009).
Parts of this work were performed by using the compute
resources from the Academic Leiden Interdisciplinary Cluster Environment (ALICE) provided by Leiden University.
 The views and opinions expressed here are solely those
of the authors and do not necessarily reflect those of the
funding institutions. Neither of the funding institutions
can be held responsible for them.

\bibliographystyle{apsrev4-2}
\bibliography{references}

\begin{thebibliography}{70}%
\makeatletter
\providecommand \@ifxundefined [1]{%
 \@ifx{#1\undefined}
}%
\providecommand \@ifnum [1]{%
 \ifnum #1\expandafter \@firstoftwo
 \else \expandafter \@secondoftwo
 \fi
}%
\providecommand \@ifx [1]{%
 \ifx #1\expandafter \@firstoftwo
 \else \expandafter \@secondoftwo
 \fi
}%
\providecommand \natexlab [1]{#1}%
\providecommand \enquote  [1]{``#1''}%
\providecommand \bibnamefont  [1]{#1}%
\providecommand \bibfnamefont [1]{#1}%
\providecommand \citenamefont [1]{#1}%
\providecommand \href@noop [0]{\@secondoftwo}%
\providecommand \href [0]{\begingroup \@sanitize@url \@href}%
\providecommand \@href[1]{\@@startlink{#1}\@@href}%
\providecommand \@@href[1]{\endgroup#1\@@endlink}%
\providecommand \@sanitize@url [0]{\catcode `\\12\catcode `\$12\catcode
  `\&12\catcode `\#12\catcode `\^12\catcode `\_12\catcode `\%12\relax}%
\providecommand \@@startlink[1]{}%
\providecommand \@@endlink[0]{}%
\providecommand \url  [0]{\begingroup\@sanitize@url \@url }%
\providecommand \@url [1]{\endgroup\@href {#1}{\urlprefix }}%
\providecommand \urlprefix  [0]{URL }%
\providecommand \Eprint [0]{\href }%
\providecommand \doibase [0]{https://doi.org/}%
\providecommand \selectlanguage [0]{\@gobble}%
\providecommand \bibinfo  [0]{\@secondoftwo}%
\providecommand \bibfield  [0]{\@secondoftwo}%
\providecommand \translation [1]{[#1]}%
\providecommand \BibitemOpen [0]{}%
\providecommand \bibitemStop [0]{}%
\providecommand \bibitemNoStop [0]{.\EOS\space}%
\providecommand \EOS [0]{\spacefactor3000\relax}%
\providecommand \BibitemShut  [1]{\csname bibitem#1\endcsname}%
\let\auto@bib@innerbib\@empty
\bibitem [{\citenamefont {Einstein}\ \emph {et~al.}(1935)\citenamefont
  {Einstein}, \citenamefont {Podolsky},\ and\ \citenamefont
  {Rosen}}]{einsteinCanQuantumMechanicalDescription1935}%
  \BibitemOpen
  \bibfield  {author} {\bibinfo {author} {\bibfnamefont {A.}~\bibnamefont
  {Einstein}}, \bibinfo {author} {\bibfnamefont {B.}~\bibnamefont {Podolsky}},\
  and\ \bibinfo {author} {\bibfnamefont {N.}~\bibnamefont {Rosen}},\ }\href
  {https://doi.org/10.1103/PhysRev.47.777} {\bibfield  {journal} {\bibinfo
  {journal} {Physical Review}\ }\textbf {\bibinfo {volume} {47}},\ \bibinfo
  {pages} {777} (\bibinfo {year} {1935})}\BibitemShut {NoStop}%
\bibitem [{\citenamefont {Bell}(1964)}]{bellEinsteinPodolskyRosen1964}%
  \BibitemOpen
  \bibfield  {author} {\bibinfo {author} {\bibfnamefont {J.~S.}\ \bibnamefont
  {Bell}},\ }\href {https://doi.org/10.1103/PhysicsPhysiqueFizika.1.195}
  {\bibfield  {journal} {\bibinfo  {journal} {Physics Physique Fizika}\
  }\textbf {\bibinfo {volume} {1}},\ \bibinfo {pages} {195} (\bibinfo {year}
  {1964})}\BibitemShut {NoStop}%
\bibitem [{\citenamefont {Clauser}\ \emph {et~al.}(1969)\citenamefont
  {Clauser}, \citenamefont {Horne}, \citenamefont {Shimony},\ and\
  \citenamefont {Holt}}]{clauserProposedExperimentTest1969}%
  \BibitemOpen
  \bibfield  {author} {\bibinfo {author} {\bibfnamefont {J.~F.}\ \bibnamefont
  {Clauser}}, \bibinfo {author} {\bibfnamefont {M.~A.}\ \bibnamefont {Horne}},
  \bibinfo {author} {\bibfnamefont {A.}~\bibnamefont {Shimony}},\ and\ \bibinfo
  {author} {\bibfnamefont {R.~A.}\ \bibnamefont {Holt}},\ }\href
  {https://doi.org/10.1103/PhysRevLett.23.880} {\bibfield  {journal} {\bibinfo
  {journal} {Physical Review Letters}\ }\textbf {\bibinfo {volume} {23}},\
  \bibinfo {pages} {880} (\bibinfo {year} {1969})}\BibitemShut {NoStop}%
\bibitem [{\citenamefont
  {Landau}(1988)}]{landauEmpiricalTwopointCorrelation1988}%
  \BibitemOpen
  \bibfield  {author} {\bibinfo {author} {\bibfnamefont {L.~J.}\ \bibnamefont
  {Landau}},\ }\href {https://doi.org/10.1007/BF00732549} {\bibfield  {journal}
  {\bibinfo  {journal} {Foundations of Physics}\ }\textbf {\bibinfo {volume}
  {18}},\ \bibinfo {pages} {449} (\bibinfo {year} {1988})}\BibitemShut
  {NoStop}%
\bibitem [{\citenamefont
  {Masanes}(2003)}]{masanesNecessarySufficientCondition2003}%
  \BibitemOpen
  \bibfield  {author} {\bibinfo {author} {\bibfnamefont {L.}~\bibnamefont
  {Masanes}},\ }\href {https://doi.org/10.48550/arXiv.quant-ph/0309137}
  {\bibinfo {title} {Necessary and sufficient condition for quantum-generated
  correlations}} (\bibinfo {year} {2003}),\ \bibinfo {note}
  {arXiv:quant-ph/0309137}\BibitemShut {NoStop}%
\bibitem [{\citenamefont {Hensen}\ \emph {et~al.}(2015)\citenamefont {Hensen},
  \citenamefont {Bernien}, \citenamefont {Dréau}, \citenamefont {Reiserer},
  \citenamefont {Kalb}, \citenamefont {Blok}, \citenamefont {Ruitenberg},
  \citenamefont {Vermeulen}, \citenamefont {Schouten}, \citenamefont
  {Abellán}, \citenamefont {Amaya}, \citenamefont {Pruneri}, \citenamefont
  {Mitchell}, \citenamefont {Markham}, \citenamefont {Twitchen}, \citenamefont
  {Elkouss}, \citenamefont {Wehner}, \citenamefont {Taminiau},\ and\
  \citenamefont {Hanson}}]{hensenLoopholefreeBellInequality2015}%
  \BibitemOpen
  \bibfield  {author} {\bibinfo {author} {\bibfnamefont {B.}~\bibnamefont
  {Hensen}}, \bibinfo {author} {\bibfnamefont {H.}~\bibnamefont {Bernien}},
  \bibinfo {author} {\bibfnamefont {A.~E.}\ \bibnamefont {Dréau}}, \bibinfo
  {author} {\bibfnamefont {A.}~\bibnamefont {Reiserer}}, \bibinfo {author}
  {\bibfnamefont {N.}~\bibnamefont {Kalb}}, \bibinfo {author} {\bibfnamefont
  {M.~S.}\ \bibnamefont {Blok}}, \bibinfo {author} {\bibfnamefont
  {J.}~\bibnamefont {Ruitenberg}}, \bibinfo {author} {\bibfnamefont {R.~F.~L.}\
  \bibnamefont {Vermeulen}}, \bibinfo {author} {\bibfnamefont {R.~N.}\
  \bibnamefont {Schouten}}, \bibinfo {author} {\bibfnamefont {C.}~\bibnamefont
  {Abellán}}, \bibinfo {author} {\bibfnamefont {W.}~\bibnamefont {Amaya}},
  \bibinfo {author} {\bibfnamefont {V.}~\bibnamefont {Pruneri}}, \bibinfo
  {author} {\bibfnamefont {M.~W.}\ \bibnamefont {Mitchell}}, \bibinfo {author}
  {\bibfnamefont {M.}~\bibnamefont {Markham}}, \bibinfo {author} {\bibfnamefont
  {D.~J.}\ \bibnamefont {Twitchen}}, \bibinfo {author} {\bibfnamefont
  {D.}~\bibnamefont {Elkouss}}, \bibinfo {author} {\bibfnamefont
  {S.}~\bibnamefont {Wehner}}, \bibinfo {author} {\bibfnamefont {T.~H.}\
  \bibnamefont {Taminiau}},\ and\ \bibinfo {author} {\bibfnamefont
  {R.}~\bibnamefont {Hanson}},\ }\href {https://doi.org/10.1038/nature15759}
  {\bibfield  {journal} {\bibinfo  {journal} {Nature}\ }\textbf {\bibinfo
  {volume} {526}},\ \bibinfo {pages} {682} (\bibinfo {year}
  {2015})}\BibitemShut {NoStop}%
\bibitem [{\citenamefont {Giustina}\ \emph {et~al.}(2015)\citenamefont
  {Giustina}, \citenamefont {Versteegh}, \citenamefont {Wengerowsky},
  \citenamefont {Handsteiner}, \citenamefont {Hochrainer}, \citenamefont
  {Phelan}, \citenamefont {Steinlechner}, \citenamefont {Kofler}, \citenamefont
  {Larsson}, \citenamefont {Abellán}, \citenamefont {Amaya}, \citenamefont
  {Pruneri}, \citenamefont {Mitchell}, \citenamefont {Beyer}, \citenamefont
  {Gerrits}, \citenamefont {Lita}, \citenamefont {Shalm}, \citenamefont {Nam},
  \citenamefont {Scheidl}, \citenamefont {Ursin}, \citenamefont {Wittmann},\
  and\ \citenamefont
  {Zeilinger}}]{giustinaSignificantLoopholeFreeTestBells2015}%
  \BibitemOpen
  \bibfield  {author} {\bibinfo {author} {\bibfnamefont {M.}~\bibnamefont
  {Giustina}}, \bibinfo {author} {\bibfnamefont {M.~A.~M.}\ \bibnamefont
  {Versteegh}}, \bibinfo {author} {\bibfnamefont {S.}~\bibnamefont
  {Wengerowsky}}, \bibinfo {author} {\bibfnamefont {J.}~\bibnamefont
  {Handsteiner}}, \bibinfo {author} {\bibfnamefont {A.}~\bibnamefont
  {Hochrainer}}, \bibinfo {author} {\bibfnamefont {K.}~\bibnamefont {Phelan}},
  \bibinfo {author} {\bibfnamefont {F.}~\bibnamefont {Steinlechner}}, \bibinfo
  {author} {\bibfnamefont {J.}~\bibnamefont {Kofler}}, \bibinfo {author}
  {\bibfnamefont {J.-A.}\ \bibnamefont {Larsson}}, \bibinfo {author}
  {\bibfnamefont {C.}~\bibnamefont {Abellán}}, \bibinfo {author}
  {\bibfnamefont {W.}~\bibnamefont {Amaya}}, \bibinfo {author} {\bibfnamefont
  {V.}~\bibnamefont {Pruneri}}, \bibinfo {author} {\bibfnamefont {M.~W.}\
  \bibnamefont {Mitchell}}, \bibinfo {author} {\bibfnamefont {J.}~\bibnamefont
  {Beyer}}, \bibinfo {author} {\bibfnamefont {T.}~\bibnamefont {Gerrits}},
  \bibinfo {author} {\bibfnamefont {A.~E.}\ \bibnamefont {Lita}}, \bibinfo
  {author} {\bibfnamefont {L.~K.}\ \bibnamefont {Shalm}}, \bibinfo {author}
  {\bibfnamefont {S.~W.}\ \bibnamefont {Nam}}, \bibinfo {author} {\bibfnamefont
  {T.}~\bibnamefont {Scheidl}}, \bibinfo {author} {\bibfnamefont
  {R.}~\bibnamefont {Ursin}}, \bibinfo {author} {\bibfnamefont
  {B.}~\bibnamefont {Wittmann}},\ and\ \bibinfo {author} {\bibfnamefont
  {A.}~\bibnamefont {Zeilinger}},\ }\href
  {https://doi.org/10.1103/PhysRevLett.115.250401} {\bibfield  {journal}
  {\bibinfo  {journal} {Physical Review Letters}\ }\textbf {\bibinfo {volume}
  {115}},\ \bibinfo {pages} {250401} (\bibinfo {year} {2015})}\BibitemShut
  {NoStop}%
\bibitem [{\citenamefont {Shalm}\ \emph {et~al.}(2015)\citenamefont {Shalm},
  \citenamefont {Meyer-Scott}, \citenamefont {Christensen}, \citenamefont
  {Bierhorst}, \citenamefont {Wayne}, \citenamefont {Stevens}, \citenamefont
  {Gerrits}, \citenamefont {Glancy}, \citenamefont {Hamel}, \citenamefont
  {Allman}, \citenamefont {Coakley}, \citenamefont {Dyer}, \citenamefont
  {Hodge}, \citenamefont {Lita}, \citenamefont {Verma}, \citenamefont
  {Lambrocco}, \citenamefont {Tortorici}, \citenamefont {Migdall},
  \citenamefont {Zhang}, \citenamefont {Kumor}, \citenamefont {Farr},
  \citenamefont {Marsili}, \citenamefont {Shaw}, \citenamefont {Stern},
  \citenamefont {Abellán}, \citenamefont {Amaya}, \citenamefont {Pruneri},
  \citenamefont {Jennewein}, \citenamefont {Mitchell}, \citenamefont {Kwiat},
  \citenamefont {Bienfang}, \citenamefont {Mirin}, \citenamefont {Knill},\ and\
  \citenamefont {Nam}}]{shalmStrongLoopholeFreeTest2015}%
  \BibitemOpen
  \bibfield  {author} {\bibinfo {author} {\bibfnamefont {L.~K.}\ \bibnamefont
  {Shalm}}, \bibinfo {author} {\bibfnamefont {E.}~\bibnamefont {Meyer-Scott}},
  \bibinfo {author} {\bibfnamefont {B.~G.}\ \bibnamefont {Christensen}},
  \bibinfo {author} {\bibfnamefont {P.}~\bibnamefont {Bierhorst}}, \bibinfo
  {author} {\bibfnamefont {M.~A.}\ \bibnamefont {Wayne}}, \bibinfo {author}
  {\bibfnamefont {M.~J.}\ \bibnamefont {Stevens}}, \bibinfo {author}
  {\bibfnamefont {T.}~\bibnamefont {Gerrits}}, \bibinfo {author} {\bibfnamefont
  {S.}~\bibnamefont {Glancy}}, \bibinfo {author} {\bibfnamefont {D.~R.}\
  \bibnamefont {Hamel}}, \bibinfo {author} {\bibfnamefont {M.~S.}\ \bibnamefont
  {Allman}}, \bibinfo {author} {\bibfnamefont {K.~J.}\ \bibnamefont {Coakley}},
  \bibinfo {author} {\bibfnamefont {S.~D.}\ \bibnamefont {Dyer}}, \bibinfo
  {author} {\bibfnamefont {C.}~\bibnamefont {Hodge}}, \bibinfo {author}
  {\bibfnamefont {A.~E.}\ \bibnamefont {Lita}}, \bibinfo {author}
  {\bibfnamefont {V.~B.}\ \bibnamefont {Verma}}, \bibinfo {author}
  {\bibfnamefont {C.}~\bibnamefont {Lambrocco}}, \bibinfo {author}
  {\bibfnamefont {E.}~\bibnamefont {Tortorici}}, \bibinfo {author}
  {\bibfnamefont {A.~L.}\ \bibnamefont {Migdall}}, \bibinfo {author}
  {\bibfnamefont {Y.}~\bibnamefont {Zhang}}, \bibinfo {author} {\bibfnamefont
  {D.~R.}\ \bibnamefont {Kumor}}, \bibinfo {author} {\bibfnamefont {W.~H.}\
  \bibnamefont {Farr}}, \bibinfo {author} {\bibfnamefont {F.}~\bibnamefont
  {Marsili}}, \bibinfo {author} {\bibfnamefont {M.~D.}\ \bibnamefont {Shaw}},
  \bibinfo {author} {\bibfnamefont {J.~A.}\ \bibnamefont {Stern}}, \bibinfo
  {author} {\bibfnamefont {C.}~\bibnamefont {Abellán}}, \bibinfo {author}
  {\bibfnamefont {W.}~\bibnamefont {Amaya}}, \bibinfo {author} {\bibfnamefont
  {V.}~\bibnamefont {Pruneri}}, \bibinfo {author} {\bibfnamefont
  {T.}~\bibnamefont {Jennewein}}, \bibinfo {author} {\bibfnamefont {M.~W.}\
  \bibnamefont {Mitchell}}, \bibinfo {author} {\bibfnamefont {P.~G.}\
  \bibnamefont {Kwiat}}, \bibinfo {author} {\bibfnamefont {J.~C.}\ \bibnamefont
  {Bienfang}}, \bibinfo {author} {\bibfnamefont {R.~P.}\ \bibnamefont {Mirin}},
  \bibinfo {author} {\bibfnamefont {E.}~\bibnamefont {Knill}},\ and\ \bibinfo
  {author} {\bibfnamefont {S.~W.}\ \bibnamefont {Nam}},\ }\href
  {https://doi.org/10.1103/PhysRevLett.115.250402} {\bibfield  {journal}
  {\bibinfo  {journal} {Physical Review Letters}\ }\textbf {\bibinfo {volume}
  {115}},\ \bibinfo {pages} {250402} (\bibinfo {year} {2015})}\BibitemShut
  {NoStop}%
\bibitem [{\citenamefont {Rosenfeld}\ \emph {et~al.}(2017)\citenamefont
  {Rosenfeld}, \citenamefont {Burchardt}, \citenamefont {Garthoff},
  \citenamefont {Redeker}, \citenamefont {Ortegel}, \citenamefont {Rau},\ and\
  \citenamefont {Weinfurter}}]{rosenfeldEventReadyBellTest2017}%
  \BibitemOpen
  \bibfield  {author} {\bibinfo {author} {\bibfnamefont {W.}~\bibnamefont
  {Rosenfeld}}, \bibinfo {author} {\bibfnamefont {D.}~\bibnamefont
  {Burchardt}}, \bibinfo {author} {\bibfnamefont {R.}~\bibnamefont {Garthoff}},
  \bibinfo {author} {\bibfnamefont {K.}~\bibnamefont {Redeker}}, \bibinfo
  {author} {\bibfnamefont {N.}~\bibnamefont {Ortegel}}, \bibinfo {author}
  {\bibfnamefont {M.}~\bibnamefont {Rau}},\ and\ \bibinfo {author}
  {\bibfnamefont {H.}~\bibnamefont {Weinfurter}},\ }\href
  {https://doi.org/10.1103/PhysRevLett.119.010402} {\bibfield  {journal}
  {\bibinfo  {journal} {Physical Review Letters}\ }\textbf {\bibinfo {volume}
  {119}},\ \bibinfo {pages} {010402} (\bibinfo {year} {2017})}\BibitemShut
  {NoStop}%
\bibitem [{\citenamefont {Storz}\ \emph {et~al.}(2023)\citenamefont {Storz},
  \citenamefont {Schär}, \citenamefont {Kulikov}, \citenamefont {Magnard},
  \citenamefont {Kurpiers}, \citenamefont {Lütolf}, \citenamefont {Walter},
  \citenamefont {Copetudo}, \citenamefont {Reuer}, \citenamefont {Akin},
  \citenamefont {Besse}, \citenamefont {Gabureac}, \citenamefont {Norris},
  \citenamefont {Rosario}, \citenamefont {Martin}, \citenamefont {Martinez},
  \citenamefont {Amaya}, \citenamefont {Mitchell}, \citenamefont {Abellan},
  \citenamefont {Bancal}, \citenamefont {Sangouard}, \citenamefont {Royer},
  \citenamefont {Blais},\ and\ \citenamefont
  {Wallraff}}]{storzLoopholefreeBellInequality2023}%
  \BibitemOpen
  \bibfield  {author} {\bibinfo {author} {\bibfnamefont {S.}~\bibnamefont
  {Storz}}, \bibinfo {author} {\bibfnamefont {J.}~\bibnamefont {Schär}},
  \bibinfo {author} {\bibfnamefont {A.}~\bibnamefont {Kulikov}}, \bibinfo
  {author} {\bibfnamefont {P.}~\bibnamefont {Magnard}}, \bibinfo {author}
  {\bibfnamefont {P.}~\bibnamefont {Kurpiers}}, \bibinfo {author}
  {\bibfnamefont {J.}~\bibnamefont {Lütolf}}, \bibinfo {author} {\bibfnamefont
  {T.}~\bibnamefont {Walter}}, \bibinfo {author} {\bibfnamefont
  {A.}~\bibnamefont {Copetudo}}, \bibinfo {author} {\bibfnamefont
  {K.}~\bibnamefont {Reuer}}, \bibinfo {author} {\bibfnamefont
  {A.}~\bibnamefont {Akin}}, \bibinfo {author} {\bibfnamefont {J.-C.}\
  \bibnamefont {Besse}}, \bibinfo {author} {\bibfnamefont {M.}~\bibnamefont
  {Gabureac}}, \bibinfo {author} {\bibfnamefont {G.~J.}\ \bibnamefont
  {Norris}}, \bibinfo {author} {\bibfnamefont {A.}~\bibnamefont {Rosario}},
  \bibinfo {author} {\bibfnamefont {F.}~\bibnamefont {Martin}}, \bibinfo
  {author} {\bibfnamefont {J.}~\bibnamefont {Martinez}}, \bibinfo {author}
  {\bibfnamefont {W.}~\bibnamefont {Amaya}}, \bibinfo {author} {\bibfnamefont
  {M.~W.}\ \bibnamefont {Mitchell}}, \bibinfo {author} {\bibfnamefont
  {C.}~\bibnamefont {Abellan}}, \bibinfo {author} {\bibfnamefont {J.-D.}\
  \bibnamefont {Bancal}}, \bibinfo {author} {\bibfnamefont {N.}~\bibnamefont
  {Sangouard}}, \bibinfo {author} {\bibfnamefont {B.}~\bibnamefont {Royer}},
  \bibinfo {author} {\bibfnamefont {A.}~\bibnamefont {Blais}},\ and\ \bibinfo
  {author} {\bibfnamefont {A.}~\bibnamefont {Wallraff}},\ }\href
  {https://doi.org/10.1038/s41586-023-05885-0} {\bibfield  {journal} {\bibinfo
  {journal} {Nature}\ }\textbf {\bibinfo {volume} {617}},\ \bibinfo {pages}
  {265} (\bibinfo {year} {2023})}\BibitemShut {NoStop}%
\bibitem [{\citenamefont {Acín}\ \emph {et~al.}(2007)\citenamefont {Acín},
  \citenamefont {Brunner}, \citenamefont {Gisin}, \citenamefont {Massar},
  \citenamefont {Pironio},\ and\ \citenamefont
  {Scarani}}]{acinDeviceIndependentSecurityQuantum2007}%
  \BibitemOpen
  \bibfield  {author} {\bibinfo {author} {\bibfnamefont {A.}~\bibnamefont
  {Acín}}, \bibinfo {author} {\bibfnamefont {N.}~\bibnamefont {Brunner}},
  \bibinfo {author} {\bibfnamefont {N.}~\bibnamefont {Gisin}}, \bibinfo
  {author} {\bibfnamefont {S.}~\bibnamefont {Massar}}, \bibinfo {author}
  {\bibfnamefont {S.}~\bibnamefont {Pironio}},\ and\ \bibinfo {author}
  {\bibfnamefont {V.}~\bibnamefont {Scarani}},\ }\href
  {https://doi.org/10.1103/PhysRevLett.98.230501} {\bibfield  {journal}
  {\bibinfo  {journal} {Physical Review Letters}\ }\textbf {\bibinfo {volume}
  {98}},\ \bibinfo {pages} {230501} (\bibinfo {year} {2007})}\BibitemShut
  {NoStop}%
\bibitem [{\citenamefont {Pironio}\ \emph {et~al.}(2009)\citenamefont
  {Pironio}, \citenamefont {Acín}, \citenamefont {Brunner}, \citenamefont
  {Gisin}, \citenamefont {Massar},\ and\ \citenamefont
  {Scarani}}]{pironioDeviceindependentQuantumKey2009}%
  \BibitemOpen
  \bibfield  {author} {\bibinfo {author} {\bibfnamefont {S.}~\bibnamefont
  {Pironio}}, \bibinfo {author} {\bibfnamefont {A.}~\bibnamefont {Acín}},
  \bibinfo {author} {\bibfnamefont {N.}~\bibnamefont {Brunner}}, \bibinfo
  {author} {\bibfnamefont {N.}~\bibnamefont {Gisin}}, \bibinfo {author}
  {\bibfnamefont {S.}~\bibnamefont {Massar}},\ and\ \bibinfo {author}
  {\bibfnamefont {V.}~\bibnamefont {Scarani}},\ }\href
  {https://doi.org/10.1088/1367-2630/11/4/045021} {\bibfield  {journal}
  {\bibinfo  {journal} {New Journal of Physics}\ }\textbf {\bibinfo {volume}
  {11}},\ \bibinfo {pages} {045021} (\bibinfo {year} {2009})}\BibitemShut
  {NoStop}%
\bibitem [{\citenamefont {Vazirani}\ and\ \citenamefont
  {Vidick}(2014)}]{vaziraniFullyDeviceIndependentQuantum2014}%
  \BibitemOpen
  \bibfield  {author} {\bibinfo {author} {\bibfnamefont {U.}~\bibnamefont
  {Vazirani}}\ and\ \bibinfo {author} {\bibfnamefont {T.}~\bibnamefont
  {Vidick}},\ }\href {https://doi.org/10.1103/PhysRevLett.113.140501}
  {\bibfield  {journal} {\bibinfo  {journal} {Physical Review Letters}\
  }\textbf {\bibinfo {volume} {113}},\ \bibinfo {pages} {140501} (\bibinfo
  {year} {2014})}\BibitemShut {NoStop}%
\bibitem [{\citenamefont {Xu}\ \emph {et~al.}(2020)\citenamefont {Xu},
  \citenamefont {Ma}, \citenamefont {Zhang}, \citenamefont {Lo},\ and\
  \citenamefont {Pan}}]{xuSecureQuantumKey2020}%
  \BibitemOpen
  \bibfield  {author} {\bibinfo {author} {\bibfnamefont {F.}~\bibnamefont
  {Xu}}, \bibinfo {author} {\bibfnamefont {X.}~\bibnamefont {Ma}}, \bibinfo
  {author} {\bibfnamefont {Q.}~\bibnamefont {Zhang}}, \bibinfo {author}
  {\bibfnamefont {H.-K.}\ \bibnamefont {Lo}},\ and\ \bibinfo {author}
  {\bibfnamefont {J.-W.}\ \bibnamefont {Pan}},\ }\href
  {https://doi.org/10.1103/RevModPhys.92.025002} {\bibfield  {journal}
  {\bibinfo  {journal} {Reviews of Modern Physics}\ }\textbf {\bibinfo {volume}
  {92}},\ \bibinfo {pages} {025002} (\bibinfo {year} {2020})}\BibitemShut
  {NoStop}%
\bibitem [{\citenamefont {Pironio}\ \emph {et~al.}(2010)\citenamefont
  {Pironio}, \citenamefont {Acín}, \citenamefont {Massar}, \citenamefont
  {De~La~Giroday}, \citenamefont {Matsukevich}, \citenamefont {Maunz},
  \citenamefont {Olmschenk}, \citenamefont {Hayes}, \citenamefont {Luo},
  \citenamefont {Manning},\ and\ \citenamefont
  {Monroe}}]{pironioRandomNumbersCertified2010}%
  \BibitemOpen
  \bibfield  {author} {\bibinfo {author} {\bibfnamefont {S.}~\bibnamefont
  {Pironio}}, \bibinfo {author} {\bibfnamefont {A.}~\bibnamefont {Acín}},
  \bibinfo {author} {\bibfnamefont {S.}~\bibnamefont {Massar}}, \bibinfo
  {author} {\bibfnamefont {A.~B.}\ \bibnamefont {De~La~Giroday}}, \bibinfo
  {author} {\bibfnamefont {D.~N.}\ \bibnamefont {Matsukevich}}, \bibinfo
  {author} {\bibfnamefont {P.}~\bibnamefont {Maunz}}, \bibinfo {author}
  {\bibfnamefont {S.}~\bibnamefont {Olmschenk}}, \bibinfo {author}
  {\bibfnamefont {D.}~\bibnamefont {Hayes}}, \bibinfo {author} {\bibfnamefont
  {L.}~\bibnamefont {Luo}}, \bibinfo {author} {\bibfnamefont {T.~A.}\
  \bibnamefont {Manning}},\ and\ \bibinfo {author} {\bibfnamefont
  {C.}~\bibnamefont {Monroe}},\ }\href {https://doi.org/10.1038/nature09008}
  {\bibfield  {journal} {\bibinfo  {journal} {Nature}\ }\textbf {\bibinfo
  {volume} {464}},\ \bibinfo {pages} {1021} (\bibinfo {year}
  {2010})}\BibitemShut {NoStop}%
\bibitem [{\citenamefont {Fyrillas}\ \emph {et~al.}(2024)\citenamefont
  {Fyrillas}, \citenamefont {Bourdoncle}, \citenamefont {Maïnos},
  \citenamefont {Emeriau}, \citenamefont {Start}, \citenamefont {Margaria},
  \citenamefont {Morassi}, \citenamefont {Lemaître}, \citenamefont {Sagnes},
  \citenamefont {Stepanov}, \citenamefont {Au}, \citenamefont {Boissier},
  \citenamefont {Somaschi}, \citenamefont {Maring}, \citenamefont {Belabas},\
  and\ \citenamefont {Mansfield}}]{fyrillasCertifiedRandomnessTight2024}%
  \BibitemOpen
  \bibfield  {author} {\bibinfo {author} {\bibfnamefont {A.}~\bibnamefont
  {Fyrillas}}, \bibinfo {author} {\bibfnamefont {B.}~\bibnamefont
  {Bourdoncle}}, \bibinfo {author} {\bibfnamefont {A.}~\bibnamefont {Maïnos}},
  \bibinfo {author} {\bibfnamefont {P.-E.}\ \bibnamefont {Emeriau}}, \bibinfo
  {author} {\bibfnamefont {K.}~\bibnamefont {Start}}, \bibinfo {author}
  {\bibfnamefont {N.}~\bibnamefont {Margaria}}, \bibinfo {author}
  {\bibfnamefont {M.}~\bibnamefont {Morassi}}, \bibinfo {author} {\bibfnamefont
  {A.}~\bibnamefont {Lemaître}}, \bibinfo {author} {\bibfnamefont
  {I.}~\bibnamefont {Sagnes}}, \bibinfo {author} {\bibfnamefont
  {P.}~\bibnamefont {Stepanov}}, \bibinfo {author} {\bibfnamefont {T.~H.}\
  \bibnamefont {Au}}, \bibinfo {author} {\bibfnamefont {S.}~\bibnamefont
  {Boissier}}, \bibinfo {author} {\bibfnamefont {N.}~\bibnamefont {Somaschi}},
  \bibinfo {author} {\bibfnamefont {N.}~\bibnamefont {Maring}}, \bibinfo
  {author} {\bibfnamefont {N.}~\bibnamefont {Belabas}},\ and\ \bibinfo {author}
  {\bibfnamefont {S.}~\bibnamefont {Mansfield}},\ }\href
  {https://doi.org/10.1103/PRXQuantum.5.020348} {\bibfield  {journal} {\bibinfo
   {journal} {PRX Quantum}\ }\textbf {\bibinfo {volume} {5}},\ \bibinfo {pages}
  {020348} (\bibinfo {year} {2024})}\BibitemShut {NoStop}%
\bibitem [{\citenamefont {Zhang}\ \emph {et~al.}(2025)\citenamefont {Zhang},
  \citenamefont {Li}, \citenamefont {Hu}, \citenamefont {Xiang}, \citenamefont
  {Li}, \citenamefont {Guo}, \citenamefont {Tura}, \citenamefont {Gong},
  \citenamefont {He},\ and\ \citenamefont
  {Liu}}]{zhangRandomnessNonlocalityMultipleInput2025}%
  \BibitemOpen
  \bibfield  {author} {\bibinfo {author} {\bibfnamefont {C.}~\bibnamefont
  {Zhang}}, \bibinfo {author} {\bibfnamefont {Y.}~\bibnamefont {Li}}, \bibinfo
  {author} {\bibfnamefont {X.-M.}\ \bibnamefont {Hu}}, \bibinfo {author}
  {\bibfnamefont {Y.}~\bibnamefont {Xiang}}, \bibinfo {author} {\bibfnamefont
  {C.-F.}\ \bibnamefont {Li}}, \bibinfo {author} {\bibfnamefont {G.-C.}\
  \bibnamefont {Guo}}, \bibinfo {author} {\bibfnamefont {J.}~\bibnamefont
  {Tura}}, \bibinfo {author} {\bibfnamefont {Q.}~\bibnamefont {Gong}}, \bibinfo
  {author} {\bibfnamefont {Q.}~\bibnamefont {He}},\ and\ \bibinfo {author}
  {\bibfnamefont {B.-H.}\ \bibnamefont {Liu}},\ }\href
  {https://doi.org/10.1103/PhysRevLett.134.090201} {\bibfield  {journal}
  {\bibinfo  {journal} {Physical Review Letters}\ }\textbf {\bibinfo {volume}
  {134}},\ \bibinfo {pages} {090201} (\bibinfo {year} {2025})}\BibitemShut
  {NoStop}%
\bibitem [{\citenamefont {Šupić}\ and\ \citenamefont
  {Bowles}(2020)}]{supicSelftestingQuantumSystems2020b}%
  \BibitemOpen
  \bibfield  {author} {\bibinfo {author} {\bibfnamefont {I.}~\bibnamefont
  {Šupić}}\ and\ \bibinfo {author} {\bibfnamefont {J.}~\bibnamefont
  {Bowles}},\ }\href {https://doi.org/10.22331/q-2020-09-30-337} {\bibfield
  {journal} {\bibinfo  {journal} {Quantum}\ }\textbf {\bibinfo {volume} {4}},\
  \bibinfo {pages} {337} (\bibinfo {year} {2020})}\BibitemShut {NoStop}%
\bibitem [{\citenamefont {Li}\ \emph {et~al.}(2025)\citenamefont {Li},
  \citenamefont {Xiang}, \citenamefont {Tura},\ and\ \citenamefont
  {He}}]{liNecessarySufficientCondition2025}%
  \BibitemOpen
  \bibfield  {author} {\bibinfo {author} {\bibfnamefont {Y.}~\bibnamefont
  {Li}}, \bibinfo {author} {\bibfnamefont {Y.}~\bibnamefont {Xiang}}, \bibinfo
  {author} {\bibfnamefont {J.}~\bibnamefont {Tura}},\ and\ \bibinfo {author}
  {\bibfnamefont {Q.}~\bibnamefont {He}},\ }\href
  {https://doi.org/10.1103/zp36-2xg4} {\bibfield  {journal} {\bibinfo
  {journal} {Physical Review Letters}\ }\textbf {\bibinfo {volume} {135}},\
  \bibinfo {pages} {060201} (\bibinfo {year} {2025})}\BibitemShut {NoStop}%
\bibitem [{\citenamefont {Baccari}\ \emph {et~al.}(2020)\citenamefont
  {Baccari}, \citenamefont {Augusiak}, \citenamefont {Šupić}, \citenamefont
  {Tura},\ and\ \citenamefont {Acín}}]{baccariScalableBellInequalities2020}%
  \BibitemOpen
  \bibfield  {author} {\bibinfo {author} {\bibfnamefont {F.}~\bibnamefont
  {Baccari}}, \bibinfo {author} {\bibfnamefont {R.}~\bibnamefont {Augusiak}},
  \bibinfo {author} {\bibfnamefont {I.}~\bibnamefont {Šupić}}, \bibinfo
  {author} {\bibfnamefont {J.}~\bibnamefont {Tura}},\ and\ \bibinfo {author}
  {\bibfnamefont {A.}~\bibnamefont {Acín}},\ }\href
  {https://doi.org/10.1103/PhysRevLett.124.020402} {\bibfield  {journal}
  {\bibinfo  {journal} {Physical Review Letters}\ }\textbf {\bibinfo {volume}
  {124}},\ \bibinfo {pages} {020402} (\bibinfo {year} {2020})}\BibitemShut
  {NoStop}%
\bibitem [{\citenamefont {Chazelle}(1993)}]{chazelleOptimalConvexHull1993}%
  \BibitemOpen
  \bibfield  {author} {\bibinfo {author} {\bibfnamefont {B.}~\bibnamefont
  {Chazelle}},\ }\href {https://doi.org/10.1007/BF02573985} {\bibfield
  {journal} {\bibinfo  {journal} {Discrete \& Computational Geometry}\ }\textbf
  {\bibinfo {volume} {10}},\ \bibinfo {pages} {377} (\bibinfo {year}
  {1993})}\BibitemShut {NoStop}%
\bibitem [{\citenamefont
  {Pitowsky}(1989)}]{pitowskyQuantumProbabilityQuantum1989}%
  \BibitemOpen
  \bibfield  {author} {\bibinfo {author} {\bibfnamefont {I.}~\bibnamefont
  {Pitowsky}},\ }\href {https://doi.org/10.1007/BFb0021186} {\emph {\bibinfo
  {title} {Quantum {Probability} — {Quantum} {Logic}}}},\ \bibinfo {series}
  {Lecture {Notes} in {Physics}}, Vol.\ \bibinfo {volume} {321}\ (\bibinfo
  {publisher} {Springer-Verlag},\ \bibinfo {address} {Berlin/Heidelberg},\
  \bibinfo {year} {1989})\BibitemShut {NoStop}%
\bibitem [{\citenamefont
  {Pitowsky}(1991)}]{pitowskyCorrelationPolytopesTheir1991}%
  \BibitemOpen
  \bibfield  {author} {\bibinfo {author} {\bibfnamefont {I.}~\bibnamefont
  {Pitowsky}},\ }\href@noop {} {\bibfield  {journal} {\bibinfo  {journal}
  {Mathematical Programming}\ }\textbf {\bibinfo {volume} {50}},\ \bibinfo
  {pages} {395} (\bibinfo {year} {1991})}\BibitemShut {NoStop}%
\bibitem [{\citenamefont {Bancal}\ \emph {et~al.}(2011)\citenamefont {Bancal},
  \citenamefont {Brunner}, \citenamefont {Gisin},\ and\ \citenamefont
  {Liang}}]{bancalDetectingGenuineMultipartite2011}%
  \BibitemOpen
  \bibfield  {author} {\bibinfo {author} {\bibfnamefont {J.-D.}\ \bibnamefont
  {Bancal}}, \bibinfo {author} {\bibfnamefont {N.}~\bibnamefont {Brunner}},
  \bibinfo {author} {\bibfnamefont {N.}~\bibnamefont {Gisin}},\ and\ \bibinfo
  {author} {\bibfnamefont {Y.-C.}\ \bibnamefont {Liang}},\ }\href
  {https://doi.org/10.1103/PhysRevLett.106.020405} {\bibfield  {journal}
  {\bibinfo  {journal} {Physical Review Letters}\ }\textbf {\bibinfo {volume}
  {106}},\ \bibinfo {pages} {020405} (\bibinfo {year} {2011})}\BibitemShut
  {NoStop}%
\bibitem [{\citenamefont
  {Froissart}(1981)}]{froissartConstructiveGeneralizationBells1981}%
  \BibitemOpen
  \bibfield  {author} {\bibinfo {author} {\bibfnamefont {M.}~\bibnamefont
  {Froissart}},\ }\href {https://doi.org/10.1007/BF02903286} {\bibfield
  {journal} {\bibinfo  {journal} {Il Nuovo Cimento B}\ }\textbf {\bibinfo
  {volume} {64}},\ \bibinfo {pages} {241} (\bibinfo {year} {1981})}\BibitemShut
  {NoStop}%
\bibitem [{\citenamefont {Pitowsky}\ and\ \citenamefont
  {Svozil}(2001)}]{pitowskyOptimalTestsQuantum2001}%
  \BibitemOpen
  \bibfield  {author} {\bibinfo {author} {\bibfnamefont {I.}~\bibnamefont
  {Pitowsky}}\ and\ \bibinfo {author} {\bibfnamefont {K.}~\bibnamefont
  {Svozil}},\ }\href {https://doi.org/10.1103/PhysRevA.64.014102} {\bibfield
  {journal} {\bibinfo  {journal} {Physical Review A}\ }\textbf {\bibinfo
  {volume} {64}},\ \bibinfo {pages} {014102} (\bibinfo {year}
  {2001})}\BibitemShut {NoStop}%
\bibitem [{\citenamefont {Tura}\ \emph
  {et~al.}(2014{\natexlab{a}})\citenamefont {Tura}, \citenamefont {Augusiak},
  \citenamefont {Sainz}, \citenamefont {Vértesi}, \citenamefont {Lewenstein},\
  and\ \citenamefont {Acín}}]{turaDetectingNonlocalityManybody2014c}%
  \BibitemOpen
  \bibfield  {author} {\bibinfo {author} {\bibfnamefont {J.}~\bibnamefont
  {Tura}}, \bibinfo {author} {\bibfnamefont {R.}~\bibnamefont {Augusiak}},
  \bibinfo {author} {\bibfnamefont {A.~B.}\ \bibnamefont {Sainz}}, \bibinfo
  {author} {\bibfnamefont {T.}~\bibnamefont {Vértesi}}, \bibinfo {author}
  {\bibfnamefont {M.}~\bibnamefont {Lewenstein}},\ and\ \bibinfo {author}
  {\bibfnamefont {A.}~\bibnamefont {Acín}},\ }\href
  {https://doi.org/10.1126/science.1247715} {\bibfield  {journal} {\bibinfo
  {journal} {Science}\ }\textbf {\bibinfo {volume} {344}},\ \bibinfo {pages}
  {1256} (\bibinfo {year} {2014}{\natexlab{a}})}\BibitemShut {NoStop}%
\bibitem [{\citenamefont {Tura}\ \emph {et~al.}(2015)\citenamefont {Tura},
  \citenamefont {Augusiak}, \citenamefont {Sainz}, \citenamefont {Lücke},
  \citenamefont {Klempt}, \citenamefont {Lewenstein},\ and\ \citenamefont
  {Acín}}]{turaNonlocalityManybodyQuantum2015}%
  \BibitemOpen
  \bibfield  {author} {\bibinfo {author} {\bibfnamefont {J.}~\bibnamefont
  {Tura}}, \bibinfo {author} {\bibfnamefont {R.}~\bibnamefont {Augusiak}},
  \bibinfo {author} {\bibfnamefont {A.~B.}\ \bibnamefont {Sainz}}, \bibinfo
  {author} {\bibfnamefont {B.}~\bibnamefont {Lücke}}, \bibinfo {author}
  {\bibfnamefont {C.}~\bibnamefont {Klempt}}, \bibinfo {author} {\bibfnamefont
  {M.}~\bibnamefont {Lewenstein}},\ and\ \bibinfo {author} {\bibfnamefont
  {A.}~\bibnamefont {Acín}},\ }\href
  {https://doi.org/10.1016/j.aop.2015.07.021} {\bibfield  {journal} {\bibinfo
  {journal} {Annals of Physics}\ }\textbf {\bibinfo {volume} {362}},\ \bibinfo
  {pages} {370} (\bibinfo {year} {2015})}\BibitemShut {NoStop}%
\bibitem [{\citenamefont {Guo}\ \emph {et~al.}(2023)\citenamefont {Guo},
  \citenamefont {Tura}, \citenamefont {He},\ and\ \citenamefont
  {Fadel}}]{guoDetectingBellCorrelations2023a}%
  \BibitemOpen
  \bibfield  {author} {\bibinfo {author} {\bibfnamefont {J.}~\bibnamefont
  {Guo}}, \bibinfo {author} {\bibfnamefont {J.}~\bibnamefont {Tura}}, \bibinfo
  {author} {\bibfnamefont {Q.}~\bibnamefont {He}},\ and\ \bibinfo {author}
  {\bibfnamefont {M.}~\bibnamefont {Fadel}},\ }\href
  {https://doi.org/10.1103/PhysRevLett.131.070201} {\bibfield  {journal}
  {\bibinfo  {journal} {Physical Review Letters}\ }\textbf {\bibinfo {volume}
  {131}},\ \bibinfo {pages} {070201} (\bibinfo {year} {2023})}\BibitemShut
  {NoStop}%
\bibitem [{\citenamefont {Marconi}\ \emph {et~al.}(2025)\citenamefont
  {Marconi}, \citenamefont {Müller-Rigat}, \citenamefont {Romero-Pallejà},
  \citenamefont {Tura},\ and\ \citenamefont
  {Sanpera}}]{marconiSymmetricQuantumStates2025}%
  \BibitemOpen
  \bibfield  {author} {\bibinfo {author} {\bibfnamefont {C.}~\bibnamefont
  {Marconi}}, \bibinfo {author} {\bibfnamefont {G.}~\bibnamefont
  {Müller-Rigat}}, \bibinfo {author} {\bibfnamefont {J.}~\bibnamefont
  {Romero-Pallejà}}, \bibinfo {author} {\bibfnamefont {J.}~\bibnamefont
  {Tura}},\ and\ \bibinfo {author} {\bibfnamefont {A.}~\bibnamefont
  {Sanpera}},\ }\href {https://doi.org/10.48550/arXiv.2506.10185} {\bibinfo
  {title} {Symmetric quantum states: a review of recent progress}} (\bibinfo
  {year} {2025}),\ \bibinfo {note} {arXiv:2506.10185 [quant-ph]}\BibitemShut
  {NoStop}%
\bibitem [{\citenamefont {Schmied}\ \emph {et~al.}(2016)\citenamefont
  {Schmied}, \citenamefont {Bancal}, \citenamefont {Allard}, \citenamefont
  {Fadel}, \citenamefont {Scarani}, \citenamefont {Treutlein},\ and\
  \citenamefont {Sangouard}}]{schmiedBellCorrelationsBoseEinstein2016}%
  \BibitemOpen
  \bibfield  {author} {\bibinfo {author} {\bibfnamefont {R.}~\bibnamefont
  {Schmied}}, \bibinfo {author} {\bibfnamefont {J.-D.}\ \bibnamefont {Bancal}},
  \bibinfo {author} {\bibfnamefont {B.}~\bibnamefont {Allard}}, \bibinfo
  {author} {\bibfnamefont {M.}~\bibnamefont {Fadel}}, \bibinfo {author}
  {\bibfnamefont {V.}~\bibnamefont {Scarani}}, \bibinfo {author} {\bibfnamefont
  {P.}~\bibnamefont {Treutlein}},\ and\ \bibinfo {author} {\bibfnamefont
  {N.}~\bibnamefont {Sangouard}},\ }\href
  {https://doi.org/10.1126/science.aad8665} {\bibfield  {journal} {\bibinfo
  {journal} {Science}\ }\textbf {\bibinfo {volume} {352}},\ \bibinfo {pages}
  {441} (\bibinfo {year} {2016})}\BibitemShut {NoStop}%
\bibitem [{\citenamefont {Engelsen}\ \emph {et~al.}(2017)\citenamefont
  {Engelsen}, \citenamefont {Krishnakumar}, \citenamefont {Hosten},\ and\
  \citenamefont {Kasevich}}]{engelsenBellCorrelationsSpinSqueezed2017}%
  \BibitemOpen
  \bibfield  {author} {\bibinfo {author} {\bibfnamefont {N.~J.}\ \bibnamefont
  {Engelsen}}, \bibinfo {author} {\bibfnamefont {R.}~\bibnamefont
  {Krishnakumar}}, \bibinfo {author} {\bibfnamefont {O.}~\bibnamefont
  {Hosten}},\ and\ \bibinfo {author} {\bibfnamefont {M.~A.}\ \bibnamefont
  {Kasevich}},\ }\href {https://doi.org/10.1103/PhysRevLett.118.140401}
  {\bibfield  {journal} {\bibinfo  {journal} {Physical Review Letters}\
  }\textbf {\bibinfo {volume} {118}},\ \bibinfo {pages} {140401} (\bibinfo
  {year} {2017})}\BibitemShut {NoStop}%
\bibitem [{\citenamefont {Wang}\ \emph {et~al.}(2018)\citenamefont {Wang},
  \citenamefont {Paesani}, \citenamefont {Ding}, \citenamefont {Santagati},
  \citenamefont {Skrzypczyk}, \citenamefont {Salavrakos}, \citenamefont {Tura},
  \citenamefont {Augusiak}, \citenamefont {Mančinska}, \citenamefont {Bacco},
  \citenamefont {Bonneau}, \citenamefont {Silverstone}, \citenamefont {Gong},
  \citenamefont {Acín}, \citenamefont {Rottwitt}, \citenamefont {Oxenløwe},
  \citenamefont {O’Brien}, \citenamefont {Laing},\ and\ \citenamefont
  {Thompson}}]{wangMultidimensionalQuantumEntanglement2018}%
  \BibitemOpen
  \bibfield  {author} {\bibinfo {author} {\bibfnamefont {J.}~\bibnamefont
  {Wang}}, \bibinfo {author} {\bibfnamefont {S.}~\bibnamefont {Paesani}},
  \bibinfo {author} {\bibfnamefont {Y.}~\bibnamefont {Ding}}, \bibinfo {author}
  {\bibfnamefont {R.}~\bibnamefont {Santagati}}, \bibinfo {author}
  {\bibfnamefont {P.}~\bibnamefont {Skrzypczyk}}, \bibinfo {author}
  {\bibfnamefont {A.}~\bibnamefont {Salavrakos}}, \bibinfo {author}
  {\bibfnamefont {J.}~\bibnamefont {Tura}}, \bibinfo {author} {\bibfnamefont
  {R.}~\bibnamefont {Augusiak}}, \bibinfo {author} {\bibfnamefont
  {L.}~\bibnamefont {Mančinska}}, \bibinfo {author} {\bibfnamefont
  {D.}~\bibnamefont {Bacco}}, \bibinfo {author} {\bibfnamefont
  {D.}~\bibnamefont {Bonneau}}, \bibinfo {author} {\bibfnamefont {J.~W.}\
  \bibnamefont {Silverstone}}, \bibinfo {author} {\bibfnamefont
  {Q.}~\bibnamefont {Gong}}, \bibinfo {author} {\bibfnamefont {A.}~\bibnamefont
  {Acín}}, \bibinfo {author} {\bibfnamefont {K.}~\bibnamefont {Rottwitt}},
  \bibinfo {author} {\bibfnamefont {L.~K.}\ \bibnamefont {Oxenløwe}}, \bibinfo
  {author} {\bibfnamefont {J.~L.}\ \bibnamefont {O’Brien}}, \bibinfo {author}
  {\bibfnamefont {A.}~\bibnamefont {Laing}},\ and\ \bibinfo {author}
  {\bibfnamefont {M.~G.}\ \bibnamefont {Thompson}},\ }\href
  {https://doi.org/10.1126/science.aar7053} {\bibfield  {journal} {\bibinfo
  {journal} {Science}\ }\textbf {\bibinfo {volume} {360}},\ \bibinfo {pages}
  {285} (\bibinfo {year} {2018})}\BibitemShut {NoStop}%
\bibitem [{\citenamefont {Wang}\ \emph {et~al.}(2025)\citenamefont {Wang},
  \citenamefont {Li}, \citenamefont {Xu}, \citenamefont {Hu}, \citenamefont
  {Chen}, \citenamefont {Wu}, \citenamefont {Zhang}, \citenamefont {Jin},
  \citenamefont {Zhu}, \citenamefont {Gao}, \citenamefont {Tan}, \citenamefont
  {Cui}, \citenamefont {Zhang}, \citenamefont {Wang}, \citenamefont {Zou},
  \citenamefont {Li}, \citenamefont {Shen}, \citenamefont {Zhong},
  \citenamefont {Bao}, \citenamefont {Zhu}, \citenamefont {Song}, \citenamefont
  {Deng}, \citenamefont {Dong}, \citenamefont {Zhang}, \citenamefont {Zhang},
  \citenamefont {Jiang}, \citenamefont {Lu}, \citenamefont {Sun}, \citenamefont
  {Li}, \citenamefont {Guo}, \citenamefont {Wang}, \citenamefont {Emonts},
  \citenamefont {Tura}, \citenamefont {Song}, \citenamefont {Wang},\ and\
  \citenamefont {Deng}}]{wangProbingManyBodyBell2025}%
  \BibitemOpen
  \bibfield  {author} {\bibinfo {author} {\bibfnamefont {K.}~\bibnamefont
  {Wang}}, \bibinfo {author} {\bibfnamefont {W.}~\bibnamefont {Li}}, \bibinfo
  {author} {\bibfnamefont {S.}~\bibnamefont {Xu}}, \bibinfo {author}
  {\bibfnamefont {M.}~\bibnamefont {Hu}}, \bibinfo {author} {\bibfnamefont
  {J.}~\bibnamefont {Chen}}, \bibinfo {author} {\bibfnamefont {Y.}~\bibnamefont
  {Wu}}, \bibinfo {author} {\bibfnamefont {C.}~\bibnamefont {Zhang}}, \bibinfo
  {author} {\bibfnamefont {F.}~\bibnamefont {Jin}}, \bibinfo {author}
  {\bibfnamefont {X.}~\bibnamefont {Zhu}}, \bibinfo {author} {\bibfnamefont
  {Y.}~\bibnamefont {Gao}}, \bibinfo {author} {\bibfnamefont {Z.}~\bibnamefont
  {Tan}}, \bibinfo {author} {\bibfnamefont {Z.}~\bibnamefont {Cui}}, \bibinfo
  {author} {\bibfnamefont {A.}~\bibnamefont {Zhang}}, \bibinfo {author}
  {\bibfnamefont {N.}~\bibnamefont {Wang}}, \bibinfo {author} {\bibfnamefont
  {Y.}~\bibnamefont {Zou}}, \bibinfo {author} {\bibfnamefont {T.}~\bibnamefont
  {Li}}, \bibinfo {author} {\bibfnamefont {F.}~\bibnamefont {Shen}}, \bibinfo
  {author} {\bibfnamefont {J.}~\bibnamefont {Zhong}}, \bibinfo {author}
  {\bibfnamefont {Z.}~\bibnamefont {Bao}}, \bibinfo {author} {\bibfnamefont
  {Z.}~\bibnamefont {Zhu}}, \bibinfo {author} {\bibfnamefont {Z.}~\bibnamefont
  {Song}}, \bibinfo {author} {\bibfnamefont {J.}~\bibnamefont {Deng}}, \bibinfo
  {author} {\bibfnamefont {H.}~\bibnamefont {Dong}}, \bibinfo {author}
  {\bibfnamefont {X.}~\bibnamefont {Zhang}}, \bibinfo {author} {\bibfnamefont
  {P.}~\bibnamefont {Zhang}}, \bibinfo {author} {\bibfnamefont
  {W.}~\bibnamefont {Jiang}}, \bibinfo {author} {\bibfnamefont
  {Z.}~\bibnamefont {Lu}}, \bibinfo {author} {\bibfnamefont {Z.-Z.}\
  \bibnamefont {Sun}}, \bibinfo {author} {\bibfnamefont {H.}~\bibnamefont
  {Li}}, \bibinfo {author} {\bibfnamefont {Q.}~\bibnamefont {Guo}}, \bibinfo
  {author} {\bibfnamefont {Z.}~\bibnamefont {Wang}}, \bibinfo {author}
  {\bibfnamefont {P.}~\bibnamefont {Emonts}}, \bibinfo {author} {\bibfnamefont
  {J.}~\bibnamefont {Tura}}, \bibinfo {author} {\bibfnamefont {C.}~\bibnamefont
  {Song}}, \bibinfo {author} {\bibfnamefont {H.}~\bibnamefont {Wang}},\ and\
  \bibinfo {author} {\bibfnamefont {D.-L.}\ \bibnamefont {Deng}},\ }\href
  {https://doi.org/10.1103/PhysRevX.15.021024} {\bibfield  {journal} {\bibinfo
  {journal} {Physical Review X}\ }\textbf {\bibinfo {volume} {15}},\ \bibinfo
  {pages} {021024} (\bibinfo {year} {2025})}\BibitemShut {NoStop}%
\bibitem [{\citenamefont {Junge}\ \emph {et~al.}(2010)\citenamefont {Junge},
  \citenamefont {Palazuelos}, \citenamefont {Pérez-García}, \citenamefont
  {Villanueva},\ and\ \citenamefont {Wolf}}]{jungeOperatorSpaceTheory2010}%
  \BibitemOpen
  \bibfield  {author} {\bibinfo {author} {\bibfnamefont {M.}~\bibnamefont
  {Junge}}, \bibinfo {author} {\bibfnamefont {C.}~\bibnamefont {Palazuelos}},
  \bibinfo {author} {\bibfnamefont {D.}~\bibnamefont {Pérez-García}},
  \bibinfo {author} {\bibfnamefont {I.}~\bibnamefont {Villanueva}},\ and\
  \bibinfo {author} {\bibfnamefont {M.~M.}\ \bibnamefont {Wolf}},\ }\href
  {https://doi.org/10.1103/PhysRevLett.104.170405} {\bibfield  {journal}
  {\bibinfo  {journal} {Physical Review Letters}\ }\textbf {\bibinfo {volume}
  {104}},\ \bibinfo {pages} {170405} (\bibinfo {year} {2010})}\BibitemShut
  {NoStop}%
\bibitem [{\citenamefont {Gómez}\ \emph {et~al.}(2021)\citenamefont {Gómez},
  \citenamefont {Uzcátegui}, \citenamefont {Machuca}, \citenamefont {Gómez},
  \citenamefont {Walborn}, \citenamefont {Lima},\ and\ \citenamefont
  {Goyeneche}}]{gomezOptimalStrategyCertify2021}%
  \BibitemOpen
  \bibfield  {author} {\bibinfo {author} {\bibfnamefont {S.}~\bibnamefont
  {Gómez}}, \bibinfo {author} {\bibfnamefont {D.}~\bibnamefont {Uzcátegui}},
  \bibinfo {author} {\bibfnamefont {I.}~\bibnamefont {Machuca}}, \bibinfo
  {author} {\bibfnamefont {E.~S.}\ \bibnamefont {Gómez}}, \bibinfo {author}
  {\bibfnamefont {S.~P.}\ \bibnamefont {Walborn}}, \bibinfo {author}
  {\bibfnamefont {G.}~\bibnamefont {Lima}},\ and\ \bibinfo {author}
  {\bibfnamefont {D.}~\bibnamefont {Goyeneche}},\ }\href
  {https://doi.org/10.1038/s41598-021-99844-2} {\bibfield  {journal} {\bibinfo
  {journal} {Scientific Reports}\ }\textbf {\bibinfo {volume} {11}},\ \bibinfo
  {pages} {20489} (\bibinfo {year} {2021})}\BibitemShut {NoStop}%
\bibitem [{\citenamefont {Li}\ \emph {et~al.}(2024)\citenamefont {Li},
  \citenamefont {Hu}, \citenamefont {Wang}, \citenamefont {Xu}, \citenamefont
  {Lu}, \citenamefont {Chen}, \citenamefont {Wu}, \citenamefont {Zhang},
  \citenamefont {Jin}, \citenamefont {Zhu}, \citenamefont {Gao}, \citenamefont
  {Cui}, \citenamefont {Zhang}, \citenamefont {Wang}, \citenamefont {Zou},
  \citenamefont {Shen}, \citenamefont {Zhong}, \citenamefont {Bao},
  \citenamefont {Zhu}, \citenamefont {Zhang}, \citenamefont {Li}, \citenamefont
  {Guo}, \citenamefont {Wang}, \citenamefont {Deng}, \citenamefont {Song},
  \citenamefont {Wang}, \citenamefont {Emonts},\ and\ \citenamefont
  {Tura}}]{liImprovedNonlocalityCertification2024a}%
  \BibitemOpen
  \bibfield  {author} {\bibinfo {author} {\bibfnamefont {W.}~\bibnamefont
  {Li}}, \bibinfo {author} {\bibfnamefont {M.}~\bibnamefont {Hu}}, \bibinfo
  {author} {\bibfnamefont {K.}~\bibnamefont {Wang}}, \bibinfo {author}
  {\bibfnamefont {S.}~\bibnamefont {Xu}}, \bibinfo {author} {\bibfnamefont
  {Z.}~\bibnamefont {Lu}}, \bibinfo {author} {\bibfnamefont {J.}~\bibnamefont
  {Chen}}, \bibinfo {author} {\bibfnamefont {Y.}~\bibnamefont {Wu}}, \bibinfo
  {author} {\bibfnamefont {C.}~\bibnamefont {Zhang}}, \bibinfo {author}
  {\bibfnamefont {F.}~\bibnamefont {Jin}}, \bibinfo {author} {\bibfnamefont
  {X.}~\bibnamefont {Zhu}}, \bibinfo {author} {\bibfnamefont {Y.}~\bibnamefont
  {Gao}}, \bibinfo {author} {\bibfnamefont {Z.}~\bibnamefont {Cui}}, \bibinfo
  {author} {\bibfnamefont {A.}~\bibnamefont {Zhang}}, \bibinfo {author}
  {\bibfnamefont {N.}~\bibnamefont {Wang}}, \bibinfo {author} {\bibfnamefont
  {Y.}~\bibnamefont {Zou}}, \bibinfo {author} {\bibfnamefont {F.}~\bibnamefont
  {Shen}}, \bibinfo {author} {\bibfnamefont {J.}~\bibnamefont {Zhong}},
  \bibinfo {author} {\bibfnamefont {Z.}~\bibnamefont {Bao}}, \bibinfo {author}
  {\bibfnamefont {Z.}~\bibnamefont {Zhu}}, \bibinfo {author} {\bibfnamefont
  {P.}~\bibnamefont {Zhang}}, \bibinfo {author} {\bibfnamefont
  {H.}~\bibnamefont {Li}}, \bibinfo {author} {\bibfnamefont {Q.}~\bibnamefont
  {Guo}}, \bibinfo {author} {\bibfnamefont {Z.}~\bibnamefont {Wang}}, \bibinfo
  {author} {\bibfnamefont {D.-L.}\ \bibnamefont {Deng}}, \bibinfo {author}
  {\bibfnamefont {C.}~\bibnamefont {Song}}, \bibinfo {author} {\bibfnamefont
  {H.}~\bibnamefont {Wang}}, \bibinfo {author} {\bibfnamefont {P.}~\bibnamefont
  {Emonts}},\ and\ \bibinfo {author} {\bibfnamefont {J.}~\bibnamefont {Tura}},\
  }\href {https://doi.org/10.48550/arXiv.2407.12347} {\bibinfo {title}
  {Improved {Nonlocality} {Certification} via {Bouncing} between {Bell}
  {Operators} and {Inequalities}}} (\bibinfo {year} {2024}),\ \bibinfo {note}
  {arXiv:2407.12347 [quant-ph]}\BibitemShut {NoStop}%
\bibitem [{\citenamefont {Deng}(2018)}]{dengMachineLearningDetection2018}%
  \BibitemOpen
  \bibfield  {author} {\bibinfo {author} {\bibfnamefont {D.-L.}\ \bibnamefont
  {Deng}},\ }\href {https://doi.org/10.1103/PhysRevLett.120.240402} {\bibfield
  {journal} {\bibinfo  {journal} {Physical Review Letters}\ }\textbf {\bibinfo
  {volume} {120}},\ \bibinfo {pages} {240402} (\bibinfo {year}
  {2018})}\BibitemShut {NoStop}%
\bibitem [{\citenamefont {Chen}\ \emph {et~al.}(2026)\citenamefont {Chen},
  \citenamefont {Hu},\ and\ \citenamefont
  {Tura}}]{chenOptimizingCertifyingMultipartite2026}%
  \BibitemOpen
  \bibfield  {author} {\bibinfo {author} {\bibfnamefont {J.-F.}\ \bibnamefont
  {Chen}}, \bibinfo {author} {\bibfnamefont {M.}~\bibnamefont {Hu}},\ and\
  \bibinfo {author} {\bibfnamefont {J.}~\bibnamefont {Tura}},\ }\href
  {https://doi.org/10.48550/arXiv.2607.08462} {\bibinfo {title} {Optimizing and
  {Certifying} {Multipartite} {Permutationally} {Invariant} {Bell}
  {Inequalities}}} (\bibinfo {year} {2026}),\ \bibinfo {note} {arXiv:2607.08462
  [quant-ph]}\BibitemShut {NoStop}%
\bibitem [{\citenamefont {Budroni}\ \emph {et~al.}(2022)\citenamefont
  {Budroni}, \citenamefont {Cabello}, \citenamefont {Gühne}, \citenamefont
  {Kleinmann},\ and\ \citenamefont
  {Larsson}}]{budroniKochenSpeckerContextuality2022}%
  \BibitemOpen
  \bibfield  {author} {\bibinfo {author} {\bibfnamefont {C.}~\bibnamefont
  {Budroni}}, \bibinfo {author} {\bibfnamefont {A.}~\bibnamefont {Cabello}},
  \bibinfo {author} {\bibfnamefont {O.}~\bibnamefont {Gühne}}, \bibinfo
  {author} {\bibfnamefont {M.}~\bibnamefont {Kleinmann}},\ and\ \bibinfo
  {author} {\bibfnamefont {J.-A.}\ \bibnamefont {Larsson}},\ }\href
  {https://doi.org/10.1103/RevModPhys.94.045007} {\bibfield  {journal}
  {\bibinfo  {journal} {Reviews of Modern Physics}\ }\textbf {\bibinfo {volume}
  {94}},\ \bibinfo {pages} {045007} (\bibinfo {year} {2022})}\BibitemShut
  {NoStop}%
\bibitem [{\citenamefont {Branciard}\ \emph {et~al.}(2015)\citenamefont
  {Branciard}, \citenamefont {Araújo}, \citenamefont {Feix}, \citenamefont
  {Costa},\ and\ \citenamefont
  {Brukner}}]{branciardSimplestCausalInequalities2015}%
  \BibitemOpen
  \bibfield  {author} {\bibinfo {author} {\bibfnamefont {C.}~\bibnamefont
  {Branciard}}, \bibinfo {author} {\bibfnamefont {M.}~\bibnamefont {Araújo}},
  \bibinfo {author} {\bibfnamefont {A.}~\bibnamefont {Feix}}, \bibinfo {author}
  {\bibfnamefont {F.}~\bibnamefont {Costa}},\ and\ \bibinfo {author}
  {\bibfnamefont {C.}~\bibnamefont {Brukner}},\ }\href
  {https://doi.org/10.1088/1367-2630/18/1/013008} {\bibfield  {journal}
  {\bibinfo  {journal} {New Journal of Physics}\ }\textbf {\bibinfo {volume}
  {18}},\ \bibinfo {pages} {013008} (\bibinfo {year} {2015})}\BibitemShut
  {NoStop}%
\bibitem [{\citenamefont {Heinrich}\ and\ \citenamefont
  {Gross}(2019)}]{heinrichRobustnessMagicSymmetries2019}%
  \BibitemOpen
  \bibfield  {author} {\bibinfo {author} {\bibfnamefont {M.}~\bibnamefont
  {Heinrich}}\ and\ \bibinfo {author} {\bibfnamefont {D.}~\bibnamefont
  {Gross}},\ }\href {https://doi.org/10.22331/q-2019-04-08-132} {\bibfield
  {journal} {\bibinfo  {journal} {Quantum}\ }\textbf {\bibinfo {volume} {3}},\
  \bibinfo {pages} {132} (\bibinfo {year} {2019})}\BibitemShut {NoStop}%
\bibitem [{\citenamefont {Brunner}\ \emph {et~al.}(2014)\citenamefont
  {Brunner}, \citenamefont {Cavalcanti}, \citenamefont {Pironio}, \citenamefont
  {Scarani},\ and\ \citenamefont {Wehner}}]{brunnerBellNonlocality2014}%
  \BibitemOpen
  \bibfield  {author} {\bibinfo {author} {\bibfnamefont {N.}~\bibnamefont
  {Brunner}}, \bibinfo {author} {\bibfnamefont {D.}~\bibnamefont {Cavalcanti}},
  \bibinfo {author} {\bibfnamefont {S.}~\bibnamefont {Pironio}}, \bibinfo
  {author} {\bibfnamefont {V.}~\bibnamefont {Scarani}},\ and\ \bibinfo {author}
  {\bibfnamefont {S.}~\bibnamefont {Wehner}},\ }\href
  {https://doi.org/10.1103/RevModPhys.86.419} {\bibfield  {journal} {\bibinfo
  {journal} {Reviews of Modern Physics}\ }\textbf {\bibinfo {volume} {86}},\
  \bibinfo {pages} {419} (\bibinfo {year} {2014})}\BibitemShut {NoStop}%
\bibitem [{\citenamefont
  {Pitowsky}(1986)}]{pitowskyRangeQuantumProbability1986}%
  \BibitemOpen
  \bibfield  {author} {\bibinfo {author} {\bibfnamefont {I.}~\bibnamefont
  {Pitowsky}},\ }\href {https://doi.org/10.1063/1.527066} {\bibfield  {journal}
  {\bibinfo  {journal} {Journal of Mathematical Physics}\ }\textbf {\bibinfo
  {volume} {27}},\ \bibinfo {pages} {1556} (\bibinfo {year}
  {1986})}\BibitemShut {NoStop}%
\bibitem [{Sup()}]{SupplementalMaterial}%
  \BibitemOpen
  \href@noop {} {\bibinfo {title} {Supplemental {Material}}}\BibitemShut
  {NoStop}%
\bibitem [{\citenamefont
  {Branciard}(2011)}]{branciardDetectionLoopholeBell2011}%
  \BibitemOpen
  \bibfield  {author} {\bibinfo {author} {\bibfnamefont {C.}~\bibnamefont
  {Branciard}},\ }\href {https://doi.org/10.1103/PhysRevA.83.032123} {\bibfield
   {journal} {\bibinfo  {journal} {Physical Review A}\ }\textbf {\bibinfo
  {volume} {83}},\ \bibinfo {pages} {032123} (\bibinfo {year}
  {2011})}\BibitemShut {NoStop}%
\bibitem [{\citenamefont {Goh}\ \emph {et~al.}(2018)\citenamefont {Goh},
  \citenamefont {Kaniewski}, \citenamefont {Wolfe}, \citenamefont {Vértesi},
  \citenamefont {Wu}, \citenamefont {Cai}, \citenamefont {Liang},\ and\
  \citenamefont {Scarani}}]{gohGeometrySetQuantum2018}%
  \BibitemOpen
  \bibfield  {author} {\bibinfo {author} {\bibfnamefont {K.~T.}\ \bibnamefont
  {Goh}}, \bibinfo {author} {\bibfnamefont {J.}~\bibnamefont {Kaniewski}},
  \bibinfo {author} {\bibfnamefont {E.}~\bibnamefont {Wolfe}}, \bibinfo
  {author} {\bibfnamefont {T.}~\bibnamefont {Vértesi}}, \bibinfo {author}
  {\bibfnamefont {X.}~\bibnamefont {Wu}}, \bibinfo {author} {\bibfnamefont
  {Y.}~\bibnamefont {Cai}}, \bibinfo {author} {\bibfnamefont {Y.-C.}\
  \bibnamefont {Liang}},\ and\ \bibinfo {author} {\bibfnamefont
  {V.}~\bibnamefont {Scarani}},\ }\href
  {https://doi.org/10.1103/PhysRevA.97.022104} {\bibfield  {journal} {\bibinfo
  {journal} {Physical Review A}\ }\textbf {\bibinfo {volume} {97}},\ \bibinfo
  {pages} {022104} (\bibinfo {year} {2018})}\BibitemShut {NoStop}%
\bibitem [{\citenamefont {Toner}\ and\ \citenamefont
  {Verstraete}(2006)}]{tonerMonogamyBellCorrelations2006}%
  \BibitemOpen
  \bibfield  {author} {\bibinfo {author} {\bibfnamefont {B.}~\bibnamefont
  {Toner}}\ and\ \bibinfo {author} {\bibfnamefont {F.}~\bibnamefont
  {Verstraete}},\ }\href {https://doi.org/10.48550/arXiv.quant-ph/0611001}
  {\bibinfo {title} {Monogamy of {Bell} correlations and {Tsirelson}'s bound}}
  (\bibinfo {year} {2006}),\ \bibinfo {note}
  {arXiv:quant-ph/0611001}\BibitemShut {NoStop}%
\bibitem [{\citenamefont {Fukuda}(2003)}]{fukudaCddlibReferenceManual2003}%
  \BibitemOpen
  \bibfield  {author} {\bibinfo {author} {\bibfnamefont {K.}~\bibnamefont
  {Fukuda}},\ }\href@noop {} {\bibinfo {title} {Cddlib reference manual}}
  (\bibinfo {year} {2003})\BibitemShut {NoStop}%
\bibitem [{\citenamefont {Wagner}\ \emph {et~al.}(2017)\citenamefont {Wagner},
  \citenamefont {Schmied}, \citenamefont {Fadel}, \citenamefont {Treutlein},
  \citenamefont {Sangouard},\ and\ \citenamefont
  {Bancal}}]{wagnerBellCorrelationsManyBody2017}%
  \BibitemOpen
  \bibfield  {author} {\bibinfo {author} {\bibfnamefont {S.}~\bibnamefont
  {Wagner}}, \bibinfo {author} {\bibfnamefont {R.}~\bibnamefont {Schmied}},
  \bibinfo {author} {\bibfnamefont {M.}~\bibnamefont {Fadel}}, \bibinfo
  {author} {\bibfnamefont {P.}~\bibnamefont {Treutlein}}, \bibinfo {author}
  {\bibfnamefont {N.}~\bibnamefont {Sangouard}},\ and\ \bibinfo {author}
  {\bibfnamefont {J.-D.}\ \bibnamefont {Bancal}},\ }\href
  {https://doi.org/10.1103/PhysRevLett.119.170403} {\bibfield  {journal}
  {\bibinfo  {journal} {Physical Review Letters}\ }\textbf {\bibinfo {volume}
  {119}},\ \bibinfo {pages} {170403} (\bibinfo {year} {2017})}\BibitemShut
  {NoStop}%
\bibitem [{\citenamefont {Tura}\ \emph {et~al.}(2017)\citenamefont {Tura},
  \citenamefont {De~Las~Cuevas}, \citenamefont {Augusiak}, \citenamefont
  {Lewenstein}, \citenamefont {Acín},\ and\ \citenamefont
  {Cirac}}]{turaEnergyDetectorNonlocality2017}%
  \BibitemOpen
  \bibfield  {author} {\bibinfo {author} {\bibfnamefont {J.}~\bibnamefont
  {Tura}}, \bibinfo {author} {\bibfnamefont {G.}~\bibnamefont {De~Las~Cuevas}},
  \bibinfo {author} {\bibfnamefont {R.}~\bibnamefont {Augusiak}}, \bibinfo
  {author} {\bibfnamefont {M.}~\bibnamefont {Lewenstein}}, \bibinfo {author}
  {\bibfnamefont {A.}~\bibnamefont {Acín}},\ and\ \bibinfo {author}
  {\bibfnamefont {J.}~\bibnamefont {Cirac}},\ }\href
  {https://doi.org/10.1103/PhysRevX.7.021005} {\bibfield  {journal} {\bibinfo
  {journal} {Physical Review X}\ }\textbf {\bibinfo {volume} {7}},\ \bibinfo
  {pages} {021005} (\bibinfo {year} {2017})}\BibitemShut {NoStop}%
\bibitem [{\citenamefont {Wang}\ \emph {et~al.}(2017)\citenamefont {Wang},
  \citenamefont {Singh},\ and\ \citenamefont
  {Navascués}}]{wangEntanglementNonlocalityInfinite2017}%
  \BibitemOpen
  \bibfield  {author} {\bibinfo {author} {\bibfnamefont {Z.}~\bibnamefont
  {Wang}}, \bibinfo {author} {\bibfnamefont {S.}~\bibnamefont {Singh}},\ and\
  \bibinfo {author} {\bibfnamefont {M.}~\bibnamefont {Navascués}},\ }\href
  {https://doi.org/10.1103/PhysRevLett.118.230401} {\bibfield  {journal}
  {\bibinfo  {journal} {Physical Review Letters}\ }\textbf {\bibinfo {volume}
  {118}},\ \bibinfo {pages} {230401} (\bibinfo {year} {2017})}\BibitemShut
  {NoStop}%
\bibitem [{\citenamefont {Hu}\ and\ \citenamefont
  {Tura}(2026)}]{huTropicalContractionTensor2026a}%
  \BibitemOpen
  \bibfield  {author} {\bibinfo {author} {\bibfnamefont {M.}~\bibnamefont
  {Hu}}\ and\ \bibinfo {author} {\bibfnamefont {J.}~\bibnamefont {Tura}},\
  }\href {https://doi.org/10.1103/r7s7-c7y2} {\bibfield  {journal} {\bibinfo
  {journal} {Physical Review Letters}\ }\textbf {\bibinfo {volume} {136}},\
  \bibinfo {pages} {100202} (\bibinfo {year} {2026})}\BibitemShut {NoStop}%
\bibitem [{\citenamefont {Hu}\ \emph {et~al.}(2026)\citenamefont {Hu},
  \citenamefont {Vallée}, \citenamefont {Seynnaeve}, \citenamefont {Emonts},
  \citenamefont {Mohammadi},\ and\ \citenamefont
  {Tura}}]{huCharacterizingTranslationinvariantBell2026a}%
  \BibitemOpen
  \bibfield  {author} {\bibinfo {author} {\bibfnamefont {M.}~\bibnamefont
  {Hu}}, \bibinfo {author} {\bibfnamefont {E.}~\bibnamefont {Vallée}},
  \bibinfo {author} {\bibfnamefont {T.}~\bibnamefont {Seynnaeve}}, \bibinfo
  {author} {\bibfnamefont {P.}~\bibnamefont {Emonts}}, \bibinfo {author}
  {\bibfnamefont {F.}~\bibnamefont {Mohammadi}},\ and\ \bibinfo {author}
  {\bibfnamefont {J.}~\bibnamefont {Tura}},\ }\href
  {https://doi.org/10.1103/mgm4-dz2x} {\bibfield  {journal} {\bibinfo
  {journal} {Physical Review A}\ }\textbf {\bibinfo {volume} {113}},\ \bibinfo
  {pages} {032421} (\bibinfo {year} {2026})}\BibitemShut {NoStop}%
\bibitem [{\citenamefont {Tura}\ \emph
  {et~al.}(2014{\natexlab{b}})\citenamefont {Tura}, \citenamefont {B~Sainz},
  \citenamefont {Vértesi}, \citenamefont {Acín}, \citenamefont {Lewenstein},\
  and\ \citenamefont
  {Augusiak}}]{turaTranslationallyInvariantMultipartite2014}%
  \BibitemOpen
  \bibfield  {author} {\bibinfo {author} {\bibfnamefont {J.}~\bibnamefont
  {Tura}}, \bibinfo {author} {\bibfnamefont {A.}~\bibnamefont {B~Sainz}},
  \bibinfo {author} {\bibfnamefont {T.}~\bibnamefont {Vértesi}}, \bibinfo
  {author} {\bibfnamefont {A.}~\bibnamefont {Acín}}, \bibinfo {author}
  {\bibfnamefont {M.}~\bibnamefont {Lewenstein}},\ and\ \bibinfo {author}
  {\bibfnamefont {R.}~\bibnamefont {Augusiak}},\ }\href
  {https://doi.org/10.1088/1751-8113/47/42/424024} {\bibfield  {journal}
  {\bibinfo  {journal} {Journal of Physics A: Mathematical and Theoretical}\
  }\textbf {\bibinfo {volume} {47}},\ \bibinfo {pages} {424024} (\bibinfo
  {year} {2014}{\natexlab{b}})}\BibitemShut {NoStop}%
\bibitem [{\citenamefont {Emonts}\ \emph {et~al.}(2024)\citenamefont {Emonts},
  \citenamefont {Hu}, \citenamefont {Aloy},\ and\ \citenamefont
  {Tura}}]{emontsEffectsTopologicalBoundary2024}%
  \BibitemOpen
  \bibfield  {author} {\bibinfo {author} {\bibfnamefont {P.}~\bibnamefont
  {Emonts}}, \bibinfo {author} {\bibfnamefont {M.}~\bibnamefont {Hu}}, \bibinfo
  {author} {\bibfnamefont {A.}~\bibnamefont {Aloy}},\ and\ \bibinfo {author}
  {\bibfnamefont {J.}~\bibnamefont {Tura}},\ }\href
  {https://doi.org/10.1103/PhysRevA.110.032201} {\bibfield  {journal} {\bibinfo
   {journal} {Physical Review A}\ }\textbf {\bibinfo {volume} {110}},\ \bibinfo
  {pages} {032201} (\bibinfo {year} {2024})}\BibitemShut {NoStop}%
\bibitem [{\citenamefont {Wang}\ and\ \citenamefont
  {Navascués}(2018)}]{wangTwodimensionalTranslationinvariantProbability2018}%
  \BibitemOpen
  \bibfield  {author} {\bibinfo {author} {\bibfnamefont {Z.}~\bibnamefont
  {Wang}}\ and\ \bibinfo {author} {\bibfnamefont {M.}~\bibnamefont
  {Navascués}},\ }\href {https://doi.org/10.1098/rspa.2017.0822} {\bibfield
  {journal} {\bibinfo  {journal} {Proceedings of the Royal Society A:
  Mathematical, Physical and Engineering Sciences}\ }\textbf {\bibinfo {volume}
  {474}},\ \bibinfo {pages} {20170822} (\bibinfo {year} {2018})}\BibitemShut
  {NoStop}%
\bibitem [{\citenamefont {Śliwa}(2003)}]{sliwaSymmetriesBellCorrelation2003}%
  \BibitemOpen
  \bibfield  {author} {\bibinfo {author} {\bibfnamefont {C.}~\bibnamefont
  {Śliwa}},\ }\href {https://doi.org/10.1016/S0375-9601(03)01115-0} {\bibfield
   {journal} {\bibinfo  {journal} {Physics Letters A}\ }\textbf {\bibinfo
  {volume} {317}},\ \bibinfo {pages} {165} (\bibinfo {year}
  {2003})}\BibitemShut {NoStop}%
\bibitem [{\citenamefont {López-Rosa}\ \emph {et~al.}(2016)\citenamefont
  {López-Rosa}, \citenamefont {Xu},\ and\ \citenamefont
  {Cabello}}]{lopez-rosaMaximumNonlocality3222016}%
  \BibitemOpen
  \bibfield  {author} {\bibinfo {author} {\bibfnamefont {S.}~\bibnamefont
  {López-Rosa}}, \bibinfo {author} {\bibfnamefont {Z.-P.}\ \bibnamefont
  {Xu}},\ and\ \bibinfo {author} {\bibfnamefont {A.}~\bibnamefont {Cabello}},\
  }\href {https://doi.org/10.1103/PhysRevA.94.062121} {\bibfield  {journal}
  {\bibinfo  {journal} {Physical Review A}\ }\textbf {\bibinfo {volume} {94}},\
  \bibinfo {pages} {062121} (\bibinfo {year} {2016})}\BibitemShut {NoStop}%
\bibitem [{\citenamefont {Bernards}\ and\ \citenamefont
  {Gühne}(2021)}]{bernardsFindingOptimalBell2021}%
  \BibitemOpen
  \bibfield  {author} {\bibinfo {author} {\bibfnamefont {F.}~\bibnamefont
  {Bernards}}\ and\ \bibinfo {author} {\bibfnamefont {O.}~\bibnamefont
  {Gühne}},\ }\href {https://doi.org/10.1103/PhysRevA.104.012206} {\bibfield
  {journal} {\bibinfo  {journal} {Physical Review A}\ }\textbf {\bibinfo
  {volume} {104}},\ \bibinfo {pages} {012206} (\bibinfo {year}
  {2021})}\BibitemShut {NoStop}%
\bibitem [{\citenamefont {Bernards}\ and\ \citenamefont
  {Gühne}(2020)}]{bernardsGeneralizingOptimalBell2020}%
  \BibitemOpen
  \bibfield  {author} {\bibinfo {author} {\bibfnamefont {F.}~\bibnamefont
  {Bernards}}\ and\ \bibinfo {author} {\bibfnamefont {O.}~\bibnamefont
  {Gühne}},\ }\href {https://doi.org/10.1103/PhysRevLett.125.200401}
  {\bibfield  {journal} {\bibinfo  {journal} {Physical Review Letters}\
  }\textbf {\bibinfo {volume} {125}},\ \bibinfo {pages} {200401} (\bibinfo
  {year} {2020})}\BibitemShut {NoStop}%
\bibitem [{\citenamefont {Gühne}\ \emph {et~al.}(2005)\citenamefont {Gühne},
  \citenamefont {Tóth}, \citenamefont {Hyllus},\ and\ \citenamefont
  {Briegel}}]{guhneBellInequalitiesGraph2005}%
  \BibitemOpen
  \bibfield  {author} {\bibinfo {author} {\bibfnamefont {O.}~\bibnamefont
  {Gühne}}, \bibinfo {author} {\bibfnamefont {G.}~\bibnamefont {Tóth}},
  \bibinfo {author} {\bibfnamefont {P.}~\bibnamefont {Hyllus}},\ and\ \bibinfo
  {author} {\bibfnamefont {H.~J.}\ \bibnamefont {Briegel}},\ }\href
  {https://doi.org/10.1103/PhysRevLett.95.120405} {\bibfield  {journal}
  {\bibinfo  {journal} {Physical Review Letters}\ }\textbf {\bibinfo {volume}
  {95}},\ \bibinfo {pages} {120405} (\bibinfo {year} {2005})}\BibitemShut
  {NoStop}%
\bibitem [{\citenamefont {Tavakoli}\ \emph {et~al.}(2022)\citenamefont
  {Tavakoli}, \citenamefont {Pozas-Kerstjens}, \citenamefont {Luo},\ and\
  \citenamefont {Renou}}]{tavakoliBellNonlocalityNetworks2022}%
  \BibitemOpen
  \bibfield  {author} {\bibinfo {author} {\bibfnamefont {A.}~\bibnamefont
  {Tavakoli}}, \bibinfo {author} {\bibfnamefont {A.}~\bibnamefont
  {Pozas-Kerstjens}}, \bibinfo {author} {\bibfnamefont {M.-X.}\ \bibnamefont
  {Luo}},\ and\ \bibinfo {author} {\bibfnamefont {M.-O.}\ \bibnamefont
  {Renou}},\ }\href {https://doi.org/10.1088/1361-6633/ac41bb} {\bibfield
  {journal} {\bibinfo  {journal} {Reports on Progress in Physics}\ }\textbf
  {\bibinfo {volume} {85}},\ \bibinfo {pages} {056001} (\bibinfo {year}
  {2022})}\BibitemShut {NoStop}%
\bibitem [{\citenamefont {Navascués}\ \emph {et~al.}(2007)\citenamefont
  {Navascués}, \citenamefont {Pironio},\ and\ \citenamefont
  {Acín}}]{navascuesBoundingSetQuantum2007}%
  \BibitemOpen
  \bibfield  {author} {\bibinfo {author} {\bibfnamefont {M.}~\bibnamefont
  {Navascués}}, \bibinfo {author} {\bibfnamefont {S.}~\bibnamefont
  {Pironio}},\ and\ \bibinfo {author} {\bibfnamefont {A.}~\bibnamefont
  {Acín}},\ }\href {https://doi.org/10.1103/PhysRevLett.98.010401} {\bibfield
  {journal} {\bibinfo  {journal} {Physical Review Letters}\ }\textbf {\bibinfo
  {volume} {98}},\ \bibinfo {pages} {010401} (\bibinfo {year}
  {2007})}\BibitemShut {NoStop}%
\bibitem [{\citenamefont {Navascués}\ \emph {et~al.}(2008)\citenamefont
  {Navascués}, \citenamefont {Pironio},\ and\ \citenamefont
  {Acín}}]{navascuesConvergentHierarchySemidefinite2008}%
  \BibitemOpen
  \bibfield  {author} {\bibinfo {author} {\bibfnamefont {M.}~\bibnamefont
  {Navascués}}, \bibinfo {author} {\bibfnamefont {S.}~\bibnamefont
  {Pironio}},\ and\ \bibinfo {author} {\bibfnamefont {A.}~\bibnamefont
  {Acín}},\ }\href {https://doi.org/10.1088/1367-2630/10/7/073013} {\bibfield
  {journal} {\bibinfo  {journal} {New Journal of Physics}\ }\textbf {\bibinfo
  {volume} {10}},\ \bibinfo {pages} {073013} (\bibinfo {year}
  {2008})}\BibitemShut {NoStop}%
\bibitem [{\citenamefont {Fadel}\ and\ \citenamefont
  {Tura}(2017)}]{fadelBoundingSetClassical2017a}%
  \BibitemOpen
  \bibfield  {author} {\bibinfo {author} {\bibfnamefont {M.}~\bibnamefont
  {Fadel}}\ and\ \bibinfo {author} {\bibfnamefont {J.}~\bibnamefont {Tura}},\
  }\href {https://doi.org/10.1103/PhysRevLett.119.230402} {\bibfield  {journal}
  {\bibinfo  {journal} {Physical Review Letters}\ }\textbf {\bibinfo {volume}
  {119}},\ \bibinfo {pages} {230402} (\bibinfo {year} {2017})}\BibitemShut
  {NoStop}%
\bibitem [{\citenamefont {Tavakoli}\ \emph {et~al.}(2024)\citenamefont
  {Tavakoli}, \citenamefont {Pozas-Kerstjens}, \citenamefont {Brown},\ and\
  \citenamefont {Araújo}}]{tavakoliSemidefiniteProgrammingRelaxations2024}%
  \BibitemOpen
  \bibfield  {author} {\bibinfo {author} {\bibfnamefont {A.}~\bibnamefont
  {Tavakoli}}, \bibinfo {author} {\bibfnamefont {A.}~\bibnamefont
  {Pozas-Kerstjens}}, \bibinfo {author} {\bibfnamefont {P.}~\bibnamefont
  {Brown}},\ and\ \bibinfo {author} {\bibfnamefont {M.}~\bibnamefont
  {Araújo}},\ }\href {https://doi.org/10.1103/RevModPhys.96.045006} {\bibfield
   {journal} {\bibinfo  {journal} {Reviews of Modern Physics}\ }\textbf
  {\bibinfo {volume} {96}},\ \bibinfo {pages} {045006} (\bibinfo {year}
  {2024})}\BibitemShut {NoStop}%
\bibitem [{\citenamefont {Designolle}\ \emph {et~al.}(2023)\citenamefont
  {Designolle}, \citenamefont {Iommazzo}, \citenamefont {Besançon},
  \citenamefont {Knebel}, \citenamefont {Gelß},\ and\ \citenamefont
  {Pokutta}}]{designolleImprovedLocalModels2023}%
  \BibitemOpen
  \bibfield  {author} {\bibinfo {author} {\bibfnamefont {S.}~\bibnamefont
  {Designolle}}, \bibinfo {author} {\bibfnamefont {G.}~\bibnamefont
  {Iommazzo}}, \bibinfo {author} {\bibfnamefont {M.}~\bibnamefont {Besançon}},
  \bibinfo {author} {\bibfnamefont {S.}~\bibnamefont {Knebel}}, \bibinfo
  {author} {\bibfnamefont {P.}~\bibnamefont {Gelß}},\ and\ \bibinfo {author}
  {\bibfnamefont {S.}~\bibnamefont {Pokutta}},\ }\href
  {https://doi.org/10.1103/PhysRevResearch.5.043059} {\bibfield  {journal}
  {\bibinfo  {journal} {Physical Review Research}\ }\textbf {\bibinfo {volume}
  {5}},\ \bibinfo {pages} {043059} (\bibinfo {year} {2023})}\BibitemShut
  {NoStop}%
\bibitem [{\citenamefont {Moore}(1920)}]{mooreReciprocalGeneralAlgebraic1920}%
  \BibitemOpen
  \bibfield  {author} {\bibinfo {author} {\bibfnamefont {E.~H.}\ \bibnamefont
  {Moore}},\ }\href@noop {} {\bibfield  {journal} {\bibinfo  {journal} {Bull.
  Amer. Math. Soc.}\ }\textbf {\bibinfo {volume} {26}},\ \bibinfo {pages} {394}
  (\bibinfo {year} {1920})}\BibitemShut {NoStop}%
\bibitem [{\citenamefont
  {Penrose}(1955)}]{penroseGeneralizedInverseMatrices1955}%
  \BibitemOpen
  \bibfield  {author} {\bibinfo {author} {\bibfnamefont {R.}~\bibnamefont
  {Penrose}},\ }\href {https://doi.org/10.1017/S0305004100030401} {\bibfield
  {journal} {\bibinfo  {journal} {Mathematical Proceedings of the Cambridge
  Philosophical Society}\ }\textbf {\bibinfo {volume} {51}},\ \bibinfo {pages}
  {406} (\bibinfo {year} {1955})}\BibitemShut {NoStop}%
\end{thebibliography}%


\begin{thebibliography}{3}%
\makeatletter
\providecommand \@ifxundefined [1]{%
 \@ifx{#1\undefined}
}%
\providecommand \@ifnum [1]{%
 \ifnum #1\expandafter \@firstoftwo
 \else \expandafter \@secondoftwo
 \fi
}%
\providecommand \@ifx [1]{%
 \ifx #1\expandafter \@firstoftwo
 \else \expandafter \@secondoftwo
 \fi
}%
\providecommand \natexlab [1]{#1}%
\providecommand \enquote  [1]{``#1''}%
\providecommand \bibnamefont  [1]{#1}%
\providecommand \bibfnamefont [1]{#1}%
\providecommand \citenamefont [1]{#1}%
\providecommand \href@noop [0]{\@secondoftwo}%
\providecommand \href [0]{\begingroup \@sanitize@url \@href}%
\providecommand \@href[1]{\@@startlink{#1}\@@href}%
\providecommand \@@href[1]{\endgroup#1\@@endlink}%
\providecommand \@sanitize@url [0]{\catcode `\\12\catcode `\$12\catcode
  `\&12\catcode `\#12\catcode `\^12\catcode `\_12\catcode `\%12\relax}%
\providecommand \@@startlink[1]{}%
\providecommand \@@endlink[0]{}%
\providecommand \url  [0]{\begingroup\@sanitize@url \@url }%
\providecommand \@url [1]{\endgroup\@href {#1}{\urlprefix }}%
\providecommand \urlprefix  [0]{URL }%
\providecommand \Eprint [0]{\href }%
\providecommand \doibase [0]{http://dx.doi.org/}%
\providecommand \selectlanguage [0]{\@gobble}%
\providecommand \bibinfo  [0]{\@secondoftwo}%
\providecommand \bibfield  [0]{\@secondoftwo}%
\providecommand \translation [1]{[#1]}%
\providecommand \BibitemOpen [0]{}%
\providecommand \bibitemStop [0]{}%
\providecommand \bibitemNoStop [0]{.\EOS\space}%
\providecommand \EOS [0]{\spacefactor3000\relax}%
\providecommand \BibitemShut  [1]{\csname bibitem#1\endcsname}%
\let\auto@bib@innerbib\@empty
\bibitem [{\citenamefont
  {Landau}(1988)}]{landauEmpiricalTwopointCorrelation1988}%
  \BibitemOpen
  \bibfield  {author} {\bibinfo {author} {\bibfnamefont {L.~J.}\ \bibnamefont
  {Landau}},\ }\href {\doibase 10.1007/BF00732549} {\bibfield  {journal}
  {\bibinfo  {journal} {Foundations of Physics}\ }\textbf {\bibinfo {volume}
  {18}},\ \bibinfo {pages} {449} (\bibinfo {year} {1988})}\BibitemShut
  {NoStop}%
\bibitem [{\citenamefont
  {Masanes}(2003)}]{masanesNecessarySufficientCondition2003}%
  \BibitemOpen
  \bibfield  {author} {\bibinfo {author} {\bibfnamefont {L.}~\bibnamefont
  {Masanes}},\ }\href {\doibase 10.48550/arXiv.quant-ph/0309137} {\enquote
  {\bibinfo {title} {Necessary and sufficient condition for quantum-generated
  correlations},}\ } (\bibinfo {year} {2003}),\ \bibinfo {note}
  {arXiv:quant-ph/0309137}\BibitemShut {NoStop}%
\bibitem [{\citenamefont {Fukuda}(2003)}]{fukudaCddlibReferenceManual2003}%
  \BibitemOpen
  \bibfield  {author} {\bibinfo {author} {\bibfnamefont {K.}~\bibnamefont
  {Fukuda}},\ }\href@noop {} {\enquote {\bibinfo {title} {Cddlib reference
  manual},}\ } (\bibinfo {year} {2003})\BibitemShut {NoStop}%
\end{thebibliography}%

\clearpage

\appendix
\section*{End Matter}
\label{appendices}

We present a gradient-based method to optimize Bell inequalities.
The key idea is to iteratively update the coefficient vector $\vec T$ in order to
minimize the quantum value $\beta_Q$, while keeping the classical bound
$\beta_C$ fixed by imposing the classical vertices saturated as
constraints.

We consider a Bell inequality specified by a coefficient vector $\vec T$ in
probability space, and set the value of the Bell functional for the white-noise
behavior $\vec P_{\mathrm{uni}}$ to zero,
$\vec T\cdot\vec P_{\mathrm{uni}}=0$.
Let
\begin{align}
S(\vec T)
=
\{\vec v_j\}_j
:=
\{\vec v\in\mathrm{vert}(\mathcal L)\mid \vec T\cdot\vec v=\beta_C\}\label{eq:SvecT}
\end{align}
denote the set of vertices saturating the classical bound $\vec T\cdot \vec v_j=\beta_C$.
All the other vertices of $\mathcal L$ satisfy $\vec T\cdot\vec v>\beta_C$.
    If the vertices in $S(\vec T)$ do not affinely span a facet, then $S(\vec T)$ does not uniquely determine the coefficient vector $\vec T$.

To explore this solution space, we arrange the saturating vertices into a
matrix $V$, with each row corresponding to one vertex $\vec v_j$. The allowed
change of the coefficient vector is then constrained by
\begin{align}
V\,\delta\vec T=0 ,
\label{eq:constraintspacebyV}
\end{align}
where $\vec T$ is treated as a column vector.
We restrict $\vec T$ and every update $\delta\vec T$ to the reduced coefficient space compatible with normalization and nonsignaling constraints.
Within this constrained subspace, we perform a projected gradient ascent to increase the quantum-to-classical
ratio $\Delta=\beta_Q/\beta_C$:
\begin{align}
\delta\vec T
=
s
\left[
I
-
V^\top(VV^\top)^+V
\right]
\frac{\partial\Delta}{\partial\vec T}.
\label{eq:updatemethod}
\end{align}
Here $s$ is the step size, $\partial\Delta/\partial\vec T$ is the gradient of
the ratio, $I$ is the identity matrix, $\top$ denotes transposition, and $+$
denotes the Moore--Penrose pseudoinverse
\cite{mooreReciprocalGeneralAlgebraic1920,penroseGeneralizedInverseMatrices1955}. If the projected
gradient vanishes, we replace it by a small random perturbation to continue the
search.

The derivative of the ratio is
\begin{align}
\frac{\partial\Delta}{\partial\vec T}
=
\frac{1}{\beta_C}
\frac{\partial\beta_Q}{\partial\vec T}
-
\frac{\beta_Q}{\beta_C^2}
\frac{\partial\beta_C}{\partial\vec T}.
\end{align}
For the Bell operator $\hat I(\boldsymbol\theta)$ in
Eq.~\eqref{eq:variationalbetaQ}, the derivative of the quantum value follows
from the Feynman--Hellmann theorem:
\begin{align}
\frac{\partial \beta_Q}{\partial \vec T}
&=
\left\langle \psi^{*}\right|
\frac{\partial}{\partial \vec T}
\hat{I}(\boldsymbol{\theta}^{*})
\left|\psi^{*}\right\rangle
+
\frac{\partial \boldsymbol{\theta}^{*}}{\partial \vec T}\,
\underbrace{
\left\langle \psi^{*}\right|
\left.
\frac{\partial}{\partial \boldsymbol{\theta}}
\hat{I}(\boldsymbol{\theta})
\right|_{\boldsymbol{\theta}=\boldsymbol{\theta}^{*}}
\left|\psi^{*}\right\rangle
}_{=0}.
\end{align}
Here $|\psi^*\rangle$ and $\boldsymbol\theta^*$ denote the optimal quantum state
and measurement settings, respectively. The second term vanishes because
$\boldsymbol\theta^*$ minimizes the quantum value $\beta_Q$.

Equation~\eqref{eq:constraintspacebyV} preserves the values of the currently identified saturating vertices. Therefore,
the classical bound $\beta_C$ remains unchanged as long as no additional vertex
becomes active. 
After updating $\vec T$, however, a new vertex may attain
the classical bound.
Once such a vertex is identified, it is appended to $V$, and the candidate
coefficient vector is projected onto the enlarged affine constraints.

\begin{figure}
    \centering
    \includegraphics[width=\linewidth]{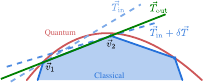}
    \caption{
    Two-dimensional projection of one constrained gradient update step in
    Algorithm~\ref{alg:finding_new_vertices}. The input coefficient vector
    $\vec T_{\mathrm{in}}$ is updated along the projected quantum-gradient direction
    to $\vec T_{\mathrm{in}}+\delta\vec T$, where a new vertex $\vec v_2$ becomes
    active. We then project $\vec T_{\mathrm{in}}+\delta\vec T$ onto the enlarged
    classical constraint subspace and obtain $\vec T_{\mathrm{out}}$. The blue region
    denotes the classical polytope, and the red curve denotes the quantum
    boundary.
    }
    \label{fig:illustrateofalgorithm1}
\end{figure}

\begin{algorithm}[H]
\caption{Constrained Gradient Update}
\label{alg:finding_new_vertices}
\begin{algorithmic}[1]
\STATE \textbf{Input:}
$s$, $\varepsilon$, $\vec{T}_{\mathrm{in}}$, $V_{\mathrm{in}}$, $\beta_{Q,\mathrm{in}}$,
$\boldsymbol{\theta}^*_{\mathrm{in}}$, $\ket{\psi^*_{\mathrm{in}}}$,
$\left.\partial\beta_Q/\partial\vec{T}\right|_{\vec{T}=\vec{T}_{\mathrm{in}}}$

\STATE \textbf{Output:}
$\vec{T}_{\mathrm{out}}$, $V_{\mathrm{out}}$, $\beta_{Q,\mathrm{out}}$,
$\boldsymbol{\theta}^*_{\mathrm{out}}$, $\ket{\psi^*_{\mathrm{out}}}$,
$\left.\partial\beta_Q/\partial\vec{T}\right|_{\vec{T}=\vec{T}_{\mathrm{out}}}$

\IF{$\mathrm{affine\_rank}(V_{\mathrm{in}})
=\dim(\mathcal{L})-1$}
  \STATE \textbf{Return}
  $(\vec{T}_{\mathrm{in}},V_{\mathrm{in}},\beta_{Q,\mathrm{in}},
  \boldsymbol{\theta}_{\mathrm{in}}^{*},
  \ket{\psi_{\mathrm{in}}^{*}},
  \left.\partial\beta_Q/\partial\vec{T}
  \right|_{\vec{T}=\vec{T}_{\mathrm{in}}})$
\ENDIF

\STATE
$\delta\vec{T}
\leftarrow
-s\left[
I-
V_{\mathrm{in}}^{\top}
(V_{\mathrm{in}}V_{\mathrm{in}}^{\top})^{+}
V_{\mathrm{in}}
\right]
\left.
{\partial\beta_Q}/{\partial\vec{T}}
\right|_{\vec{T}=\vec{T}_{\mathrm{in}}}$

\STATE
$\vec{T}\leftarrow\vec{T}_{\mathrm{in}}+\delta\vec{T}$

\STATE Evaluate the classical bound at $\vec{T}$ and identify a new minimizing vertex
$V_{\mathrm{add}}$

\IF{$V_{\mathrm{add}}\neq\emptyset$}

  \STATE
  $V\leftarrow\mathrm{merge}(V_{\mathrm{in}},V_{\mathrm{add}})$
\STATE Project $\vec T$ onto the affine constraint  $V\vec{T}_{\mathrm{out}}=-\boldsymbol{1}$:
$$\vec{T}_{\mathrm{out}}
  \leftarrow
  \vec{T}
  -
  V^{\top}(VV^{\top})^{+}
  \bigl(V\vec{T}+\boldsymbol{1}\bigr)$$
  \STATE Evaluate $\beta_C(\vec{T}_{\mathrm{out}})$ at $\vec{T}_{\mathrm{out}}$
  and set
  $V_{\mathrm{out}}\leftarrow S(\vec{T}_{\mathrm{out}})$
\IF{$\beta_C(\vec{T}_{\mathrm{out}})<-1-\varepsilon$}
    \STATE Decrease $s$ and repeat the update
  \ENDIF
\ELSE
  \STATE
$\vec{T}_{\mathrm{out}}\leftarrow\vec{T}$

  \STATE
  $V_{\mathrm{out}}\leftarrow V_{\mathrm{in}}$
\ENDIF
\STATE Evaluate $\beta_{Q,\mathrm{out}}$,
$\boldsymbol{\theta}^*_{\mathrm{out}}$,
$\ket{\psi^*_{\mathrm{out}}}$, and
$\left.\partial\beta_Q/\partial\vec{T}
\right|_{\vec{T}=\vec{T}_{\mathrm{out}}}$
at $\vec{T}_{\mathrm{out}}$

\STATE \textbf{Return}
$(\vec{T}_{\mathrm{out}},V_{\mathrm{out}},\beta_{Q,\mathrm{out}},
\boldsymbol{\theta}^*_{\mathrm{out}},
\ket{\psi^*_{\mathrm{out}}},
\left.\partial\beta_Q/\partial\vec{T}
\right|_{\vec{T}=\vec{T}_{\mathrm{out}}})$

\end{algorithmic}
\end{algorithm}

Algorithm~\ref{alg:finding_new_vertices} summarizes one step of the
constrained gradient update. We normalize the coefficients such that $\beta_C=-1$, and the ratio becomes $\Delta=\beta_Q/\beta_C=-\beta_Q$, so
maximizing the quantum-to-classical ratio is equivalent to minimizing
$\beta_Q$.  Starting from the input coefficient vector
$\vec T_{\mathrm{in}}$ and the currently active classical vertices $V_{\mathrm{in}}$, the
algorithm first checks whether these vertices span 
a facet of the projected local polytope. If
$
\operatorname{affine\_rank}(V_{\mathrm{in}})=\dim(\mathcal L)-1,
$
then the current inequality is already tight, and the algorithm returns the
current coefficient vector and quantum optimum.

Otherwise, the coefficients are updated along the projected negative gradient
of the optimized quantum value according to Eq.~\eqref{eq:updatemethod}.
The projection ensures that the update stays within the null space of
$V_{\mathrm{in}}$, and therefore preserves the current constraints.
The intermediate coefficient vector is
$\vec T=\vec T_{\mathrm{in}}+\delta\vec T$.
We evaluate $\beta_C(\vec T)$ at $\vec T$. 
If
$\beta_C(\vec T)<-1$, a new minimizing vertex $\vec v_{\mathrm{add}}\notin V_{\mathrm{in}}$ is identified, 
and we set $V_{\mathrm{add}}=\{\vec v_{\mathrm{add}}\}$. 
\space We then update the active vertex set by
$V=\operatorname{merge}(V_{\mathrm{in}},V_{\mathrm{add}})$ and project
the candidate $\vec T$ onto the affine constraint
$V\vec T_{\mathrm{out}}=-\boldsymbol 1$.
By construction, the affine projection ensures that all vertices in
$V=\operatorname{merge}(V_{\mathrm{in}},V_{\mathrm{add}})$ attain the
value $-1$ at $\vec T_{\mathrm{out}}$.
We then reevaluate the classical bound at $\vec T_{\mathrm{out}}$.
If $\beta_C(\vec T_{\mathrm{out}})\geq-1-\varepsilon$, the update is
accepted and we set
$V_{\mathrm{out}}=S(\vec T_{\mathrm{out}})$ according to
Eq.~\eqref{eq:SvecT}. If
$\beta_C(\vec T_{\mathrm{out}})<-1-\varepsilon$, the step size $s$ is
reduced and the update is repeated. Here, $\varepsilon>0$ is a numerical
tolerance used to verify that the normalized classical bound remains
$\beta_C=-1$.
If no new minimizing vertex is found, we set
$\vec T_{\mathrm{out}}=\vec T$ and
$V_{\mathrm{out}}=V_{\mathrm{in}}$.

Finally, the quantum optimization is repeated at the new coefficients
$\vec T_{\mathrm{out}}$. The optimized quantum value
$\beta_{Q,\mathrm{out}}$, the optimal measurement settings
$\boldsymbol\theta^*_{\mathrm{out}}$, the optimal state
$\ket{\psi^*_{\mathrm{out}}}$, and the new gradient
$\left.\partial\beta_Q/\partial\vec T\right|_{\vec T=\vec T_{\mathrm{out}}}$
are all recomputed. These quantities are then used as the input for the next
iteration.
If the
projected update $\delta\vec T$ vanishes, a small random perturbation within the
constraint subspace can be used to continue the search. The constrained gradient update rule can also be applied to optimize correlator coefficients
$\vec\alpha$.

\end{document}


\begin{CJK*}{UTF8}{gbsn} 
\crefname{equation}{Eq.}{Eqs.}
\crefname{figure}{Fig.}{Fig.}
\crefname{appendix}{Appendix}{Appendix}

\title{Supplemental Material: Optimizing quantum violation yields facet Bell inequalities}

\author{Jin-Fu Chen (陈劲夫)\orcidlink{0000-0002-7207-969X}}
\email{jinfuchen@lorentz.leidenuniv.nl}
\affiliation{Instituut-Lorentz, Universiteit Leiden, P.O. Box 9506, 2300 RA Leiden, The Netherlands}
\affiliation{$\langle aQa^L\rangle$ Applied Quantum Algorithms, Universiteit Leiden, The Netherlands}
\author{Mengyao Hu (胡梦瑶)
\orcidlink{0000-0003-2621-3365}}
\email{mengyao@lorentz.leidenuniv.nl}
\affiliation{Instituut-Lorentz, Universiteit Leiden, P.O. Box 9506, 2300 RA Leiden, The Netherlands}
\affiliation{$\langle aQa^L\rangle$ Applied Quantum Algorithms, Universiteit Leiden, The Netherlands}
\affiliation{Centre for Quantum Technologies, National University of Singapore, 3 Science Drive 2, Singapore 117543, Singapore}
\author{Jordi Tura
\orcidlink{0000-0002-6123-1422}}
\email{tura@lorentz.leidenuniv.nl}
\affiliation{Instituut-Lorentz, Universiteit Leiden, P.O. Box 9506, 2300 RA Leiden, The Netherlands}
\affiliation{$\langle aQa^L\rangle$ Applied Quantum Algorithms, Universiteit Leiden, The Netherlands}
\date{\today}

\maketitle
\end{CJK*}

\onecolumngrid

\renewcommand{\thefigure}{S\arabic{figure}}
\renewcommand{\theequation}{S\arabic{equation}}
\renewcommand{\bibnumfmt}[1]{[S#1]}
\renewcommand{\citenumfont}[1]{S#1}
\renewcommand{\thetable}{S\Roman{table}}

\tableofcontents

The supplementary material is devoted to providing detailed derivations in the main text. 
In Section~\ref{sec:prooftheorem1}, we prove Theorem 1 of the main text.
In Section~\ref{sec:222}, we visualize the local polytope and the quantum set for the symmetric bipartite scenario.
In Section~\ref{sec:Nm2}, we introduce the basics of permutation-invariant (PI) Bell inequalities for the $(N,m,2)$ scenario with up to two-body correlators, where each of the $N$ parties performs one of the $m$ possible measurements with binary outcomes $\pm1$. 
In Section~\ref{sec:N22}, we enumerate the facet Bell inequalities of the PI Bell polytopes for $m=2$ and $m=3$ with small numbers of parties $N$.

\section{Proof of Theorem 1}\label{sec:prooftheorem1}
We prove Theorem~1 of the main text in the following. We consider the \((N,m,d)\) Bell scenario with \(N\) parties, \(m\) measurement settings per party, and \(d\) outcomes per measurement. The linear functional is 
\begin{equation}\label{eq:bell-functional-full}
I(\vec P)=\vec T\cdot \vec P
=\sum_{\boldsymbol a,\boldsymbol x} T_{\boldsymbol x}^{\boldsymbol a}\,P(\boldsymbol a\vert \boldsymbol x),
\end{equation}
where $\vec{T}$ and $\vec{P}$ denote the coefficient and the probability vectors, respectively.

Normalization and nonsignaling impose linear relations among the components
of $\vec P$. For example, consider the bipartite scenario with
$N=2$, $m=2$, and $d=2$, where
$a,b,x,y\in\{0,1\}$. Any nonsignaling behavior is completely specified by $\left\{P(00\vert xy)\right\}_{x,y}$, $\left\{P_A(0\vert x)\right\}_{x}$, and $\left\{P_B(0\vert y)\right\}_{y}$.
The remaining probabilities are determined by $P(01\vert xy)
=P_A(0\vert x)-P(00\vert xy)$, $P(10\vert xy)
=P_B(0\vert y)-P(00\vert xy)$, and $P(11\vert xy)
=1-P_A(0\vert x)-P_B(0\vert y)+P(00\vert xy)$.
Only eight independent coordinates are required instead of the
sixteen components of the full probability table.
The coefficient vectors that
differ by combinations of the normalization and nonsignaling relations
define the same linear functional 
up to an additive constant on the nonsignaling set, and the constant is fixed by Eq.~\eqref{eq:uniform-gauge} below. These
redundancies can be removed by using reduced probability coordinates or a
correlator representation. To keep the proof independent of any particular
representation, we work with the probability representation and the
original coefficients. Throughout the proof, coefficient vectors that agree
on all nonsignaling behaviors are regarded as equivalent.

Furthermore, we choose the uniform behavior
$P_\mathrm{uni}(\boldsymbol a\vert\boldsymbol x)=1/d^N$
as the reference point and set the linear functional to zero
\begin{equation}\label{eq:uniform-gauge}
I(\vec P_\mathrm{uni})
=
\sum_{\boldsymbol a,\boldsymbol x}
T_{\boldsymbol x}^{\boldsymbol a}/d^N
=0.
\end{equation}
For a coefficient vector $\vec T$, we define the classical bound, the quantum value, and the quantum-to-classical ratio as
\begin{equation}
\beta_C(\vec T)=\min_{\vec P\in\mathcal L}\vec T\cdot\vec P,
\qquad
\beta_Q(\vec T)=\min_{\vec P\in\mathcal Q}\vec T\cdot\vec P,
\qquad
\Delta(\vec T)=\frac{\beta_Q(\vec T)}{\beta_C(\vec T)} .
\end{equation}
The ratio characterizes robustness to white noise. For a
quantum behavior $\vec P_Q$ attaining $\beta_Q$, let
$\vec P_v=v\vec P_Q+(1-v)\vec P_{\mathrm{uni}}$. The critical visibility
for violating the Bell inequality is $v_c=1/\Delta$.
We note that $\vec P_{\mathrm{uni}}$ is an interior point of the local polytope
$\mathcal L$ relative to its affine hull. Any nonzero linear functional satisfying~\eqref{eq:uniform-gauge}
takes both positive and negative values on $\mathcal L$,
hence $\beta_Q(\vec T)\leq\beta_C(\vec T)<0$.
After quotienting the normalization and nonsignaling redundancies, the dual polytope is given by
\begin{equation}
\mathcal T=\{\vec T:\vec T\cdot\vec v\geq -1,\ \forall\,\vec v\in\mathrm{vert}(\mathcal L)\},
\end{equation}
where the constraints include all vertices $\vec v$ of the local polytope. Consequently, $\mathcal T$ is bounded and closed, and is therefore
compact in the finite-dimensional reduced coefficient space.
Because both bounds are positively homogeneous in $\vec T$, we fix the scale by working on the boundary
\begin{equation}\label{eq:normalized-slice}
\partial\mathcal T
=
\{\vec T:\beta_C(\vec T)=-1\},
\end{equation}
where the inequality is saturated by at least one vertex, and 
the scaling freedom is removed. 
Maximizing $\Delta(\vec T)$ over $\partial\mathcal T$ is equivalent to minimizing $\beta_Q(\vec T)$ over $\partial\mathcal T$. A vector
$\vec T\in \partial \mathcal{T}$ is tight if $\vec{T}\cdot \vec P\geq -1$ defines a facet of $\mathcal{L}$.
The extreme points of $\mathcal{T}$ are in one-to-one correspondence with the facets of $\mathcal{L}$. 

Define the active vertices of $\vec T \in \partial \mathcal T$ as
\begin{equation}
S(\vec T)
=
\left\{
\vec v\in\mathrm{vert}(\mathcal L):
\vec T\cdot\vec v=-1
\right\}.
\label{eq:active-vertices}
\end{equation}
For $\vec T\in\partial\mathcal T$, define
\begin{equation}
F(\vec T)
=
\left\{
\vec T'\in\mathcal T:
\vec T'\cdot\vec v=-1,\ 
\forall\,\vec v\in S(\vec T)
\right\}.
\label{eq:smallest-face}
\end{equation}
Thus, $F(\vec T)$ consists of all coefficient vectors in $\mathcal T$
for which every vertex active at $\vec T$ remains active. The condition
$\vec T'\in\mathcal T$ also ensures
$\vec T'\cdot\vec w\geq-1$ for every other local vertex
$\vec w\notin S(\vec T)$. Therefore, $F(\vec T)$ is the intersection of
$\mathcal T$ with all supporting hyperplanes active at $\vec T$, which is the smallest face containing $\vec T$.
We further define
\begin{equation}
\delta F(\vec T)
=
\left\{
\delta\vec T:
\delta\vec T\cdot\vec v=0,\ 
\forall\,\vec v\in S(\vec T)
\right\}.
\label{eq:face-direction-space}
\end{equation}
It is the linear
subspace parallel to the affine hull of $F(\vec T)$. For
$\vec T^\star$, we write
$F^\star=F(\vec T^\star)$ and
$\delta F^\star=\delta F(\vec T^\star)$.
We now state and prove the theorem.
\begin{theorem}[Nondegenerate local maximum implies tightness]
\label{thm:local-opt-tight}
Let $\vec T^\star\in\partial\mathcal T$ be a local maximizer of $\Delta$ on $\partial\mathcal T$.

\textnormal{(i)} If there is no nonzero
direction $\delta\vec T\in\delta F^\star$
along which $\Delta$ is locally constant, then the hyperplane
$\vec T^\star\cdot\vec P=-1$ defines a facet of $\mathcal L$.

\textnormal{(ii)} If $\vec T^\star$ is non-tight, then
$\Delta$ is locally constant along every
direction $\delta\vec T\in\delta F^\star$.

\textnormal{(iii)} If $\vec T^\star$ is non-tight, then every extreme point
$\vec T_{\mathrm f}$ of the smallest face $F^\star$ defines a
facet of $\mathcal L$, and $\Delta$ is nondecreasing along the segment from
$\vec T^\star$ to $\vec T_{\mathrm f}$. In particular,
$\Delta(\vec T_{\mathrm f})\geq\Delta(\vec T^\star)$.
\end{theorem}
\begin{proof}
We first prove part~\textnormal{(ii)}. 
Suppose that $\vec T^\star$ is non-tight. Since the extreme points of
$\mathcal T$ are in one-to-one correspondence with the facets of
$\mathcal L$, $\vec T^\star$ is not an extreme point of $\mathcal T$.
Hence the smallest face $F^\star$ containing $\vec T^\star$ has positive
dimension, and therefore $\delta F^\star$ contains a nonzero direction.

Let $\delta\vec T\in\delta F^\star$ be arbitrary and define
\begin{equation}
\vec T(t)=\vec T^\star+t\,\delta\vec T.
\label{eq:coefficient-variation}
\end{equation}
By the definition of $\delta F^\star$, all active vertices continue to
attain
$\vec T(t)\cdot\vec v
=
\vec T^\star\cdot\vec v
+t\,\delta\vec T\cdot\vec v
=
-1$.
Every inactive vertex has a value strictly greater than $-1$ at
$\vec T^\star$. Since $\mathcal L$ has finitely many vertices, this remains
true for all sufficiently small $|t|$. Therefore, there exists
$\epsilon>0$ such that
\begin{equation}
\beta_C(\vec T(t))=-1,
\qquad
|t|<\epsilon.
\label{eq:constant-classical-bound}
\end{equation}
Hence, for $|t|<\epsilon$, we have
$\vec T(t)\in\partial\mathcal T$ according to Eq.~\eqref{eq:normalized-slice}.

For the quantum value $\beta_Q$, we restrict it to the
line $\vec T(t)$ and define
the single-variable function as
\begin{equation}
h(t)=\beta_Q(\vec T(t)).
\label{eq:h-definition}
\end{equation}
Within the interval $|t|<\epsilon$, since $\beta_C=-1$,
the ratio is 
\begin{equation}
\Delta(\vec T(t))=-h(t).
\label{eq:ratio-along-variation}
\end{equation}
The function $h(t)$ is concave because
$\beta_Q(\vec T)=\min_{\vec P\in\mathcal Q}\vec T\cdot\vec P$.
Since $\vec T^\star$ is a local maximizer of $\Delta$
on $\partial\mathcal T$, $t=0$ is a local minimizer of
$h(t)$. For small enough $\epsilon$, the local minimizer leads to
\begin{equation}
h(t)\geq h(0),
\qquad
h(-t)\geq h(0),
\qquad
|t|<\epsilon.
\label{eq:local-minimum-h}
\end{equation}
On the other hand, concavity gives
\begin{equation}
h(0)
\geq
\frac{h(t)+h(-t)}{2}.
\label{eq:concavity-chain}
\end{equation}
The above two inequalities must therefore be equalities, implying
\begin{equation}
h(t)=h(0),								
\qquad
|t|<\epsilon.
\label{eq:local-flatness}
\end{equation}
Thus, $\beta_Q$ and $\Delta$ are locally constant along
$\delta\vec T$. Since $\delta\vec T\in\delta F^\star$ was arbitrary,
they are locally constant along
every direction in $\delta F^\star$.
This proves part~\textnormal{(ii)}.

In particular, every non-tight local maximizer has a nonzero flat
direction. Part~\textnormal{(i)} is proven by taking the contrapositive.

Finally, we prove part~\textnormal{(iii)}.
Since $\vec T^\star$ is non-tight, $F^\star$ has nonzero dimension.
Since $\mathcal T$ is a compact polytope, its nonempty face
$F^\star$ is also compact and therefore has extreme points.
Let $\vec T_{\mathrm f}$ be an arbitrary extreme point of $F^\star$ and define
\begin{equation}
\delta\vec T=\vec T_{\mathrm f}-\vec T^\star.
\end{equation}
Because every vertex in $S(\vec T^\star)$ remains active throughout
$F^\star$, we have $\delta\vec T\in\delta F^\star$, and the segment
$\vec T(t)=\vec T^\star+t\,\delta\vec T$ with
$0\leq t\leq1$
lies entirely in $F^\star$.

On this segment, the classical bound is $\beta_C=-1$,
and we define
$h(t)=\beta_Q(\vec T(t))$. Therefore,
$\Delta(\vec T(t))=-h(t)$ is convex as a function of $t$. By
part~\textnormal{(ii)}, for sufficiently small $\epsilon$,
\begin{equation}
h(t)=h(0),
\qquad
0\leq t\leq\epsilon.
\end{equation}
For any $0<\epsilon<t_1<t_2\leq1$, the successive secant
slopes of the concave function $h$ satisfy
\begin{equation}
0
=
\frac{h(\epsilon)-h(0)}{\epsilon}
\geq
\frac{h(t_1)-h(\epsilon)}{t_1-\epsilon}
\geq
\frac{h(t_2)-h(t_1)}{t_2-t_1}.
\label{eq:concave-secant-slopes}
\end{equation}
The first inequality implies
$h(t_1)\leq h(\epsilon)$, and the second then implies
$h(t_2)\leq h(t_1)$. 
Therefore, $\Delta(\vec T(t) )=-h(t)$ is nondecreasing on the interval $0\leq t \leq 1$, and satisfies
\begin{equation}
\Delta(\vec T_{\mathrm f})=\Delta(\vec T(1))
\geq
\Delta(\vec T^\star)=\Delta(\vec T (0)).
\end{equation}
Since an extreme point of a face of $\mathcal T$ is also an
extreme point of $\mathcal T$, $\vec T_{\mathrm f}$ defines a facet of
$\mathcal L$. Since $\vec T_{\mathrm f}$ was arbitrary, this proves
part~\textnormal{(iii)}.
\end{proof}

The proof relies only on the concavity of the quantum value, viewed
as the pointwise minimum of functions linear in $\vec T$. Therefore, it remains
valid when $\beta_Q$ is obtained by minimizing over any restricted family of
states and measurements. 
If the coefficient optimization is restricted
to a linear subspace, tightness refers to facets of the corresponding projected
local polytope. For the $(N,m,2)$ scenarios studied in the main text, the
projected local polytope is full-dimensional in the chosen correlator space.

For our algorithm shown in the main text, nontriviality is preserved and termination occurs after finitely many steps. Trivial positivity facets satisfy
$\Delta=1$. In practice,
with a sufficiently small step size $s$, each projected-gradient update does not
decrease $\Delta$. Therefore, if $\Delta(\vec T^\star)>1$, any facet reached
by the algorithm is necessarily a nontrivial Bell inequality.
Whenever the algorithm moves along a segment to the boundary of the current smallest face $F^\star$, one new minimizing vertex outside the affine span of the current active set $S(\vec T^\star)$ becomes active. Therefore, each such update strictly increases the affine rank of the active vertices. Since $\operatorname{vert}(\mathcal L)$ is finite, the algorithm reaches a facet after finitely many steps.

\section{Local polytope and Quantum Set for $(2,2,2)$ scenario}\label{sec:222}

We first consider the $(2,2,2)$ scenario as a benchmark to introduce and validate our optimization framework.
Our gradient optimization converges to the CHSH inequality in this scenario.
We consider the linear functional  restricted to two-body terms, given explicitly by 
\begin{align}
    I = \alpha_{0,0}\expval{A_0 B_0} + \alpha_{0,1}\expval{A_0 B_1} + \alpha_{1,0}\expval{A_1 B_0} + \alpha_{1,1}\expval{A_1 B_1},
\end{align}
where $A_k=A_{k}^{(1)}$ and $B_l=A_{l}^{(2)}$ denote the measurements performed by  Alice and   Bob. The two-body correlator satisfies
\begin{align}
    -1\leq\expval{A_k B_l}\leq 1.
\end{align}
The local hidden variable model gives the following constraints on the correlators
\begin{align}
    \begin{aligned}
        -2&\leq-\expval{A_{0}B_{0}}+\expval{A_{0}B_{1}}+\expval{A_{1}B_{0}}+\expval{A_{1}B_{1}}\leq2,\\-2&\leq\expval{A_{0}B_{0}}-\expval{A_{0}B_{1}}+\expval{A_{1}B_{0}}+\expval{A_{1}B_{1}}\leq2,\\-2&\leq\expval{A_{0}B_{0}}+\expval{A_{0}B_{1}}-\expval{A_{1}B_{0}}+\expval{A_{1}B_{1}}\leq2,\\-2&\leq\expval{A_{0}B_{0}}+\expval{A_{0}B_{1}}+\expval{A_{1}B_{0}}-\expval{A_{1}B_{1}}\leq2.
    \end{aligned}
\end{align}
The quantum set of the two-body correlations is represented by \cite{landauEmpiricalTwopointCorrelation1988,masanesNecessarySufficientCondition2003}
\begin{align}
\begin{aligned}
    -\pi&\leq-\arcsin(\expval{A_{0}B_{0}})+\arcsin(\expval{A_{0}B_{1}})+\arcsin(\expval{A_{1}B_{0}})+\arcsin(\expval{A_{1}B_{1}})\leq\pi,\\-\pi&\leq\arcsin(\expval{A_{0}B_{0}})-\arcsin(\expval{A_{0}B_{1}})+\arcsin(\expval{A_{1}B_{0}})+\arcsin(\expval{A_{1}B_{1}})\leq\pi,\\-\pi&\leq\arcsin(\expval{A_{0}B_{0}})+\arcsin(\expval{A_{0}B_{1}})-\arcsin(\expval{A_{1}B_{0}})+\arcsin(\expval{A_{1}B_{1}})\leq\pi,\\-\pi&\leq\arcsin(\expval{A_{0}B_{0}})+\arcsin(\expval{A_{0}B_{1}})+\arcsin(\expval{A_{1}B_{0}})-\arcsin(\expval{A_{1}B_{1}})\leq\pi.\label{eq:quantum_convex_set_original}
\end{aligned} 
\end{align}
Let us consider the symmetric Bell inequality with $\alpha_{0,1}=\alpha_{1,0}$. We introduce 
\begin{align}
    \langle\mathcal{S}_{0,0}\rangle=\expval{A_{0}B_{0}},\;\langle\mathcal{S}_{0,1}\rangle=(\expval{A_{0}B_{1}}+\expval{A_{1}B_{0}})/2,\;\langle\mathcal{S}_{1,1}\rangle=\expval{A_{1}B_{1}},
\end{align}
and obtain the symmetric bipartite Bell inequality
\begin{align}
I=\alpha_{0,0}\langle\mathcal{S}_{0,0}\rangle+2\alpha_{0,1}\langle\mathcal{S}_{0,1}\rangle+\alpha_{1,1}\langle\mathcal{S}_{1,1}\rangle\geq\beta_{C}.
\end{align}

\begin{figure}
    \centering
\includegraphics[width=0.8\linewidth]{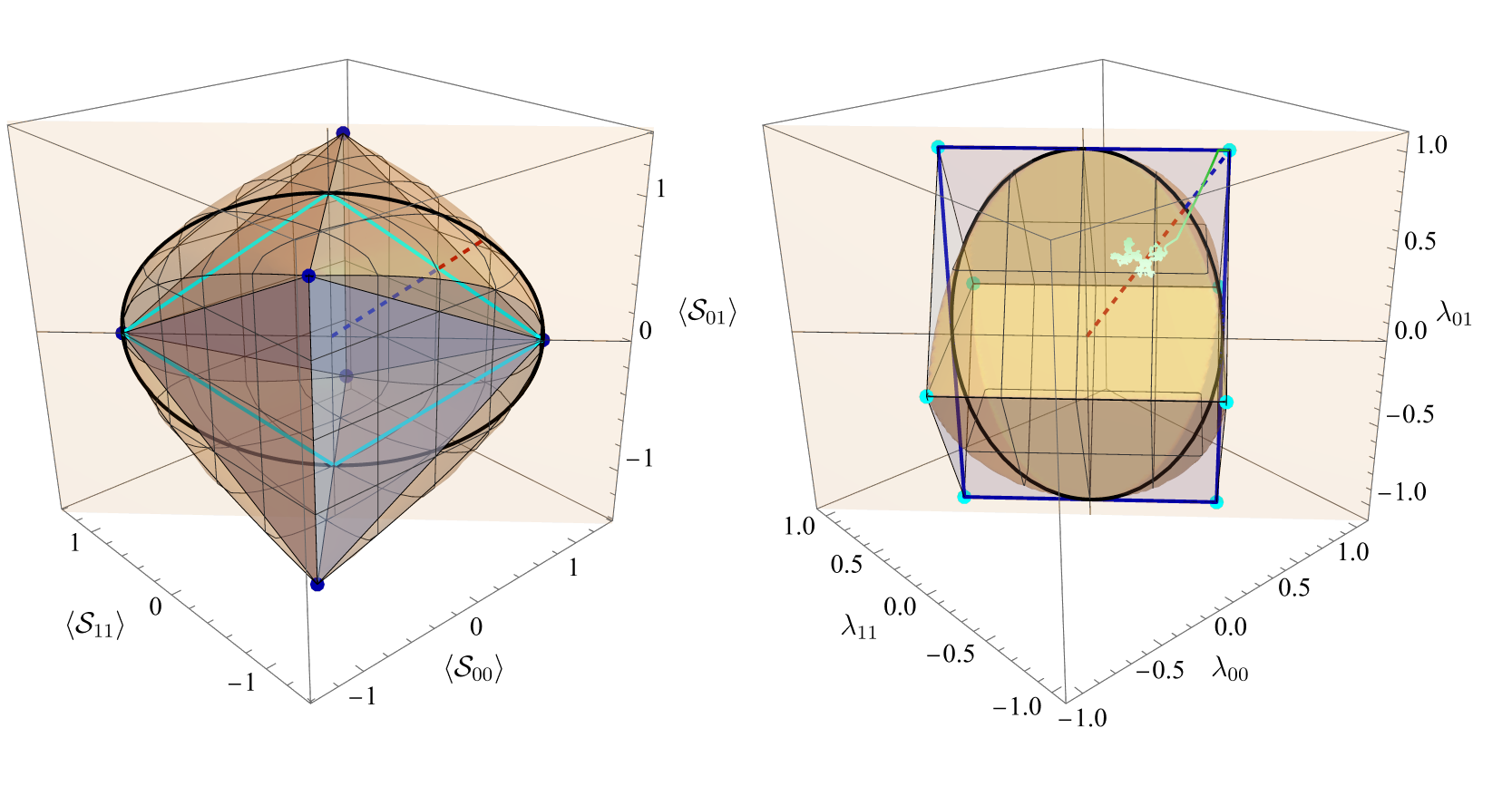}
    \caption{Local polytope and quantum set for bipartite system. (a) The original local polytope  (inside) and quantum set (outside). Four CHSH inequalities correspond to the facets indicated by the cyan lines. (b) The dual polytope (outside) and the dual quantum set (inside). The vertices of the polytope are tight Bell inequalities. 
    The light-green trajectory shows the convergence of the inequality to one of the CHSH inequalities during the optimization.
    In both subfigures, the solid curves show a 2D cross section where the largest ratio is obtained. }
    \label{fig:illustration of the optimization}
\end{figure}

Figure \ref{fig:illustration of the optimization}(a) shows the local polytope (inner octahedron) and the quantum set (outer convex body) for the symmetric $(2,2,2)$ scenario. The local polytope is simplified into
\begin{align}
|\langle\mathcal{S}_{0,1}\rangle|\leq1-|\langle\mathcal{S}_{0,0}\rangle-\langle\mathcal{S}_{1,1}\rangle|/2,
\end{align}
which has six vertices $(\mathcal{S}_{0,0},\mathcal{S}_{0,1},\mathcal{S}_{1,1})=\pm(1,\pm1,1)$ and $\pm(1,0,-1)$ (blue dots). 
The quantum set by Eq.~\eqref{eq:quantum_convex_set_original} is simplified into
\begin{align}
|\langle\mathcal{S}_{0,1}\rangle(\langle\mathcal{S}_{0,0}\rangle-\langle\mathcal{S}_{1,1}\rangle)|\leq(1-\langle\mathcal{S}_{0,1}\rangle^{2})^{1/2}[(1-\langle\mathcal{S}_{0,0}\rangle^{2})^{1/2}+(1-\langle\mathcal{S}_{1,1}\rangle^{2})^{1/2}].
\end{align}
Four CHSH inequalities with the largest ratio correspond to the facets of the local polytope. In a 2D cross section, the CHSH inequalities are shown by the cyan lines, and the black curves show the quantum value. 

Figure \ref{fig:illustration of the optimization}(b) shows the polar transformation of both the local polytope (outer hexahedron) and the quantum set (inner convex body) by solving the dual variables $\lambda_{0,0}$, $\lambda_{0,1}$, and $\lambda_{1,1}$ from
\begin{align}
\lambda_{0,0}\langle\mathcal{S}_{0,0}\rangle+\lambda_{0,1}\langle\mathcal{S}_{0,1}\rangle+\lambda_{1,1}\langle\mathcal{S}_{1,1}\rangle\geq-1.
\end{align}
Each tight Bell inequality corresponds to a cyan vertex of the polytope. 
Four CHSH vertices are 
$(\lambda_{0,0},\lambda_{0,1},\lambda_{1,1})=\pm(1/2,\pm1,-1/2)$. The other four vertices are $(\lambda_{0,0},\lambda_{0,1},\lambda_{1,1})=\pm(1,0,0)$ and $\pm(0,0,1)$. The dual quantum set is given by $\sqrt{(\lambda_{0,0}-\lambda_{1,1})^{2}+\lambda_{0,1}^{2}}\leq1$ and $(|\lambda_{0,0}+\lambda_{1,1}|+|\lambda_{0,1}|)\leq1$.
We also show the two-dimensional cross section that contains the maximum ratio achieved by the CHSH inequalities represented by the four vertices in the cross section.  We show the trajectory during optimization in Fig.~\ref{fig:illustration of the optimization}(b) by the green curve; the step size is chosen as $s=0.01$. 
In the initial stage, if the gradient projected onto the constraint subspace vanishes, we replace the gradient term with a small random perturbation.
We remark that the two-dimensional affine projection and the cross section in both subfigures correspond to Fig.~1 (a) and (b) in the main text.
 
\section{Permutation-invariant Bell Inequalities for $(N,m,2)$ scenario with up to two-body correlators}\label{sec:Nm2}
We consider the \((N,m,2)\) Bell scenario, in which each of the \(N\) parties can choose one of \(m\) measurements, each yielding two outcomes. The linear functional is
\begin{align}
I
=
\sum_{\ell=1}^{N}
\sum_{\substack{
1\le i_1<\cdots<i_\ell\le N\\
k_1,\ldots,k_\ell=0,\ldots,m-1
}}
\alpha^{\,i_1,\ldots,i_\ell}_{k_1,\ldots,k_\ell}
\left\langle
A_{k_1}^{(i_1)}\cdots A_{k_\ell}^{(i_\ell)}
\right\rangle .
\end{align}
Here we focus on permutation-invariant Bell expressions, namely those whose coefficients satisfy
$\alpha_{k_{1},\ldots,k_{\ell}}^{\,i_{1},\ldots,i_{\ell}}=\alpha_{k_{1},\ldots,k_{\ell}}$.
In this setting, it is convenient to introduce the collective observables
\begin{align}
\mathcal{S}_{k_{1},\ldots,k_{\ell}}=\sum_{(i_{1},\ldots,i_{\ell})}A_{k_{1}}^{(i_{1})}\cdots A_{k_{\ell}}^{(i_{\ell})},
\end{align}
where $(i_{1},\ldots,i_{\ell})$ denotes the sum over all ordered $\ell$-tuples of pairwise distinct indices. The linear functional can then be rewritten as
\begin{align}
I=\sum_{\ell=1}^{N}\sum_{0\le k_{1}\le\cdots\le k_{\ell}\le m-1}\alpha_{k_{1},...,k_{\ell}}\frac{\langle\mathcal{S}_{k_{1},...,k_{\ell}}\rangle}{\prod_{v=0}^{m-1}n_{v}!},
\label{eq:PIBell}
\end{align}
where \(n_v\) denotes the multiplicity of the setting \(v\) in the multiset \(\{k_1,\ldots,k_\ell\}\), so that \(\prod_{v=0}^{m-1} n_v!\) is the denominator of the corresponding multinomial coefficient. For example, in the \((N,2,2)\) scenario, a general permutation-invariant linear functional containing correlators up to three-body terms can be written as
\begin{align}
\begin{aligned} I=&\alpha_{0}\,\langle\mathcal{S}_{0}\rangle+\alpha_{1}\,\langle\mathcal{S}_{1}\rangle+\alpha_{0,0}\frac{\langle\mathcal{S}_{0,0}\rangle}{2!}+\alpha_{0,1}\langle\mathcal{S}_{0,1}\rangle+\alpha_{1,1}\frac{\langle\mathcal{S}_{1,1}\rangle}{2!}\\&+\alpha_{0,0,0}\frac{\langle\mathcal{S}_{0,0,0}\rangle}{3!}+\alpha_{0,0,1}\frac{\langle\mathcal{S}_{0,0,1}\rangle}{2!}+\alpha_{0,1,1}\frac{\langle\mathcal{S}_{0,1,1}\rangle}{2!}+\alpha_{1,1,1}\frac{\langle\mathcal{S}_{1,1,1}\rangle}{3!}.
\end{aligned}
\end{align}

In the following, we consider the linear functional with up to two-body correlators 
\begin{align}
I=\sum_{k=0}^{m-1}\alpha_{k}\left\langle \mathcal{S}_{k}\right\rangle +\sum_{k,l=0}^{m-1}\alpha_{k,l}\frac{\left\langle \mathcal{S}_{k,l}\right\rangle }{2},\label{eq:Iwithuptotwobody}
\end{align}
where $\alpha_{k,l}=\alpha_{l,k}$. It is convenient to collect the coefficients into
\begin{align}
\vec{\alpha}=(\alpha_{0},\alpha_{1},\dots,\alpha_{m-1},\alpha_{0,0},\alpha_{0,1},\dots,\alpha_{m-2,m-1},\alpha_{m-1,m-1}),
\end{align}
and the correlators into
\begin{align}
\langle\vec{\mathcal{S}}\rangle=(\langle\mathcal{S}_{0}\rangle,\langle\mathcal{S}_{1}\rangle,\dots,\langle\mathcal{S}_{m-1}\rangle,\frac{1}{2}\langle\mathcal{S}_{0,0}\rangle,\langle\mathcal{S}_{0,1}\rangle,\dots,\langle\mathcal{S}_{m-2,m-1}\rangle,\frac{1}{2}\langle\mathcal{S}_{m-1,m-1}\rangle).
\end{align}
We rewrite \eqref{eq:Iwithuptotwobody} more compactly as 
\begin{align}
    I=\vec{\alpha}\cdot \langle\vec{\mathcal{S}}\rangle.
\end{align}

\subsection{Deterministic local strategy and classical bound}\label{sec:Deterministic local strategy and classical bound}

The deterministic local strategy for a single party is specified by a binary outcome assigned in advance to each of the \(m\) measurements. We denote such a strategy by
\begin{equation}
\boldsymbol{A}=(A_0,\dots,A_{m-1})\in\{\pm1\}^m,
\end{equation}
where \(A_k\) is the predetermined outcome for measurement \(k\).

Since each party must follow one such deterministic local strategy, an \(N\)-party deterministic configuration can be fully characterized, in the permutation-invariant setting, by the occupation numbers
\begin{equation}
N_{\boldsymbol{A}}\in\mathbb{Z}_{\ge 0},
\qquad
\sum_{\boldsymbol{A}\in\{\pm1\}^m} N_{\boldsymbol{A}}=N,
\end{equation}
where \(N_{\boldsymbol{A}}\) counts the number of parties adopting the strategy \(\boldsymbol{A}\). Therefore, for finite \(m\) and finite \(N\), the deterministic local polytope is generated by all integer occupation vectors \(\{N_{\boldsymbol{A}}\}_{\boldsymbol{A}\in\{\pm1\}^m}\) satisfying the above constraints.

All collective one-body and same-site two-body quantities can then be written as
\begin{equation}
\left\langle \mathcal{S}_{k}\right\rangle =\sum_{\boldsymbol{A}}N_{\boldsymbol{A}}\,A_{k},\qquad\left\langle \mathcal{Z}_{k,l}\right\rangle =\sum_{\boldsymbol{A}}N_{\boldsymbol{A}}\,A_{k}A_{l}.\label{eq:SkandZklwithNa}
\end{equation}
Accordingly, the permutation-invariant two-body correlators are
\begin{equation}
\left\langle \mathcal{S}_{k,l}\right\rangle =\left\langle \mathcal{S}_{k}\mathcal{S}_{l}\right\rangle -\left\langle \mathcal{Z}_{k,l}\right\rangle .
\end{equation}

We can rewrite the linear functional \eqref{eq:Iwithuptotwobody} with the deterministic local strategy as 
\begin{align}
I&=\sum_{x,y=0}^{m-1}\alpha_{x,y}\frac{\left\langle \mathcal{S}_{x}\mathcal{S}_{y}\right\rangle }{2}+\sum_{x=0}^{m-1}\alpha_{x}\left\langle \mathcal{S}_{x}\right\rangle -\sum_{x,y=0}^{m-1}\alpha_{x,y}\frac{\left\langle \mathcal{Z}_{x,y}\right\rangle }{2}\\&=\sum_{\boldsymbol{A},\boldsymbol{B}}N_{\boldsymbol{A}}N_{\boldsymbol{B}}\sum_{x,y=0}^{m-1}\alpha_{x,y}\frac{A_{x}B_{y}}{2}+\sum_{\boldsymbol{A}}N_{\boldsymbol{A}}\left(\sum_{x=0}^{m-1}\alpha_{x}A_{x}-\sum_{x,y=0}^{m-1}\alpha_{x,y}\frac{A_{x}A_{y}}{2}\right).
\label{eq:I_Na_form}
\end{align}

Using $AB=1-|A-B|=-1+|A+B|$ for any $A,B\in\{\pm1\}$, we obtain in the classical case
\begin{align}
\left\langle \mathcal{Z}_{k,l}\right\rangle &=N-\sum_{i=1}^{N}\bigl|\langle A_{k}^{(i)}\rangle -\langle A_{l}^{(i)}\rangle \bigl|=-N+\sum_{i=1}^{N}\bigl|\langle A_{k}^{(i)}\rangle +\langle A_{l}^{(i)}\rangle \bigl|.
\label{eq:Zkl222}
\end{align}
Applying the triangle inequality to Eq.~\eqref{eq:Zkl222}, one finds that for each fixed pair \((k,l)\), the corresponding same-site correlator is bounded by
\begin{align}
-N+\bigl|\langle\mathcal S_k\rangle+\langle\mathcal S_l\rangle\bigr|
\le
\langle\mathcal Z_{k,l}\rangle
\le
N-\bigl|\langle\mathcal S_k\rangle-\langle\mathcal S_l\rangle\bigr|.
\label{ineq:Zkl}
\end{align}
For every pair \((k,l)\), the triple
\((\langle\mathcal S_k\rangle,\langle\mathcal S_l\rangle,\langle\mathcal Z_{k,l}\rangle)\)
lies inside a regular tetrahedron, and its coordinates take integer values with the same parity as $N$.

However, Eq.~\eqref{ineq:Zkl} gives only pairwise necessary constraints and
does not fully characterize the classical correlator set. The one-body
correlators $\{\langle\mathcal S_k\rangle\}_k$ and the same-site two-body
quantities $\{\langle\mathcal Z_{k,l}\rangle\}_{k,l}$ must be mutually
compatible: they arise from the same deterministic-strategy occupation
vector $\{N_{\boldsymbol{A}}\}_{\boldsymbol A}$. Therefore, we enumerate the
admissible occupation vectors and map them to correlator vectors
$\langle\vec{\mathcal S}\rangle$, which generate the projected local polytope
$\mathcal L$ in correlator space. Since $I$ is linear in
$\langle\vec{\mathcal S}\rangle$, the classical bound $\beta_C$ is attained at
an extreme point of $\mathcal L$ and can be obtained by optimizing over these
occupation-number configurations.

\subsection{Bell operator and quantum value}\label{sec:Bell operator and quantum value}
For the quantum case, we adopt the symmetric measurement setting \begin{align}
\hat{A}^{(i)}_k=\cos\theta_k\hat \sigma_z^{(i)}+\sin\theta_k\hat \sigma_x^{(i)},
\end{align}
where $\hat{\sigma}^{(i)}_{z}$ and $\hat{\sigma}^{(i)}_{x}$ are Pauli operators of site $i$, and $\theta_k$ is site-independent. We restrict the Hilbert space to the symmetric subspace of $N$ qubits, corresponding to the spin-$N/2$ irreducible representation of SU(2). The value obtained by minimizing in the symmetric subspace gives a variational upper bound on the exact minimum $\beta_Q$.
The observables $\hat{\mathcal{S}}_k=\sum_{i=1}^N \hat{A}_{k}^{(i)}$  can be represented by the collective spin operators
\begin{align}
\hat{\mathcal{S}}_{k}=2\cos\theta_{k}\hat{S}_{z}+2\sin\theta_{k}\hat{S}_{x},\label{eq:Shatk}
\end{align}
where $\hat{S}_{x}$, $\hat{S}_{y}$, and $\hat{S}_{z}$ are the collective spin operators. 
In the \(\hat S_z\) eigenbasis 
\(\lvert M\rangle\), with 
\(M=N/2,N/2-1,\ldots,-N/2\), these matrices are given by
\begin{align}
(\hat S_z)_{M,M'}&=M\delta_{M,M'},\\
(\hat S_x)_{M,M'}&=
\frac{1}{2}\left[
\sqrt{\left(\tfrac{N}{2}-M'\right)\left(\tfrac{N}{2}+M'+1\right)}
\delta_{M,M'+1}
+
\sqrt{\left(\tfrac{N}{2}+M'\right)\left(\tfrac{N}{2}-M'+1\right)}
\delta_{M,M'-1}
\right].
\end{align}
From Eq.~\eqref{eq:Shatk}, the operator $\hat{\mathcal{Z}}_{k,l}$ is explicitly given by
\begin{align}
    \hat{\mathcal{Z}}_{k,l}=N \cos(\theta_k -\theta _l ). \label{eq: hatZkl}
\end{align}
Substituting $\hat{\mathcal{S}}_k$ and $\hat{\mathcal{Z}}_{k,l}$ into Eq.~\eqref{eq:Iwithuptotwobody} gives the Bell operator 
\begin{align}
    \hat{I}(\boldsymbol{\theta})=\sum_{k=0}^{m-1}\alpha_{k}\hat{\mathcal{S}}_{k}+\frac{1}{2}\sum_{k,l=0}^{m-1}\alpha_{k,l}(\hat{\mathcal{S}}_{k}\hat{\mathcal{S}}_{l}-\hat{\mathcal{Z}}_{k,l}).\label{eq:Bell_operator I_theta}
\end{align}
The Bell operator $\hat{I}(\boldsymbol{\theta})$ depends on the coefficients $\vec{\alpha}$ and on the measurement settings $\boldsymbol{\theta}=(\theta_{0},\theta_{1},...,\theta_{m-1})$. 

Generally, the Bell operator is in the form
\begin{align}
\hat{I}(\boldsymbol{\theta})=J+J_{x}\hat{S}_{x}+J_{z}\hat{S}_{z}+\frac{1}{2}J_{x,x}\hat{S}_{x}^{2}+\frac{1}{2}J_{x,z}(\hat{S}_{x}\hat{S}_{z}+\hat{S}_{z}\hat{S}_{x})+\frac{1}{2}J_{z,z}\hat{S}_{z}^{2},
\end{align}
where all the coefficients $J$, $J_x$, $J_z$, $J_{x,x}$, $J_{x,z}$, and $J_{z,z}$ depend on both the coefficients $\vec{\alpha}$ and the measurement settings $\boldsymbol{\theta}$. 
For given coefficients $\vec\alpha$, the variational quantum value is $\beta_{Q}:=\min_{\ket{\psi},\boldsymbol{\theta}}\bra{\psi}\hat{I}(\boldsymbol{\theta})\ket{\psi}=\min_{\boldsymbol{\theta}}\bra{\psi^{*}}\hat{I}(\boldsymbol{\theta})\ket{\psi^{*}}$, where $\ket{\psi^*(\boldsymbol{\theta})}$ is a ground state of the Bell operator for fixed measurement angles $\boldsymbol{\theta}$.

\section{Facet Bell inequalities for $(N,2,2)$ and $(N,3,2)$ scenarios with small $N$}\label{sec:N22}

In the following, we focus on finding tight PI Bell inequalities involving up to two-body correlators in the $(N,m,2)$ scenario with $m=2$ or $m=3$. 

\subsection{$(N,2,2)$}
For $m=2$, the linear functional \eqref{eq:Iwithuptotwobody} is explicitly
\begin{align}
I=\alpha_{0}\langle\mathcal{S}_{0}\rangle+\alpha_{1}\langle\mathcal{S}_{1}\rangle+\frac{1}{2}\alpha_{0,0}(\langle\mathcal{S}_{0}^{2}\rangle-N)+\alpha_{0,1}(\langle\mathcal{S}_{0}\mathcal{S}_{1}-\mathcal{Z}_{0,1}\rangle)+\frac{1}{2}\alpha_{1,1}(\langle\mathcal{S}_{1}^{2}\rangle-N),
\end{align}
where $\mathcal{S}_0$, $\mathcal{S}_1$ and $\mathcal{Z}_{0,1}$ are classical random variables.
To search for tight Bell inequalities, we enumerate the vertices of the local polytope corresponding to local deterministic strategies according to Sec. \ref{sec:Deterministic local strategy and classical bound}. 
The classical bound $\beta_C$ is obtained as the minimum value of $I$ over the vertices. 

To evaluate the quantum value, the Bell operator according to Eq.~\eqref{eq:Bell_operator I_theta} is explicitly
\begin{align}
\hat{I}(\boldsymbol{\theta})=\alpha_{0}\hat{\mathcal{S}}_{0}+\alpha_{1}\hat{\mathcal{S}}_{1}+\frac{1}{2}\alpha_{0,0}(\mathcal{\hat{S}}_{0}^{2}-N)+\frac{1}{2}\alpha_{0,1}(\mathcal{\hat{S}}_{0}\mathcal{\hat{S}}_{1}+\mathcal{\hat{S}}_{1}\mathcal{\hat{S}}_{0}-2\hat{\mathcal{Z}}_{0,1})+\frac{1}{2}\alpha_{1,1}(\mathcal{\hat{S}}_{1}^{2}-N).\label{eq:IhatN22}
\end{align}
Plugging the measurement settings (Eqs.~\eqref{eq:Shatk} and \eqref{eq: hatZkl}) into Eq.~\eqref{eq:IhatN22}, we obtain
\begin{align}
\begin{aligned}
\hat{I}(\boldsymbol{\theta})&=2(\alpha_{0,0}\cos^{2}\theta_{0}+2\alpha_{0,1}\cos\theta_{0}\cos\theta_{1}+\alpha_{1,1}\cos^{2}\theta_{1})\hat{S}_{z}^{2}+2(\alpha_{0,0}\sin^{2}\theta_{0}+2\alpha_{0,1}\sin\theta_{0}\sin\theta_{1}+\alpha_{1,1}\sin^{2}\theta_{1})\hat{S}_{x}^{2}\\&+2[\alpha_{0,0}\cos\theta_{0}\sin\theta_{0}+\alpha_{1,1}\cos\theta_{1}\sin\theta_{1}+\alpha_{0,1}(\sin\theta_{0}\cos\theta_{1}+\cos\theta_{0}\sin\theta_{1})](\hat{S}_{x}\hat{S}_{z}+\hat{S}_{z}\hat{S}_{x})\\&+2(\alpha_{0}\cos\theta_{0}+\alpha_{1}\cos\theta_{1})\hat{S}_{z}+2(\alpha_{0}\sin\theta_{0}+\alpha_{1}\sin\theta_{1})\hat{S}_{x}-\frac{1}{2}\alpha_{0,0}N-\alpha_{0,1}N\cos(\theta_{0}-\theta_{1})-\frac{1}{2}\alpha_{1,1}N.
\end{aligned}
\end{align}
The minimal quantum value is obtained by searching over $\theta_0$ and $\theta _1$ to minimize the ground state energy of $\hat I(\boldsymbol{\theta}) $.

\begin{center}
\begin{longtable}{c|c|c|c}
    \caption{The ratio and the corresponding coefficients for nontrivial facet Bell inequalities for $N=4$, $m=2$.}\label{table:ratio_N=4} \\
    \hline
    $\Delta_{4,2}$  & $\beta_Q$ & $\beta_C$ & $(\alpha_0,\alpha_1,\alpha_{0,0},\alpha_{0,1},\alpha_{1,1})$ \\
    \hline
    \endfirsthead

    \hline
    \multicolumn{4}{l}{\textit{(continued from previous page)}} \\
    \hline
    $\Delta_{4,2}$  & $\beta_Q$ & $\beta_C$ & $(\alpha_0,\alpha_1,\alpha_{0,0},\alpha_{0,1},\alpha_{1,1})$ \\
    \hline
    \endhead

    \hline
    \multicolumn{4}{r}{\textit{(continued on next page)}} \\
    \endfoot

    \hline
    \endlastfoot
{ $\dfrac{6\sqrt{65}-25}{21}$}
& { $\dfrac{150-36\sqrt{65}}{7}$}
& $-18$ & $(0, 0, 6, 2, -1)$ \\
$1.02579$ & $-30.77363$ & $-30$ & $(12, 3, 6, 2, -1)$ \\
$1.01391$ & $-54.75094$ & $-54$ & $(12, 6, 12, 8, -1)$ \\
$1.01339$ & $-42.56235$ & $-42$ & $(12, 9, 6, 6, 1)$ \\
\end{longtable}
\end{center}

\begin{center}
\begin{longtable}{c|c|c|c}
    \caption{The ratio and the corresponding coefficients for nontrivial facet Bell inequalities for $N=5$, $m=2$. }\label{table:ratio_N=5}\\
    \hline
    $\Delta_{5,2}$  & $\beta_Q$ & $\beta_C$ & $(\alpha_0,\alpha_1,\alpha_{0,0},\alpha_{0,1},\alpha_{1,1})$ \\
    \hline
    \endfirsthead

    \hline
    \multicolumn{4}{l}{\textit{(continued from previous page)}} \\
    \hline
    $\Delta_{5,2}$  & $\beta_Q$ & $\beta_C$ & $(\alpha_0,\alpha_1,\alpha_{0,0},\alpha_{0,1},\alpha_{1,1})$ \\
    \hline
    \endhead

    \hline
    \multicolumn{4}{r}{\textit{(continued on next page)}} \\
    \endfoot

    \hline
    \endlastfoot
$1.05904$ & $-84.72301$ & $-80$ & $(20, 4, 20, 5, -2)$ \\
$1.01769$ & $-24.42458$ & $-24$ & $(6, 6, 1, 2, 1)$ \\
$1.01515$ & $-10.15146$ & $-10$ & $(2, 0, 1, 1, 1)$ \\
$1.01117$ & $-70.78216$ & $-70$ & $(20, 14, 5, 5, 1)$ \\
$1.00805$ & $-201.61020$ & $-200$ & $(60, 24, 30, 15, -2)$ \\
$1.00747$ & $-130.97171$ & $-130$ & $(36, 28, 8, 10, 3)$ \\
$1.00679$ & $-161.08648$ & $-160$ & $(60, 12, 20, 5, -2)$ \\
$1.00138$ & $-400.55077$ & $-400$ & $(60, 36, 60, 45, 2)$ \\
\end{longtable}
\end{center}

\begin{center}
\begin{longtable}{c|c|c|c}
    \caption{The ratio and the corresponding coefficients for nontrivial facet Bell inequalities for $N=6$, $m=2$.}\label{table:ratio_N=6}\\
    \hline
    $\Delta_{6,2}$  & $\beta_Q$ & $\beta_C$ & $(\alpha_0,\alpha_1,\alpha_{0,0},\alpha_{0,1},\alpha_{1,1})$ \\
    \hline
    \endfirsthead

    \hline
    \multicolumn{4}{l}{\textit{(continued from previous page)}} \\
    \hline
    $\Delta_{6,2}$  & $\beta_Q$ & $\beta_C$ & $(\alpha_0,\alpha_1,\alpha_{0,0},\alpha_{0,1},\alpha_{1,1})$ \\
    \hline
    \endhead

    \hline
    \multicolumn{4}{r}{\textit{(continued on next page)}} \\
    \endfoot

    \hline
    \endlastfoot
$1.05884$ & $-63.53040$ & $-60$ & $(0, 0, 15, 3, -1)$ \\
$1.02444$ & $-92.19967$ & $-90$ & $(30, 5, 15, 3, -1)$ \\
$1.02153$ & $-12.25837$ & $-12$ & $(2, 0, 1, 1, 1)$ \\
$1.01628$ & $-16.26044$ & $-16$ & $(4, 2, 1, 1, 1)$ \\
$1.01616$ & $-42.67892$ & $-42$ & $(9, 9, 1, 2, 1)$ \\
$1.00870$ & $-121.04430$ & $-120$ & $(30, 20, 5, 5, 1)$ \\
$1.00804$ & $-226.80917$ & $-225$ & $(54, 41, 8, 10, 3)$ \\
$1.00740$ & $-157.15448$ & $-156$ & $(37, 17, 23, 12, 1)$ \\
$1.00616$ & $-112.69001$ & $-112$ & $(26, 12, 17, 9, 1)$ \\
$1.00544$ & $-48.26112$ & $-48$ & $(13, 5, 4, 3, 2)$ \\
$1.00517$ & $-165.85362$ & $-165$ & $(40, 20, 20, 12, 1)$ \\
$1.00471$ & $-24.11309$ & $-24$ & $(5, 1, 3, 2, 1)$ \\
$1.00408$ & $-39.15927$ & $-39$ & $(12, 3, 4, 2, 1)$ \\
$1.00381$ & $-195.74390$ & $-195$ & $(60, 20, 20, 8, -1)$ \\
$1.00290$ & $-240.69671$ & $-240$ & $(45, 15, 45, 18, -1)$ \\
$1.00224$ & $-70.15691$ & $-70$ & $(18, 12, 3, 3, 1)$ \\
$1.00216$ & $-180.38922$ & $-180$ & $(60, 10, 15, 3, -1)$ \\
$1.00138$ & $-300.41277$ & $-300$ & $(0, 0, 45, 27, 1)$ \\
$1.00065$ & $-129.08436$ & $-129$ & $(24, 20, 12, 12, 5)$ \\
$1.00041$ & $-273.11285$ & $-273$ & $(60, 32, 36, 24, 5)$ \\
\end{longtable}
\end{center}
For $m=2$, the \texttt{cddlib} \cite{fukudaCddlibReferenceManual2003} is used to enumerate facets of the local polytope, and the ratio $\Delta_{N,2}$ is evaluated for each facet Bell inequality. Our gradient optimization converges to a facet Bell inequality in this scenario. 
For $N=3$, there is a single nontrivial facet Bell inequality $(6, 2, 6, 3, -2)$ with $\beta_Q=(150-36\sqrt{65})/{7}=-20.03447$ and  $\Delta_{3,2}^{\mathrm{max}}=(6\sqrt{65}-25)/21=1.11303$.
Up to exchanges of the measurement settings and outcome relabelings,
Tables~\ref{table:ratio_N=4}, \ref{table:ratio_N=5}, and
\ref{table:ratio_N=6} present one representative of each inequivalent
nontrivial facet Bell inequality. We find $4$, $8$, and $20$ inequivalent
classes with $\Delta_{N,2}>1$ for $N=4$, $5$, and $6$, respectively.

\subsection{$(N,3,2)$}

For the $(N,3,2)$ scenario, the linear functional \eqref{eq:Iwithuptotwobody} is 
\begin{align}
\begin{aligned}
    I&=\alpha_{0}\langle\mathcal{S}_{0}\rangle+\alpha_{1}\langle\mathcal{S}_{1}\rangle+\alpha_{2}\langle\mathcal{S}_{2}\rangle+\frac{1}{2}\alpha_{0,0}(\langle\mathcal{S}_{0}^{2}\rangle-N)+\frac{1}{2}\alpha_{1,1}(\langle\mathcal{S}_{1}^{2}\rangle-N)+\frac{1}{2}\alpha_{2,2}(\langle\mathcal{S}_{2}^{2}\rangle-N)\\&+\alpha_{0,1}(\langle\mathcal{S}_{0}\mathcal{S}_{1}-\mathcal{Z}_{0,1}\rangle)+\alpha_{0,2}(\langle\mathcal{S}_{0}\mathcal{S}_{2}-\mathcal{Z}_{0,2}\rangle)+\alpha_{1,2}(\langle\mathcal{S}_{1}\mathcal{S}_{2}-\mathcal{Z}_{1,2}\rangle),
\end{aligned}
\end{align}
and the Bell operator $\hat{I}(\boldsymbol{\theta})$ is constructed similarly to Eq.~\eqref{eq:IhatN22} by introducing $\hat{\mathcal{S}}_{0}=2\cos\theta_{0}\hat{S}_{z}+2\sin\theta_{0}\hat{S}_{x}$, $\hat{\mathcal{S}}_{1}=2\cos\theta_{1}\hat{S}_{z}+2\sin\theta_{1}\hat{S}_{x}$, $\hat{\mathcal{S}}_{2}=2\cos\theta_{2}\hat{S}_{z}+2\sin\theta_{2}\hat{S}_{x}$, $\hat{\mathcal{Z}}_{0,1}=N\cos(\theta_{0}-\theta_{1})$, $\hat{\mathcal{Z}}_{0,2}=N\cos(\theta_{0}-\theta_{2})$, and $\hat{\mathcal{Z}}_{1,2}=N\cos(\theta_{1}-\theta_{2})$.

We list the ratios over all the facet Bell inequalities with $N\leq 4$ using \texttt{cddlib}~\cite{fukudaCddlibReferenceManual2003}. The results of the ratio and the nontrivial facet Bell inequalities are given in Tables~\ref{table:ratio_N=2m=3} and~\ref{table:ratio_N=3m=3} for $N=2$ and $3$. For $N=2$, the maximum $m=3$ ratio reduces to the CHSH value, while for $N=3$ and $N=4$ the three-measurement scenario contains facets with larger ratios than the corresponding two-measurement scenario. For $N = 4$, the ten largest ratios and their corresponding coefficients are shown in Table~\ref{table:ratio_N=4m=3}. The Bell inequality $(0, 0, 0, 9, 3, 2, 3, -2, 0)$ 
has a larger ratio $\Delta_{4,3}^{\max}=1.15997$ compared to $\Delta_{4,2}^{\max}=1.11303$ by $(0,0,6,2,-1)$.
Since \texttt{cddlib}~\cite{fukudaCddlibReferenceManual2003} becomes infeasible for larger values of $N$, we instead employ our gradient-based method to search for Bell inequalities and obtain the inequalities listed in Table II of the main text.

\begin{center}
\begin{longtable}{c|c|c|c}
    \caption{The ratio and the corresponding coefficients for nontrivial facet Bell inequalities for $N=2$, $m=3$.}
    \label{table:ratio_N=2m=3} \\
    \hline
    $\Delta_{2,3}$  & $\beta_Q$& $\beta_C$& $(\alpha_0,\alpha_1,\alpha_2, \alpha_{0,0},\alpha_{0,1},\alpha_{0,2},\alpha_{1,1},\alpha_{1,2},\alpha_{2,2})$ \\
    \hline
    \endfirsthead

    \hline
    \multicolumn{4}{l}{\textit{(continued from previous page)}} \\
    \hline
    $\Delta_{2,3}$  & $\beta_Q$& $\beta_C$& $(\alpha_0,\alpha_1,\alpha_2, \alpha_{0,0},\alpha_{0,1},\alpha_{0,2},\alpha_{1,1},\alpha_{1,2},\alpha_{2,2})$ \\
    \hline
    \endhead

    \hline
    \multicolumn{4}{r}{\textit{(continued on next page)}} \\
    \endfoot

    \hline
    \endlastfoot

$\sqrt{2}$ & $-4\sqrt{2}$ & $-4$ & $(0,0,0,2,1,1,0,-1,0)$ \\

$\sqrt{2}$ & $-4\sqrt{2}$ & $-4$ & $(0,0,0,0,1,1,0,1,-2)$ \\

$\sqrt{2}$ & $-2\sqrt{2}$ & $-2$ & $(0,0,0,1,1,0,-1,0,0)$ \\

$5/4$ & $-10$ & $-8$ & $(2,1,1,2,2,2,-2,1,-2)$ \\

$5/4$ & $-5$ & $-4$ & $(1,1,0,1,1,1,1,-1,0)$ \\
\end{longtable}
\end{center}

\begin{center}
\begin{longtable}{c|c|c|c}
    \caption{The ratio and the corresponding coefficients for nontrivial facet Bell inequalities for $N=3$, $m=3$.}
    \label{table:ratio_N=3m=3} \\
    \hline
    $\Delta_{3,3}$  & $\beta_Q$ & $\beta_C$ & $(\alpha_0,\alpha_1,\alpha_2, \alpha_{0,0},\alpha_{0,1},\alpha_{0,2},\alpha_{1,1},\alpha_{1,2},\alpha_{2,2})$ \\
    \hline
    \endfirsthead

    \hline
    \multicolumn{4}{l}{\textit{(continued from previous page)}} \\
    \hline
    $\Delta_{3,3}$  & $\beta_Q$ & $\beta_C$ & $(\alpha_0,\alpha_1,\alpha_2, \alpha_{0,0},\alpha_{0,1},\alpha_{0,2},\alpha_{1,1},\alpha_{1,2},\alpha_{2,2})$ \\
    \hline
    \endhead

    \hline
    \multicolumn{4}{r}{\textit{(continued on next page)}} \\
    \endfoot

    \hline
    \endlastfoot
$1.14792$ & $-13.77509$ & $-12$ & $(2, 2, 0, 2, 2, 1, 2, -1, 0)$ \\

$1.14261$ & $-11.42606$ & $-10$ & $(2, 2, 0, 2, 1, 1, 2, -1, 0)$ \\

$1.13258$ & $-13.59093$ & $-12$ & $(4, 2, 0, 3, 1, 1, 1, -1, 0)$ \\

$1.13244$ & $-20.38393$ & $-18$ & $(6, 2, 0, 6, 2, 1, 0, -1, 0)$ \\

$1.11776$ & $-40.23940$ & $-36$ & $(12, 2, 2, 12, 4, 2, -1, -2, 1)$ \\

$\dfrac{6\sqrt{65}-25}{21}$
& $\dfrac{150-36\sqrt{65}}{7}$
& $-18$ & $(6, 2, 0, 6, 3, 0, -2, 0, 0)$ \\

$\dfrac{6\sqrt{65}-25}{21}$
& $\dfrac{150-36\sqrt{65}}{7}$
& $-18$ & $(4, 2, 2, 2, 2, 2, 0, 1, -2)$ \\

$1.10156$ & $-33.04687$ & $-30$ & $(8, 2, 2, 8, 4, 2, -1, -2, 3)$ \\

$1.10033$ & $-46.21398$ & $-42$ & $(6, 6, 4, 6, 6, 3, 0, 3, -4)$ \\

$1.09124$ & $-22.91614$ & $-21$ & $(6, 2, 0, 7, 3, 1, -1, -1, 1)$ \\

$1.08496$ & $-15.18940$ & $-14$ & $(4, 4, 0, 2, 1, 1, 2, -1, 0)$ \\

$1.06941$ & $-38.49887$ & $-36$ & $(6, 4, 2, 6, 6, 3, -2, 2, -2)$ \\

$1.06475$ & $-22.35971$ & $-21$ & $(6, 2, 0, 6, 3, 1, -1, -1, 2)$ \\

$1.05625$ & $-47.53139$ & $-45$ & $(12, 6, 2, 12, 3, -3, 2, 5, 1)$ \\

$1.05099$ & $-63.05926$ & $-60$ & $(6, 6, 4, 12, 9, 6, -2, 4, -4)$ \\

$1.05051$ & $-18.90921$ & $-18$ & $(4, 2, 2, 2, 2, 2, 3, -2, 1)$ \\

$1.05047$ & $-22.05980$ & $-21$ & $(6, 2, 0, 6, 1, 2, 1, -2, 0)$ \\

$1.03970$ & $-11.43666$ & $-11$ & $(2, 2, 0, 2, 0, 1, 2, -1, -1)$ \\

$1.03910$ & $-28.05576$ & $-27$ & $(6, 4, 2, 8, 3, -2, 0, 2, 1)$ \\

$1.03825$ & $-37.37713$ & $-36$ & $(4, 2, 2, 8, 5, 5, -2, 2, -2)$ \\

$1.03806$ & $-37.37011$ & $-36$ & $(12, 4, 4, 6, 3, 3, -2, 1, -2)$ \\

$1.02985$ & $-30.89545$ & $-30$ & $(6, 2, 0, 3, 3, 3, -2, 3, 3)$ \\

$1.02831$ & $-30.84942$ & $-30$ & $(6, 2, 0, 4, 3, 4, -2, 2, 2)$ \\

$1.02767$ & $-43.16210$ & $-42$ & $(10, 8, 4, 3, 4, 4, 7, -4, 2)$ \\

$1.02745$ & $-24.65877$ & $-24$ & $(8, 4, 2, 3, 2, 2, 2, -2, 1)$ \\

$1.02498$ & $-30.74945$ & $-30$ & $(4, 2, 2, 0, 4, 2, 6, 3, -2)$ \\

$1.01967$ & $-21.41316$ & $-21$ & $(4, 4, 2, 5, 2, -2, 1, 2, 1)$ \\

$1.00967$ & $-18.17402$ & $-18$ & $(4, 2, 2, 5, 2, -1, -1, 1, 2)$ \\

$1.00897$ & $-36.32307$ & $-36$ & $(6, 2, 0, 2, 3, 4, -2, 4, 6)$ \\

$1.00879$ & $-18.15819$ & $-18$ & $(4, 2, 2, 4, 2, -1, -1, 1, 3)$ \\

$1.00819$ & $-36.29491$ & $-36$ & $(6, 6, 4, 9, 3, -3, 1, 4, 2)$ \\
\end{longtable}
\end{center}

\begin{center}
\scriptsize
\begin{longtable}{c|c|c|c}
\caption{The ten largest ratios and the corresponding coefficients for nontrivial facet Bell inequalities for $N=4$, $m=3$.
}\label{table:ratio_N=4m=3}\\
\hline
$\Delta_{4,3}$  & $\beta_Q$ & $\beta_C$ & $(\alpha_0,\alpha_1,\alpha_2, \alpha_{0,0},\alpha_{0,1},\alpha_{0,2},\alpha_{1,1},\alpha_{1,2},\alpha_{2,2})$ \\
\hline
\endfirsthead
\hline
\multicolumn{4}{l}{\textit{(continued from previous page)}} \\
\hline
$\Delta_{4,3}$  & $\beta_Q$ & $\beta_C$ & $(\alpha_0,\alpha_1,\alpha_2, \alpha_{0,0},\alpha_{0,1},\alpha_{0,2},\alpha_{1,1},\alpha_{1,2},\alpha_{2,2})$ \\
\hline
\endhead
\hline
\multicolumn{4}{r}{\textit{(continued on next page)}} \\
\endfoot
\hline
\endlastfoot
$1.15997$ & $-41.75878$ & $-36$ & $(0, 0, 0, 9, 3, 2, 3, -2, 0)$ \\

$1.14651$ & $-18.34417$ & $-16$ & $(0, 0, 0, 4, 1, 1, 2, -1, 0)$ \\

$1.13495$ & $-27.23888$ & $-24$ & $(0, 0, 0, 4, 3, 1, 2, -1, 0)$ \\

$1.13197$ & $-18.11159$ & $-16$ & $(1, 1, 0, 3, 1, 1, 3, -1, 0)$ \\

$\dfrac{83+16\sqrt{19}}{135}$
& $-\dfrac{332+64\sqrt{19}}{15}$
& $-36$ & $(0, 0, 0, 12, 3, 1, 0, -1, 0)$ \\

$1.12289$ & $-26.94937$ & $-24$ & $(3, 3, 0, 3, 3, 1, 3, -1, 0)$ \\

$1.12100$ & $-80.71223$ & $-72$ & $(0, 0, 0, 21, 6, 3, -1, -3, 4)$ \\

$1.11762$ & $-118.46796$ & $-106$ & $(2, 1, 0, 6, 4, 4, -2, -9, 31)$ \\

$1.11760$ & $-33.52793$ & $-30$ & $(0, 0, 0, 3, 3, 2, 3, 0, -1)$ \\

$1.11575$ & $-120.50049$ & $-108$ & $(0, 0, 0, 36, 9, 3, -2, -3, 2)$ \\
\end{longtable}
\end{center}

\bibliographystyle{apsrev4-1}
\bibliography{references}